%% file: main_arxiv.tex
\documentclass{article}

\usepackage{iclr_conference_arxiv,times}
\iclrfinalcopy  %

\usepackage[utf8]{inputenc} %
\usepackage[T1]{fontenc}    %
\usepackage{hyperref}       %
\usepackage{url}            %
\usepackage{booktabs}       %
\usepackage{amsfonts}       %
\usepackage{nicefrac}       %
\usepackage{microtype}      %
\usepackage{xcolor}         %
\usepackage{svg}
\usepackage[abs]{overpic}
\usepackage{longtable}
\usepackage{nameref}
\usepackage[shortlabels]{enumitem}
\usepackage{textgreek}
\usepackage{float}
\usepackage{graphicx}
\usepackage{color}
\usepackage{amsmath}
\usepackage{amssymb}
\usepackage{amsthm}
\usepackage{mathtools}
\usepackage{thmtools}
\usepackage[mathscr]{eucal}
\usepackage{bm}
\usepackage{bbm}
\usepackage{relsize}
\usepackage{graphics}
\usepackage{wrapfig}
\usepackage{multirow}
\usepackage{multicol}
\usepackage{enumitem} 
\usepackage{array}
\usepackage{adjustbox}
\usepackage{tocloft}
\usepackage{pgf}
\usepackage{xinttools}
\usepackage{tikz}
\usepackage{comment}
\usepackage{bbding}
\usepackage{pifont}
\usetikzlibrary{arrows.meta}
\usepackage{textcomp}
\usetikzlibrary{calc,shapes,arrows,positioning,automata,trees}
\usepackage{placeins} 
\usepackage{tabularx}
\usepackage{graphicx}
\usepackage{subcaption} 
\usepackage[para]{threeparttable}
\usepackage{listings}
\usepackage{xargs,lipsum,caption,changepage,ifthen}
\usepackage{tikz} 
\usepackage{makecell}
\usepackage{siunitx}
\usepackage{footmisc}
\usepackage{sidecap}
\usepackage{ml-defs}
\usepackage{multirow}
\usepackage{acronym}  

\definecolor{mLSTM}{HTML}{3073AD}    %
\definecolor{sLSTM}{HTML}{4B9D7A}    %
\definecolor{m_s_Hybrid}{HTML}{4FA5AA}  %
\definecolor{Mamba}{HTML}{DF8953}    %
\definecolor{Hyena}{HTML}{E86A61}    %
\definecolor{S4}{HTML}{D275AB}       %
\definecolor{Transformers}{HTML}{B6B6B6} %

\definecolor{xLSTMRed}{RGB}{229, 46, 102}
\definecolor{xLSTMBlue}{RGB}{22, 91, 137}

\definecolor{linkcol}{HTML}{3073AD}  %
\definecolor{citecol}{HTML}{3073AD} %
\definecolor{urlcol}{HTML}{3073AD} %

\definecolor{lightblue}{RGB}{32, 194, 217}
\definecolor{jku_red}{RGB}{217, 92, 76}
\definecolor{jku_blue}{RGB}{0, 132, 187}
\definecolor{jku_green}{RGB}{91, 167, 85} %
\definecolor{jku_yellow}{RGB}{241, 188, 63}
\definecolor{jku_cyan}{RGB}{79,176,191}
\definecolor{jku_grey}{RGB}{125,130,140}
\definecolor{jku_lightgreen}{RGB}{191,206,82}
\definecolor{jku_violett}{RGB}{174,97,157}

\hypersetup{
   colorlinks=true,
   linkcolor=linkcol,
   citecolor=citecol,
   urlcolor=urlcol,
}

\newcommand{\stoptocwriting}{%
  \addtocontents{toc}{\protect\setcounter{tocdepth}{-5}}}
\newcommand{\resumetocwriting}{%
  \addtocontents{toc}{\protect\setcounter{tocdepth}{\arabic{tocdepth}}}}

\newcommand{\gcol}{\rowcolor{jku_grey!10}}

\acrodef{rnn}[RNN]{recurrent neural network}
\acrodef{ssm}[SSM]{state-space model}
\acrodef{llm}[LLM]{large language model}
\acrodef{mlm}[MLM]{masked language modeling}
\acrodef{clm}[CLM]{causal language modeling}
\acrodef{icl}[ICL]{in-context learning}
\acrodef{fim}[FIM]{fill-in the middle}
\acrodef{ntp}[NTP]{next token prediction}
\acrodef{smiles}[SMILES]{Simplified Molecular Input Line Entry System}
\acrodef{bp}[bp]{base-pair}
\acrodef{fcd}[FCD]{Fréchet ChemNet Distance}
\acrodef{rc}[RC]{reverse complement}
\acrodef{ph}[PH]{post-hoc conjoining}
\acrodef{ps}[PS]{parameter sharing}
\acrodef{ar}[AR]{autoregressive}
\acrodef{rope}[RoPE]{Rotary Position Encodings}
\acrodef{ppl}[ppl]{perplexity}
\acrodef{mcc}[MCC]{Matthews correlation coefficient}

\title{Bio-xLSTM: Generative modeling, representation
and in-context learning
of biological and chemical sequences}

\author{%
    Niklas Schmidinger$^{1}$ \quad
    Lisa Schneckenreiter$^{1}$ \quad
    Philipp Seidl$^{1}$ \quad
    Johannes Schimunek$^{1}$ \\
    \textbf{Pieter-Jan Hoedt}$^{1}$ \quad
    \textbf{Johannes Brandstetter}$^{1,2}$ \quad
    \textbf{Andreas Mayr}$^{1}$ \\
    \textbf{Sohvi Luukkonen}$^{1}$ \quad
    \textbf{Sepp Hochreiter}$^{1,2}$ \quad
    \textbf{Günter Klambauer}$^{1,2}$ \\ \\
    $^{1}$ ELLIS Unit Linz and LIT AI Lab, Institute for Machine Learning,\\
    Johannes Kepler University, Linz, Austria\\
    $^{2}$ NXAI GmbH, Linz, Austria
}

\begin{document}
\maketitle

\stoptocwriting

\begin{abstract}
Language models for biological and chemical sequences 
enable crucial applications such as drug discovery, protein engineering, and precision medicine. 
Currently, these language models are predominantly based on Transformer architectures. 
While Transformers have yielded impressive results, their quadratic runtime dependency on the sequence length complicates their use for long genomic sequences and in-context learning on proteins and chemical sequences. 
Recently, the recurrent {xLSTM} architecture has been shown to perform favorably compared to Transformers and modern \ac{ssm} architectures
in the natural language domain. 
Similar to \acp{ssm}, xLSTMs have a linear runtime dependency on the sequence length and allow for constant-memory decoding at inference time, 
which makes them prime candidates for modeling long-range dependencies in biological and chemical sequences.
In this work, we tailor xLSTM towards these domains and propose a suite of architectural variants called Bio-xLSTM. 
Extensive experiments in three large domains, genomics, proteins, and chemistry, were performed to assess xLSTM's ability to model biological and chemical sequences. 
The results show that models based on Bio-xLSTM 
a) can serve as proficient generative models for 
DNA, protein, and chemical sequences, 
b) learn rich representations for those modalities, and
c) can perform in-context learning for proteins and small molecules.
\end{abstract}

\acresetall

\section{Introduction}
\begin{figure}
    \centering
    \includegraphics[width=1.0\textwidth]{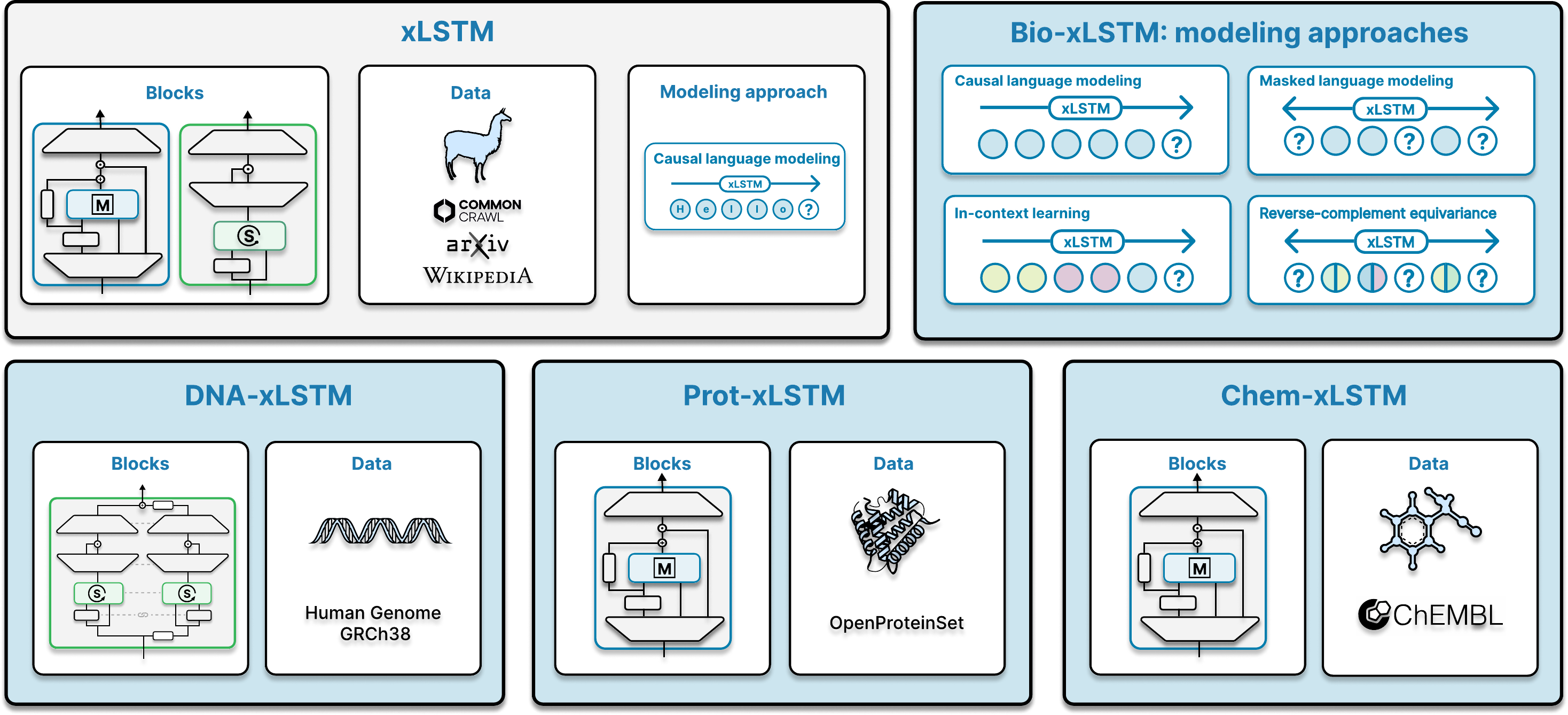}
    \caption{Overview of Bio-xLSTM. 
    \textbf{Top~left:} xLSTM for natural language processing tasks. 
    \textbf{Top~right:} Considered modeling approaches 
    for biological sequences: 
        masked language modeling, 
        equivariance to reverse complementary sequence, 
        and in-context learning. 
    \textbf{Bottom left:} DNA-xLSTM models are trained on genomic DNA sequences and then fine-tuned on downstream tasks.
    \textbf{Bottom~center:} Prot-xLSTM models are trained in a causal modeling setting with a fill-in-the-middle objective and use homologous proteins for in-context learning.
      \textbf{Bottom~right:} Chem-xLSTM models are trained to generate small molecules. 
      For an in-context learning setting, Chem-xLSTM models use 
      molecules with known properties.
    }
    \label{fig:overview}
\end{figure}

\textbf{Accurate computational models for biological sequences are essential for translating data into actionable insights in modern biology.} Biological sequences like DNA, RNA, and proteins 
are central to molecular biology, genomics, and drug discovery. 
Major projects like the Human Genome Project \citep{lander2001initial} 
and the 1000 Genomes Project \citep{1000genomes2010} have driven large-scale data collection efforts. 
Modeling these sequences is key to advancing life sciences \citep{benegas2023dna,karollus2024species}, 
interacting with biological systems \citep{hopf2017mutation,riesselman2018deep,yang2019protein} 
or predicting phenotypes from genetic variants 
\citep{ashley2016towards,brandes2023genome,acosta2022multimodal}. 
Similar efforts exist for protein sequences \citep{uniprot2023uniprot} and 
small molecules \citep{kim2023pubchem,zdrazil2023chembl}, 
used for tasks like protein engineering \citep{arnold2018directed,yang2019protein}, 
predicting 3D structures \citep{jumper2021highly}, and drug discovery \citep{zhavoronkov2019deep}. 
\Acp{llm} \citep{brown20gpt3,bubeck2023sparks} have emerged as prime candidates 
for modeling biological sequences and serving as foundation models for 
molecular biology and chemistry \citep{ji2021dnabert,schiff2024caduceus,nguyen2023hyenadna,rives2021biological,lin2023evolutionary}.

\textbf{Large language models for biological sequences must handle long sequences and incorporate context.} 
The rise of \acp{llm} \citep{radford2018gpt,brown20gpt3,bubeck2023sparks} 
has revolutionized numerous fields, including life sciences. 
Most \acp{llm} are based on the Transformer architecture \citep{vaswani2017attention}, 
which excels at predicting the next or missing token using self-attention. 
However, the self-attention mechanism scales quadratically with sequence length, 
making long-sequence modeling computationally expensive. 
As a result, most biological sequence models use short contexts \citep{rives2021biological,ji2021dnabert,dalla2023nucleotide}. 
However, biological sequences require long context windows for accurate modeling because 
of their important long-range interactions due to 3D folding \citep{anfinsen1973principles}, 
or gene regulation in DNA \citep{bouwman2015getting}. 
The human genome spans around three billion \acp{bp}, 
far exceeding the context limits of Transformer-based models.
Furthermore, long contexts also benefit models that exploit homologous proteins \citep{truong2023poet, sgarbossa2024protmamba} and molecular context for small molecules \citep{papadatos2010lead, schimunek2023context}. 
The emergence of \acp{ssm}, like S4 \citep{gu2022s4}, Hyena \citep{poli2023hyena}, and Mamba \citep{gu2024mamba}, enables handling longer contexts in biological domains \citep{nguyen2023hyenadna, schiff2024caduceus, sgarbossa2024protmamba}. 
However, the recently proposed xLSTM architecture \citep{beck2024xlstm}, 
a recurrent neural network, 
has outperformed \ac{ssm} architectures in natural language processing \citep{beck2024xlstm}. 
For further related work, see Appendix Section~\ref{sec:related_work}.

\textbf{The recently proposed xLSTM is a powerful architecture 
for sequence modeling and a promising candidate for biological and chemical sequences.} 
The xLSTM architecture \citep{beck2024xlstm} introduces enhanced 
memory structures and exponential gates that boost its 
performance, particularly in natural language modeling. 
Despite these enhancements over traditional LSTM, xLSTM retains the efficiency of a recurrent neural network 
and can handle varying sequence lengths effectively \citep{beck2024xlstm}, 
while maintaining expressivity \citep{merrill2024illusion} 
and scalability \citep{katharopoulos2020transformers,choromanski2021rethinking}. 
These features make xLSTM ideal for modeling:
i) DNA sequences, which are inherently long and for which long-range interactions between distant parts of the sequence have been observed,
ii) protein sequences, where modeling
strongly benefits from contextual information 
of evolutionary-related proteins \citep{rives2021biological}, and 
iii) small molecules represented as chemical sequences, such as \ac{smiles} \citep{weininger1988smiles}, 
for which \ac{icl} abilities are an option to generate new molecules with desired properties or from a particular molecular domain \citep{segler2018generating,schimunek2023context}.
However, it remains unclear 
how to best tailor xLSTM for biological and chemical sequences
and how xLSTM compares to other domain-specific \acp{llm} architectures. 

We introduce: 
a) DNA-xLSTM, an architectural variant tailored for DNA sequences with reverse-complement equivariant blocks, and evaluate its performance on long-context generative modeling, representation learning, and downstream tasks. 
b) Prot-xLSTM, a variant for homology-aware protein language modeling with in-context learning for both generative modeling and in-painting, which we benchmark on generative modeling, conditioned protein design and protein variant fitness prediction tasks. 
c) Chem-xLSTM, an architectural variant for \ac{smiles} representations 
of small molecules for which we demonstrate \ac{icl} capabilities.
An overview of Bio-xLSTM is shown in Fig.~\ref{fig:overview}.

\section{Background and Notation}
\label{sec:background}

xLSTM \citep{beck2024xlstm} consists of two types of layers: 
sLSTM (see Section~\ref{sec:sLSTM})
and mLSTM  (see Section~\ref{sec:mLSTM})
which are the main components within
block structures  (see Section~\ref{sec:block})
of its multi-layer architectures.
We consider a series of 
input vectors $\Bx_t \in \reals^{D}$ 
given at a certain time step $t \in \{1,\ldots,T\}$. 
$\BX=\BX_{1:T}=(\Bx_1,\Bx_2,\ldots,\Bx_T) \in \reals^{D \times T}$
denotes the matrix of stacked input vectors from all time steps. 
Both sLSTM and mLSTM are recurrent neural networks, which either map a 
state $(\Bh_{t-1}, \Bc_{t-1}, \Bn_{t-1})$ to 
a successor state $(\Bh_{t}, \Bc_{t}, \Bn_{t})$ 
given an input $\Bx_{t-1}$ (sLSTM) or 
a state $(\Bh_{t-1}, \BC_{t-1}, \Bn_{t-1})$ to a 
successor state $(\Bh_{t}, \BC_{t}, \Bn_{t})$ given 
an input $\Bx_{t-1}$ (mLSTM). 
Here, $\Bh_{t} \in \reals^{d}$ denotes a hidden state, 
$\Bc_t \in \reals^{d}$ and $\BC_t \in \reals^{d \times d}$ 
denote cell states
responsible for long-term memory
and, $\Bn_{t}  \in \reals^d$ denotes a normalizer state. 
sLSTM and mLSTM utilize several adjustable 
weight matrices and bias vectors (detailed equations below) 
and employ input-, output-, and forget-gates, 
activated by exponential ($\exp$) or the sigmoid functions ($\sigma$). 
For cell inputs in sLSTM, the hyperbolic tangent function 
($\tanh$, abbreviated as~$\varphi$) is used as an activation function.

\subsection{sLSTM}
\label{sec:sLSTM}

The forward pass of sLSTM in the vectorized version is defined as follows:
\begin{align}
\Bc_t \ &= \  \bff_t \odot \Bc_{t-1} \ + \ \bfi_t \odot \Bz_t &  & &\text{cell state} \\
\Bn_t \ &= \  \bff_t \odot \Bn_{t-1} \ + \ \bfi_t &  & &\text{normalizer state} \\
\Bh_t  \ &= \ \bfo_t \odot \tilde{\Bh}_t \ , 
  & \tilde{\Bh}_t \ &= \ \Bc_t  \odot \Bn_t ^{-1}
  &\text{hidden state} \\
\Bz_t \ &= \ \varphi \left( \tilde{\Bz}_t \right) \ , 
  &\tilde{\Bz}_t \ &=  \ \BW_{\Bz} \ \Bx_t \ + \
  \BR_{\Bz}  \ \Bh_{t-1} \ + \  \Bb_{\Bz} \ \
  &\text{cell input} \\
\bfi_t\ &= \  \exp \left( \tilde{\bfi}_t  \right) \ , 
  &\tilde{\bfi}_t \ &= \ \BW_{\bfi} \ \Bx_t \ + \
  \BR_{\bfi}  \ \Bh_{t-1} \ + \  \Bb_{\bfi} \ \
  &\text{input gate} \\
\bff_t \ &= \ \exp \left( \tilde{\bff}_t \right) \ \text{OR} \ \sigma \left(  \tilde{\bff}_t \right) \ , 
  &\tilde{\bff}_t \ &= \ \BW_{\bff} \ \Bx_t  \ + \
  \BR_{\bff}  \ \Bh_{t-1} \ + \  \Bb_{\bff} \ \
  &\text{forget gate} \\
\bfo_t \ &= \ \sigma \left( \tilde{\bfo}_t \right) \ , 
  &\tilde{\bfo}_t  \ &= \ \BW_{\bfo} \ \Bx_t \ + \
  \BR_{\bfo}  \ \Bh_{t-1} \ + \  \Bb_{\bfo} \ \
  &\text{output gate},
\end{align}
where $\bfi_t ,\bfo_t, \bff_t \in \reals^d$ are the input, 
output and forget gate, respectively,
$\BW_{\Bz}, \BW_{\bfi},\BW_{\bff},\BW_{\bfo} \in \reals^{d \times D}$, 
$\BR_{\Bz}, \BR_{\bfi},\BR_{\bff},\BR_{\bfo} \in \reals^{d \times d}$,
and $\Bb_{\Bz}, \Bb_{\bfi}, \Bb_{\bff}, \Bb_{\bfo} \in \reals^d$
are trainable weight matrices and biases.

\subsection{mLSTM}
\label{sec:mLSTM}

The forward pass of the mLSTM is defined as follows:
\begin{align}
\label{eq:mlstm_recurrent_begin}
\BC_t \ &= \  \Rf_t \ \BC_{t-1} \ + \ 
  \Ri_t \ \Bv_t \ \Bk_t^\top & &  &\text{cell state} \\
\Bn_t \ &= \  \Rf_t \ \Bn_{t-1} \ + \ 
\Ri_t \ \Bk_t & &  &\text{normalizer state} \\
\Bh_t  \ &= \ \bfo_t \ \odot \ \tilde{\Bh}_t
\ , \qquad \qquad 
\tilde{\Bh}_t \ = \   \BC_t \Bq_t \ / \ 
  \max \left\{ \ABS{\Bn_t^\top \Bq_t}, 1 \right\} 
   &&&\text{hidden state} \\
\Bq_t \ &= \ \BW_q \ \Bx_t \ + \ \Bb_q  & &  &\text{query input} \\
\Bk_t \ &= \ \frac{1}{\sqrt{d}} \BW_k \ \Bx_t \ + \ \Bb_k  & &  &\text{key input} \\
\Bv_t \ &= \ \BW_v \ \Bx_t \ + \ \Bb_v  & &  &\text{value input} \\
\Ri_t \ &= \ \exp \left( \tilde{\Ri}_t  \right) \ , 
 \qquad \qquad \ \ \ \ \,
  \tilde{\Ri}_t \ = \ \Bw^\top_{\Ri} \ \Bx_t \ + \  b_{\Ri} \ \ \ 
  &&&\text{input gate} \\
\Rf_t \ &= \ \sigma \!  \left(  \tilde{\Rf}_t \right) \ \text{OR} \ \exp \!  \! \left(  \tilde{\Rf}_t \right) , \ \ \: \:
\tilde{\Rf}_t \ = \ \Bw^\top_{\Rf} \ \Bx_t  \ + \
  b_{\Rf}
  &&& \text{forget gate} \\
\label{eq:mlstm_recurrent_end}
\bfo_t \ &= \ \sigma \left( \tilde{\bfo}_t \right) \ , \qquad \qquad \quad \ \ \ \,
  \tilde{\bfo}_t  \ = \ \BW_{\bfo} \ \Bx_t \ + \
  \Bb_{\bfo}
  &&&\text{output gate} 
\end{align}
where $\Ri_t ,\Ro_t, \Rf_t \in \reals$ are the input, 
output and forget gate, respectively, 
$\Bq_t,\Bk_t, \Bv_t \in \reals^d$ are query, key and 
value inputs with trainable weight matrices $\BW_q,\BW_k,\BW_v \in \reals^{d \times D}$,
$\Bw_{\Ri}, \Bw_{\Rf} \in \reals^D$ 
are input and forget gate weights and 
the respective $b_{\Ri},b_{\Rf} \in \reals$ biases.
All other quantities are identical to sLSTM. 

\subsection{Block Structures}
\label{sec:block}
The sLSTM and mLSTM layers are integrated into 
larger residual backbones \citep{srivastava2015highway,he2016resnet}, 
which incorporate layer normalization \citep{ba2016layernorm}, 
pre- or post-up projection layers \citep{vaswani2017attention, dao2023flashattention2},
with short causal convolutions 
and group normalization \citep{wu2020groupnorm}. 
Figure \ref{fig:xlstm_block_diagram} depicts sLSTM and mLSTM blocks, 
as well as, a bidirectional mLSTM configuration with weight-tied blocks.
For more details, refer to \citep[Sec.2.4]{beck2024xlstm}.
For Bio-xLSTM we retain these basic building blocks but adjust them to the respective domains. 
The entire architecture, including all layers, normalization, blocks, 
and other components, defines a mapping from an input sequence of length $t$ to an output sequence. This mapping is denoted as $\mathrm{xLSTM}: \reals^{D \times t} \mapsto \reals^{D \times t}$, 
where $\mathrm{xLSTM}$ transforms the stacked inputs up to time step $t$, i.e., $\BX_{1:t} \coloneqq (\Bx_1,\Bx_2,\ldots,\Bx_t) \in \reals^{D \times t}$, to the corresponding stacked outputs of sequence length $t$,  i.e., $\BY_{1:t} \coloneqq (\By_1,\By_2,\ldots,\By_t) \in \reals^{D \times t}$ \footnote{Here $\Bx_i$ and $\By_i$ represent the inputs to and outputs from a particular model from an instance of an xLSTM architecture, rather than the inputs and outputs of a specific sLSTM or mLSTM layer}. 
The $i$-th sequence element is denoted with the subscript $i$, e.g. the $i$-th element from $\BX_{1:t}$ would be $(\BX_{1:t})_{i}$.
Similarly to the mapping $\mathrm{xLSTM}$, we also define mappings for the sequence-wise input-/output behaviour of layers themselves for an sLSTM layer ($\mathrm{sLSTM} : \reals^{D \times t} \mapsto \reals^{D \times t}$) or an mLSTM  layer ($\mathrm{mLSTM} : \reals^{D \times t} \mapsto \reals^{D \times t}$). 
If the specific parameters used for the mapping are unclear, we will denote this by including a second argument in the function, separated by a semicolon.
For details on the block structures, see Appendix Section~\ref{app-sec:blocks_details}.

\subsection{Modes of Operation: Parallel, Chunkwise, and Recurrent}
\label{sec:xlstm_modes_of_operation}

The recurrent forms of sLSTM and mLSTM, introduced in Sections \ref{sec:sLSTM} and \ref{sec:mLSTM}, provide efficient, constant-memory decoding during inference. This eliminates the need for expensive key-value caching, which represents a major challenge for Transformer models in long-range settings. Like Transformers, mLSTM allows for parallelization across the sequence length which significantly speeds up training. Additionally, similar to linear attention variants \citep{katharopoulos2020transformers, yang2024gated}, mLSTM supports chunkwise parallel processing, blending recurrent and parallel modes 
\citep{beck2025unlocking}. This approach is especially advantageous for long-sequence training and prompt encoding. For further details, refer to Appendix \ref{sec:mlstm_forms}.

\section{Bio-xLSTM: Longe-Range Modeling of Biological and Chemical Sequences}
\label{sec:bio-xlstm}

Bio-xLSTM introduces three xLSTM-based architectural variants tailored 
specifically to DNA (Section~\ref{sec:dna-xlstm}), 
proteins (Section~\ref{sec:prot-xlstm}) and
small molecules (Section~\ref{sec:chem-xlstm}). 
For these application domains, we extend xLSTM from \acf{clm} 
(Section~\ref{sec:clm}) to new modeling approaches such as \acf{fim} , \acf{icl} 
and \acf{mlm} (Section~\ref{sec:MLM}).

\subsection{Causal Language Modeling and Next-Token Prediction}\label{sec:clm}

\Acf{clm} uses the 

\begin{align} \label{eq:clm_xlstm}
   \text{CLM loss:  } & \mathcal L^{\mathrm{CLM}}= \EXP_{\BX \sim p_{\BX}} \ \EXP_{t \sim [[1,T-1]]} \ \mathrm{CE} \left(\Bx_{t+1}, \mathrm{xLSTM} (\BX_{1:t})_{t} \right) ,
\end{align}

where $\mathrm{CE}$ is the cross-entropy loss (with logits), 
$p_{\BX}$ is the data distribution, and $[[1,T-1]]$ is the discrete
uniform distribution from $1$ to $T-1$. The objective
measures how well a particular sequence token $\Bx_{t+1}$
can be predicted based on the previous tokens $\BX_{1:t}$ 
by the model
$\mathrm{xLSTM}: \reals^{D \times t} \mapsto \reals^{D \times t}$.
Therefore, this type of modeling is sometimes 
also called \emph{\ac{ntp}}, \emph{uni-directional modeling} or 
\emph{\ac{ar} modeling} and the loss is also called \emph{\ac{ntp} loss}.

\textbf{\Acf{fim}} \label{sec:FIM}
\citep{bavarian2022efficienttraininglanguagemodels}
is a modeling paradigm 
that integrates aspects of both \ac{clm} and \ac{mlm}. 
In this approach, parts of the sequence are replaced with mask tokens, 
which are then appended to the end of the sequence. 
This allows the model to utilize the entire context to predict the 
masked tokens while maintaining an \ac{ar} training framework.
This strategy, allows the model to perform both a) generative modeling and b) inpainting with \ac{clm}.

\textbf{\Acf{icl}} \label{sec:ICL}
is a capability of language models to learn and perform tasks by leveraging additional information provided as the contextual input without updating their parameters \citep{brown20gpt3, min2022rethinking}. 
This approach allows models to learn from analogy, drawing insights from patterns in the context to adapt their behavior. 
When the contextual input sequence is denoted as $\BZ \in \reals^{D \times S}$ and the conventional input sequence is denoted as $\BX \in \reals^{D \times T}$, then with an \ac{icl} model, we obtain $\BY' = \mathrm{xLSTM} ([\BZ,\BX])_{(S+1) : (S+T)}$, where $[\BZ,\BX]$ indicates concatenation of contextual and conventional inputs, and, the subscript ${(S+1):(S+T)}$ indicates that the last output tokens (those corresponding to the $\BX$ tokens) are selected. 
For natural language processing tasks, the context $\BZ$ often contains the solution to a similar problem or exemplary solutions that guide the input and inform the model. 
For biological and chemical sequences, the context $\BZ$ could represent similar genetic regions, homologous proteins, or molecules with desired properties, enabling the model to make informed predictions or generate outputs based on these analogies.

\subsection{Masked Language Modeling (MLM)} \label{sec:MLM}
Bio-xLSTM extends xLSTM to masked modeling of biological sequences,
for which the typical de-masking
or de-noising objective \citep{vincent2010denoisingae,devlin2019bert}
is used, concretely the 

\begin{align} \label{eq:mlm_xlstm} 
   \text{MLM loss:  } & \mathcal L^{\mathrm{MLM}}= \EXP_{\BX \sim p_{\BX}} \ \EXP_{t \sim [[1,T]]} \ \EXP_{\BM \sim p_{\BM}} \ \mathrm{CE} \left(\Bx_t, \mathrm{xLSTM} (\BX \odot \BM)_t \right) ,
\end{align}

where $\BM \in \{0,1\}^{D \times T}$ is a random matrix with binary entries which are
usually drawn from a Bernoulli distribution $p_{\BM}$, and
$\odot$ is element-wise multiplication. The objective
measures how well the original sequence $\BX$
can be reconstructed from a noisy version $\BX \odot \BM$ 
by the model
$\mathrm{xLSTM}: \reals^{D \times T} \mapsto \reals^{D \times T}$.
This modeling paradigm has also been called \emph{bidirectional modeling}.
It has been highly successful in learning
representations of proteins at evolutionary scale \citep{rives2021biological},
which has powered many subsequent applications such as protein
engineering and machine-learning guided directed evolution
\citep{yang2019protein}. For details on how xLSTM is extended to the MLM setting, we refer to Appendix Section \ref{sec:bidirectional_modelling}.

\subsection{\Acf{rc} equivariance} 
\label{sec:RC}
We develop an xLSTM block that is equivariant
to the \ac{rc} of an input sequence, a property particularly relevant to DNA-based applications.
In double-helix DNA structures, both strands are semantically equivalent,
with one strand being the \ac{rc} of the other.
The \ac{rc} strand is oriented in the opposite direction of the \emph{forward} strand, 
with base pairs converted from \texttt{A}~to~\texttt{T} and \texttt{C}~to~\texttt{G}.
\citet{shrikumar2017reverse} show that a data-driven approach to 
learn the equivalence between \ac{rc} sequences can fail. 
Therefore, \citet{schiff2024caduceus}
propose to enforce \ac{rc}-equivariance by design, 
making use of two different inductive 
biases,  \acf{ph} \citep{zhou2022towards} and \acf{ps}, in the architecture.
In \ac{ph} architectures, the backbone is trained to handle 
both DNA sequences and their \acp{rc} by applying \ac{rc} augmentations during pre-training.
For downstream tasks, \ac{ph} architectures are applied 
to both the original sequence and its \ac{rc}, and 
their outputs are summed to reach overall RC invariance.
In contrast, \ac{ps} architectures  
integrate \ac{rc}-equivariant xLSTM blocks with equivariant 
word embeddings and language model heads similar to \citet{schiff2024caduceus}. 
For additional details, see Appendix~Section~\ref{sec:app_dna_reverse_complement}.

\subsection{DNA-xLSTM} \label{sec:dna-xlstm}

For the DNA domain, we propose the DNA-xLSTM architecture
to enhance sequence modeling capabilities, particularly for varying context lengths. 
We introduce three model configurations based on DNA-xLSTM: 
two sLSTM-based configurations trained with a context window of 1,024 tokens (DNA-xLSTM-500k and DNA-xLSTM-2M), 
and an mLSTM-based configuration trained with a context window of 32,768 tokens (DNA-xLSTM-4M). 
The short-context configuration, DNA-xLSTM-500k, has an embedding dimension of 128, 5 sLSTM blocks, 
an up-projection ratio of 1.25:1 to match the baseline model parameter count, 
and a total parameter count of 500k, while DNA-xLSTM-2M has an embedding dimension of 256, 
6 sLSTM blocks, a 1:1 up-projection ratio, and 2M parameters.
The long-context configuration, DNA-xLSTM-4M, has an embedding 
dimension of 256, 9 mLSTM blocks, a 2:1 up-projection ratio, and 
is augmented with \ac{rope} \citep{su2021roformer} to 
handle long-range dependencies effectively, with a total of 4M parameters. 
All three configurations are trained with both \ac{clm} and \ac{mlm}. 
Furthermore, we introduce \ac{rc}-equivariant versions, xLSTM-PH and xLSTM-PS, which 
use the original sequence and its \acf{rc}. 
We trained models with these configurations on the human genome and benchmarked them 
against state-of-the-art DNA models, such as 
Transformers, DNA-Mamba (Caduceus) \citep{schiff2024caduceus}, and HyenaDNA \citep{nguyen2023hyenadna},
showing competitive or better performance on pre-training and downstream classification tasks 
(see Section~\ref{sec:results_dna}).

\subsection{Prot-xLSTM} \label{sec:prot-xlstm}
For the protein domain, we propose the architectural variant 
Prot-xLSTM to address the complexities of protein sequence data, 
particularly in capturing long-range dependencies to enable homology-conditioned modeling. 
We introduce two configurations: Prot-xLSTM-26M and Prot-xLSTM-102M, with~26M and~102M parameters, respectively. Both 
configurations consist of 16~mLSTM blocks, with embedding dimensions of~512 for Prot-xLSTM-26M and~1,024 for Prot-
xLSTM-102M and maintaining a consistent 2:1~projection ratio across both configurations, 
and are trained with increasing context sizes ranging from~2,048 to~262,144 tokens.
To effectively manage the wide range of protein sequence lengths and context sizes, 
\acp{rope} are implemented for Prot-xLSTM. 
The according Prot-xLSTM models are trained with \ac{clm} using a \ac{fim} 
strategy on non-aligned homologous sequences,
enabling them to perform \ac{icl} at inference time in two modes: a) generative and b) inpainting.
Both approaches can be used for protein design, 
with the latter also suited for residue-based predictions, such as mutant fitness estimation.
Prot-xLSTM shows better performance than similarly configured Mamba- and Transformer-based models 
and shows promising results for homology-conditioned sequence generation (see Section~\ref{sec:results_prot}).

\subsection{Chem-xLSTM} \label{sec:chem-xlstm}
For the chemical sequence domain, we introduce Chem-xLSTM to enhance 
generative modeling of SMILES strings \citep{weininger1988smiles}, 
enabling domain-conditioned generation of small molecules without fine-tuning. 
We introduce two models: an unconditional generative model 
trained with a context length of 100 tokens (Chem-xLSTM-15M) 
and a domain-conditioned model trained with a 4,096-token context 
for in-context learning tasks (Chem-xLSTM-15M-icl). 
The latter can generate molecules within a specific domain without fine-tuning only based
on examples provided as context, a highly sought-after capability in drug discovery. 
Both models are configured to have 15M parameters, consist of 9 mLSTM blocks with an 
embedding dimension of 512 and a 1.3:1 projection ratio. 
The models have been benchmarked against other generative models 
for \ac{smiles} and at their \ac{icl} capabilities (see Section~\ref{sec:results_chem}).

\section{Experiments and Results}

\subsection{DNA Sequences} \label{sec:results_dna}

For the DNA-xLSTM experiments, we followed the experimental protocol outlined in \citet{schiff2024caduceus} and \citet{nguyen2023hyenadna} for both pre-training and downstream adaptation.

 \begin{figure}[h]
        \centering
        \includegraphics[width=0.49\textwidth]{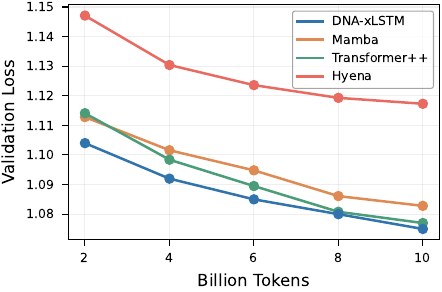}
        \includegraphics[width=0.49\textwidth]{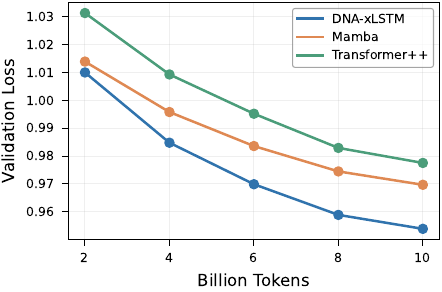}
        \caption{Pre-training of 2M-parameter DNA models on the human reference genome (GRCh38).
        Models are trained at single-nucleotide resolution with a context length of 1024 bases. \textbf{Left: \acl{clm}.} Learning curves display \textbf{\ac{ntp} loss} ($\downarrow$) on a test set, plotted against the number of tokens processed. \textbf{Right: \acl{mlm}.} Learning curves showing \textbf{\ac{mlm} loss} ($\downarrow$) on a test set across the number of tokens seen for various models. In both tasks, the xLSTM-based models consistently achieve the lowest loss values across all update steps.}
        \label{fig:dna_pretraining:ntp}
    \end{figure}
    
\textbf{Pre-training}. The training data for both the \ac{clm} and \ac{mlm} tasks was sourced from the human reference genome \citep{church2011modernizing}, with context lengths set to 1,024 and 32k tokens. Our baseline models included HyenaDNA \citep{nguyen2023hyenadna} and Caduceus \citep{schiff2024caduceus}, which is based on the Mamba architecture . Additionally, we trained Transformer++ baselines, building on the Llama architecture \citep{touvron2023llama}. Similar to Caduceus, we experimented with both PH- and PS-equivariant xLSTM configurations, benchmarking them against the corresponding Mamba baselines. All models that did not use PS-equivariance were trained with RC augmentation. Hyperparameters were selected using a separate validation set. 
Figure~\ref{fig:dna_pretraining:ntp} presents the test losses for 2M parameter CLM and MLM models trained with RC augmentation, i.e. non-PS models, and a context size of 1,024 tokens. In the CLM setting, DNA-xLSTM-2M achieved the best performance, surpassing Transformer++, Mamba, and HyenaDNA. The performance gap became even more pronounced on the MLM task, where DNA-xLSTM-2M outperformed both Transformer-based models and Mamba. 
Additionally, we extended DNA-xLSTM-2M to the PS equivariant setting and trained smaller RC-equivariant DNA-xLSTM-500k models. The resulting PH and PS models were subsequently used for downstream fine-tuning. In Appendix Section~\ref{sec:app_dna}, we present additional pre-training results including comparisons for large-context and PS equivariant models. We found that xLSTM-DNA matches or outperforms strong baselines in all pre-training settings.

\textbf{Downstream fine-tuning.}
To evaluate the learned representations, we fine-tuned the pre-trained DNA-xLSTM-2M and DNA-xLSTM-500K (both PH and PS) on two genomic classification benchmarks: the Genomic benchmark \citep{grevsova2023genomic} and
the Nucleotide Transformers Tasks \citep{dalla2023nucleotide}, 
which span 18~datasets from five studies.
DNA-xLSTM-2M-PH and DNA-xLSTM-2M-PS models pre-trained with context size 1,024 were compared against HyenaDNA and Mamba-PS and Mamba-PH. 
DNA-xLSTM performed best (see Table~\ref{tab:dna_nucleotide_downstream_table}), 
outperforming baseline models in the under 2M parameter range on 12~out~of~18 tasks, and was comparable to 
the much larger Nucleotide Transformer (500M parameters), winning 8~tasks. 
The comparisons with larger Transformer models and xLSTM-DNA-500k performance on the Genomics benchmark are presented in Appendix Section~\ref{sec:app_dna}.

    \begin{table}[h]
        \centering
        \caption{Downstream adaption of DNA models.
            The performance of 2M~parameter models fine-tuned on Nucleotide Transformer classification tasks on the test set is shown. \ac{ps} or \ac{ph} indicate models trained to be \ac{rc} equivariant. 
            Performance is averaged over 10~random seeds and error bars indicate the difference between maximum and minimum values across the 10~runs. The best values are highlighted in green.
            DNA-xLSTM outperforms both Mamba and Hyena on 12~out~of~18 tasks. Scores for Mamba- and Hyena-based models were obtained from \citet{schiff2024caduceus}. 
        }
        \label{tab:dna_nucleotide_downstream_table}
        \begin{threeparttable}
        \resizebox{0.95\columnwidth}{!}{%

\input{tables/dna_NTv2_smallModels_v2}

        }
        \begin{tablenotes}
        \item[a] this method is also called Caduceus \citep{schiff2024caduceus}.
        \end{tablenotes}
        \end{threeparttable}
    \end{table}

\subsection{Protein Sequences} \label{sec:results_prot}

We followed the experimental protocols from \citet{sgarbossa2024protmamba} for protein sequences.

\begin{figure}
            \centering
            \includegraphics[width=\textwidth]{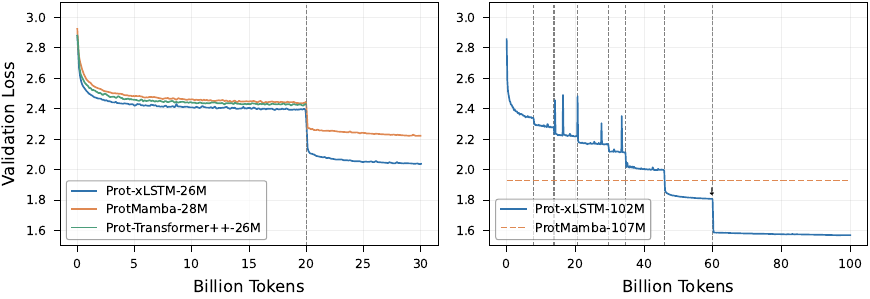}
            \caption{
                Generative pre-training of protein language models.
                The learning curves show the validation loss of homology-aware protein language models during training.
                \textbf{Left:} Small models trained for 20B~tokens with a context size of~$2^{11}$ and fine-tuned for~10B with a context of $2^{17}$~tokens. Transformer++ can only be run for a small context size.
                \textbf{Right:} Prot-xLSTM-102M model 
                trained with increasing context sizes from~$2^{11}$ to~$2^{18}$.
                Vertical gray dashed lines mark the points where context size was increased.
                The arrow at 60B tokens indicates the model used for downstream tasks.
                The orange dashed line corresponds to the validation loss of ProtMamba-107M trained up to a context size of~$2^{17}$ for a total of 195B~tokens. 
                Prot-xLSTM consistently outperforms other models and sets a new state-of-the-art at 
                homology-aware generation.
            }
            \label{fig:learning-curves-prot}
        \end{figure}

\textbf{Homology-aware training.}
Training data was sourced from the filtered OpenProteinSet 
\citep{ahdritz2023openproteinset}, consisting of 270k UniClust30 
clusters (508M sequences, 110B residues). Using the ProtMamba 
pipeline, we constructed homology-aware, alignment-free inputs by concatenating unaligned homologous sequences and mask patches for training with the \ac{fim} strategy. 
We trained two xLSTM-based models: Prot-xLSTM-26M and Prot-xLSTM-102M. For comparison, we also trained a smaller ProtMamba (ProtMamba-28M) and Transformer-based (Prot-Transformer++-26M) \citep{touvron2023llama} model and used the \textit{ProtMamba Long Foundation} (ProtMamba-107M) provided by \citet{sgarbossa2024protmamba}. The initial training followed a context length scheduling strategy, with models gradually increasing context from $2^{11}$ to $2^{17}$ tokens. Finally, Prot-xLSTM-102M was further trained with $T=2^{18}$.\footnote{The protein downstream tasks were evaluated using the model trained with $T$ up to $2^{17}$ tokens.}
We evaluated the models using negative log-likelihood and perplexity, calculated for different parts of the concatenated-\ac{fim} sequences. As shown in Fig.~\ref{fig:learning-curves-prot} and Tab.~\ref{tab:prot-testset-perf}, Prot-xLSTM outperformed the other architectures. Its advantage becomes even more pronounced with longer contexts, which Prot-Transformer++ cannot handle, and where Prot-xLSTM significantly outperforms ProtMamba. Furthermore, Prot-xLSTM-102M outperforms ProtMamba-107M, despite being trained on only a quarter of the total training tokens used for ProtMamba-107M.
Further details are provided in Appendix Section ~\ref{app-sec:prot_pretrain}.

    \begin{table}
        \centering
        \small
        \caption{Performance comparison of protein language models
        at homology-conditioned generation.
        Test set \textbf{perplexity ($\downarrow$)} of different models with a context size of $2^{17}$ is shown across different token subsets. 
        The average and 95\%~confidence interval values are computed across the ~test set clusters. 
        Prot-xLSTM outperforms ProtMamba, especially when using a long context.
        }  
        \input{tables/proteins/protein_testset_results}

        \label{tab:prot-testset-perf}

    \end{table}

\textbf{Homology-conditioned protein generation.} We generated 2,500 protein sequences each for 19 clusters using different parameters and score them using multiple metrics. Hamming distance, HMMER score, and structural scores correlate well with sequence perplexity, with an average absolute Pearson correlation of 0.57 across clusters for the large Prot-xLSTM model (Table~\ref{tab:prot-ppl-r}). Table~\ref{tab:prot-generation} shows Kolmogorov-Smirnov test statistics, which quantifies how well the score distributions of the generated proteins match those of the real proteins. For each cluster, we compared scores between 100 random real proteins and the 100 generated proteins with the lowest perplexity. For further details see Appendix Section~\ref{app-sec:prot_generation}.

    \begin{table}[h]
        \begin{threeparttable}[b]
        \footnotesize
        \caption{Homology-conditioned protein generation. Average \textbf{Kolmogorov-Smirnov statistic ($\downarrow$)} between scores of natural and generated sequences with 95\% confidence intervals across 19 homology clusters. For three of five metrics, score distributions of Prot-xLSTM-generated sequences are closest to natural sequences.}
        \label{tab:prot-generation}
        \input{tables/proteins/protein_sequence_evaluation_ks}
        \end{threeparttable}
    \end{table}

\textbf{Protein variant fitness prediction.} We assessed Prot-xLSTM's predictive power for mutational effects by leveraging its inpainting capability from the FIM training objective and using homologous sequences as context on the ProteinGym DMS zero-shot substitutions benchmark \citep{notin2023proteingym}. Table \ref{tab:prot-gym-perf} presents the performance comparison of Prot-xLSTM, other well-known protein models, and the current top performers on the ProteinGym leaderboard. In summary, Prot-xLSTM outperforms larger unconditional protein language models like ESM-2 \citep{lin2023evolutionary} and ProGen2 \citep{nijkamp2023progen2}, and matches or surpasses ProtMamba's performance. However, it falls short compared to models that directly use the MSA, such as TranceptEVE \citep{notin2022trancepteve}, or use structural information, like ProSST \citep{li2024prosst}. Further details and results are provided in Appendix Section~\ref{app-sec:prot_fitness}.

\subsection{Chemical Sequences} \label{sec:results_chem}

\textbf{Unconditional molecule generation} aims to produce valid small organic molecules without imposing specific constraints, such as being from a particular molecular domain. 
Following the setup of \citet{ozccelik2024chemical}, we trained models to generate 
\ac{smiles} strings using a \ac{clm} approach on a dataset derived from ChEMBL with a context length of 100 tokens.
We compared our Chem-xLSTM architecture with several architectures, including LSTM, GPT, S4, and Mamba, where all models contain approximately 15 million parameters. 
The evaluation focused on two primary metrics: perplexity and \ac{fcd} \citep{preuer2018frechet}. 
Chem-xLSTM achieved the lowest \ac{fcd} of~0.13 and a competitive perplexity score of~1.68, indicating its strong ability to generate realistic chemical structures (see Table~\ref{tab:design_smiles}). 
All models produced valid, unique, and novel 
molecules, showcasing their effectiveness in this task. Further details and results are provided in Appendix Section~\ref{app-sec:chem-unconditional}.

\begin{table}[h]
    \centering
    \caption{%
        Unconditional generation of molecules with 15M~parameter models. 
        102,400~\ac{smiles} sequences have been generated and evaluated. 
        Error bars represent standard deviations 
        across bootstrap resampling.%
        Green cells highlight the best values per row. 
        Chem-xLSTM yields the best \ac{fcd} and \ac{smiles}-GPT the best perplexity.
        \label{tab:design_smiles}
    }
    \begin{threeparttable}
    \footnotesize
    \begin{tabular}{lccccc}
    \toprule
     & SMILES-LSTM\tnote{a}   & SMILES-GPT\tnote{b} & SMILES-S4\tnote{c} & Chem-Mamba\tnote{d} & Chem-xLSTM \\
     \midrule
     \rowcolor{jku_grey!10} FCD ↓ &  $0.46^{\pm{<0.01}}$ & $0.15^{\pm{<0.01}}$ & $0.28^{\pm{<0.01}}$ &  $0.21^{\pm{<0.01}}$ & \cellcolor{jku_green!20}$0.13^{\pm{<0.01}}$ \\
     Perplexity ↓ & $1.88^{\pm{3.8}}$ & \cellcolor{jku_green!20}$1.65^{\pm{0.6}}$ & $1.73^{\pm{2.4}}$ & $1.74^{\pm{0.5}}$ & $1.68^{\pm{1.0}}$\\
    \bottomrule
    \end{tabular}
    \begin{tablenotes}
        \item[a] \citet{segler2018generating} \hspace{2mm} \textsuperscript{b}\,\citet{adilov2021generative} \hspace{2mm}
        \textsuperscript{c}\,\citet{ozccelik2024chemical} \hspace{2mm}
        \textsuperscript{d}\,Adapted to \ac{smiles} in this work.
    \end{tablenotes}
    \end{threeparttable}
\end{table}

\begin{figure}[h]
        \centering
        \includegraphics[width=0.49\textwidth]{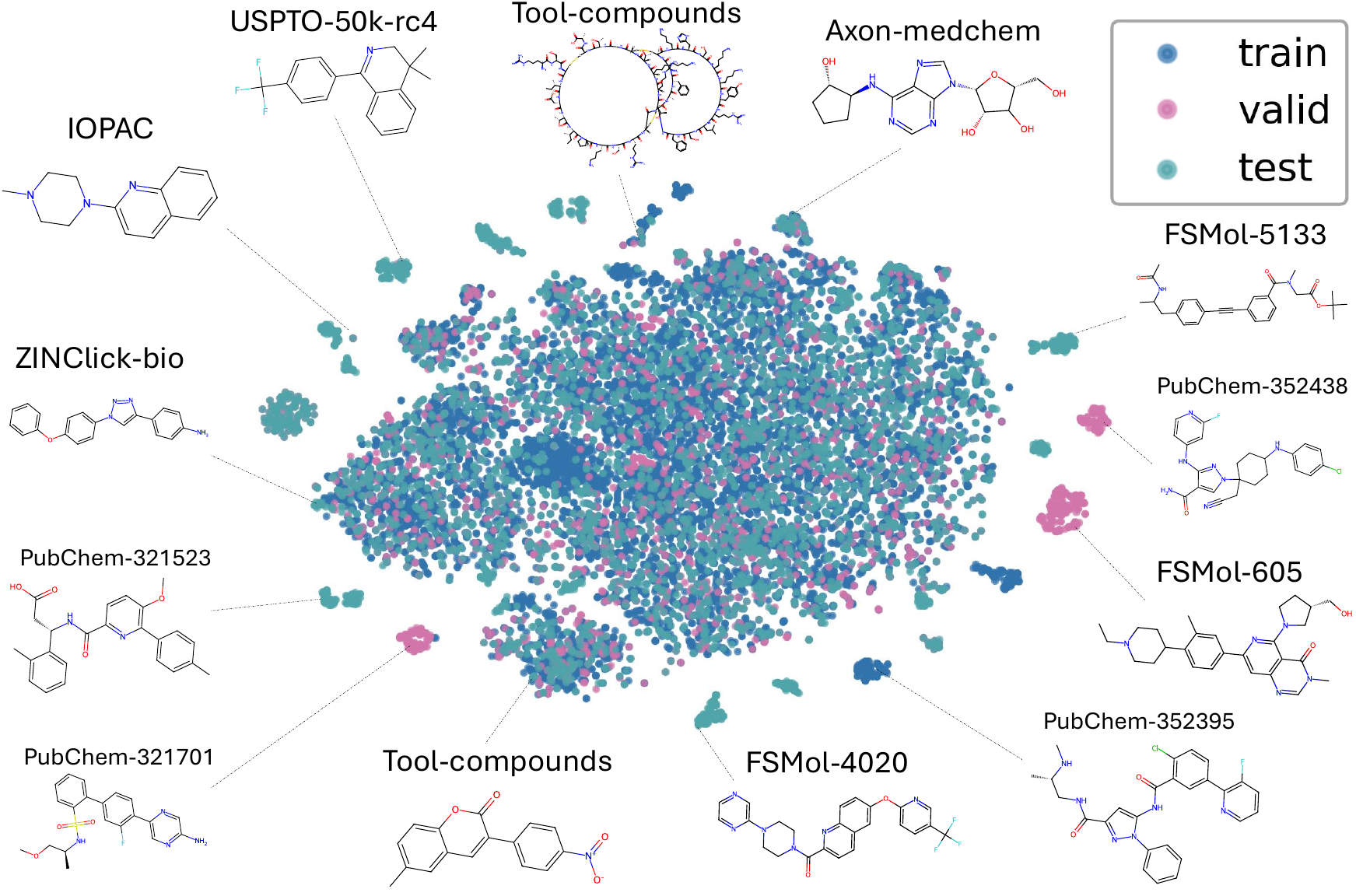} %
        \includegraphics[width=0.49\textwidth]{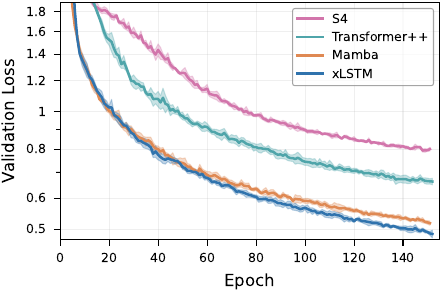}
        \caption{Conditional generation of molecules via \ac{icl} and 15M parameter models. \textbf{Left:} Visualization of different molecular domains contained in the \emph{molecular domains} dataset. A t-SNE down-projection of molecules from 
        different domains is shown. Clusters on the exterior represent highly specific molecular domains. The validation 
        and test sets contain molecules from highly specific,  unseen molecular 
        domains.
        \textbf{Right:} Generative training of chemistry language models on the \emph{molecular domains} dataset. Learning curves showing mean \textbf{\ac{clm} loss} ($\downarrow$) on a validation set across %
        the training epochs. Shaded areas represent the standard-deviation over runs.
        The Chem-xLSTM achieved the lowest loss at conditional generation of molecules using \ac{icl}. }
        \label{fig:cond_mol_gen}
    \end{figure}

For \textbf{conditional molecule generation}, the objective 
is to generate molecules belonging to a specific 
molecular domain or possessing desired properties. 
Here, we focused on generating molecules from a particular domain using the in-context learning abilities of \acp{llm}.
To achieve this, we assembled a dataset, referred to as the \emph{molecular domains} dataset, which comprises a diverse range of molecular domains: 
natural products, click-chemistry, proteolysis-targeting chimera (PROTACs), DNA-encoded chemical libraries, approved and failed drugs, and bioactive compounds from various bioassays.
Molecules from the same domain, are concatenated as a long sequence, and augmented through permutation during training.  
We split the dataset into training, validation, and test domains, following an 8:1:1 ratio (Figure~\ref{fig:cond_mol_gen}, left). The validation and test sets contained molecules from unseen domains, enabling us to evaluate the models' conditional generation capabilities through in-context learning.
We trained Chem-xLSTM, Mamba, Transformer++, and S4-based models with the \ac{clm} approach on the \emph{molecular domains} with an increased context length of 4,096 tokens. The context length for S4 models was restricted to 2,048 due to memory constraints. We evaluated the models based on NTP loss across unseen domains.
The trained model 
Chem-xLSTM-15M-icl shows promising results in this 
conditional setting, outperforming the other benchmarked model classes (Figure~\ref{fig:cond_mol_gen}, right). 
This demonstrates Chem-xLSTM's capability to generate
molecules from an unseen chemical domain when provided with only a few exemplary molecules without fine-tuning. Further details and results are provided in Appendix Section~\ref{app-sec:chem_conditional}.

\subsection{Compute demand and resources.}
The experiments were conducted on multiple GPU servers with A100 GPUs. Model training was performed in both single-node and multi-node setups, utilizing 1–8 A100 GPUs per node. Prot-xLSTM-102M training with a context length of $2^{17}$ was completed on a node with 8 H200 GPUs. The largest models were trained across up to four nodes using distributed data parallelism. Some experiments leveraged compute resources provided by EuroHPC Joint Undertaking clusters, including Karolina at IT4Innovations, Leonardo at CINECA, and MeluXina at LuxProvide.  The total amount of GPU hours required for the experiments in this paper is approximately 50k.

\section{Discussion}
In this work, we demonstrated the potential of Bio-xLSTM variants as prime candidates to model biological and chemical sequences. We have provided clarity in two key areas: a) how to tailor xLSTM for biological and chemical sequences, and b) comparing xLSTM-based models to other domain-specific \acp{llm}, showcasing their robust performance in DNA, protein, and chemical sequence modeling tasks.
Despite certain limitations, DNA-xLSTM showed strong performance in DNA sequence modeling, excelling in both masked and causal language tasks across different context sizes. In protein modeling, Prot-xLSTM proved particularly effective at handling long-range dependencies, positioning it as a promising tool for generating homologous proteins. In small molecule modeling, Chem-xLSTM achieved the best \ac{fcd} scores for unconditional generation and demonstrated strong \ac{icl} capabilities.
Our findings underscore the potential of xLSTM as a prime candidate for foundational models in molecular biology. The models we have introduced, trained, and made available can be used for example to generate rich learned representations for DNA sequences and homology- and chemical domain-conditioned generation of proteins and molecules without the need for fine-tuning.

While Bio-xLSTM has shown strong performance across DNA, protein, and chemical sequence modeling, it has several limitations. 
The manual hyperparameter selection process, which was due to limited 
computational resources, may prevent optimal model configurations. We 
will explore the hyperparameter space further in the future, 
which might yield even better models. 
For DNA, the reliance on character-level tokenization might also 
restrict the performance and scaling to larger context sizes. Also for 
proteins, amino acid level tokenization without explicit structural 
information might limit it's performance.
The DNA-xLSTM, Prot-xLSTM, and Chem-xLSTM models are currently 
constrained by the training dataset and their generalizability across 
organisms and chemical domains needs further exploration.
Across all three domains, the training datasets contain 
biases -- whether it is population biases in the genomic data, sequence distribution biases in protein datasets, 
or chemical exploration biases in molecular datasets. 
These biases could influence the model's predictions and limit its generalizability in real-world applications. 
In line with many works, we consider the perplexity metric, for example, next token perplexity, or the related cross-entropy losses as a proxy for performance on downstream tasks. However, this metric might not capture the capacities of biological and chemical language models appropriately. 
Future work could address these limitations by expanding the training datasets and downstream evaluations of Bio-xLSTM. Finally, assessing Bio-xLSTM's performance in parameter regimes beyond the billion scale remains an open question.

\section*{Ethics Statement}
The development of our large language model for biological 
sequences, including DNA, proteins, and small molecules, 
has the potential to significantly advance biomedical 
research and therapeutics. In creating this model, we have 
taken care to train exclusively on publicly available 
data, such as the human reference genome, OpenProteinSet, 
and publicly available small molecule databases. Note 
that no human subjects are involved in the studies, since 
the human reference genome does not represent a particular 
individual but a type of average human genome.
As common with machine learning methods, potential
danger lies in the possibility that users rely too much on 
our new approach and use it without reflecting
on the outcomes. However, the full pipeline, in which our 
method would be used, includes wet lab tests
after its application, to verify and investigate the 
results, which decreases the danger of misuse or overly
relying on the predictions. To further mitigate the risk of misuse, 
we provide model limitations and mention potential biases. 
Users are encouraged to approach the model's predictions critically and 
consider them as one component of a broader decision-making process.

\section*{Code and Data Availability}

To ensure reproducibility and facilitate future research, we provide three standalone code repositories for \href{https://github.com/ml-jku/DNA-xLSTM}{DNA-xLSTM}, \href{https://github.com/ml-jku/Prot-xLSTM}{Prot-xLSTM}, and \href{https://github.com/ml-jku/Chem-xLSTM}{Chem-xLSTM}, each containing the respective implementations, training scripts, evaluation procedures, and pre-processed datasets.

\section*{Acknowledgements}
The ELLIS Unit Linz, the LIT AI Lab, the Institute for Machine Learning, are supported by the Federal State Upper Austria. We thank the projects INCONTROL-RL (FFG-881064), PRIMAL (FFG-873979), S3AI (FFG-872172), DL for GranularFlow (FFG-871302), EPILEPSIA (FFG-892171), FWF AIRI FG 9-N (10.55776/FG9), AI4GreenHeatingGrids (FFG-899943), INTEGRATE (FFG-892418), ELISE (H2020-ICT-2019-3 ID: 951847), Stars4Waters (HORIZON-CL6-2021-CLIMATE-01-01). We thank NXAI GmbH, Audi.JKU Deep Learning Center, TGW LOGISTICS GROUP GMBH, Silicon Austria Labs (SAL), FILL Gesellschaft mbH, Anyline GmbH, Google, ZF Friedrichshafen AG, Robert Bosch GmbH, UCB Biopharma SRL, Merck Healthcare KGaA, Verbund AG, GLS (Univ. Waterloo), Software Competence Center Hagenberg GmbH, Borealis AG, TÜV Austria, Frauscher Sensonic, TRUMPF and the NVIDIA Corporation.

\bibliographystyle{apalike}
\bibliography{refs}

\clearpage 

\appendix
\resumetocwriting
\renewcommand{\thetable}{A\arabic{table}} 
\renewcommand{\thefigure}{A\arabic{figure}} 
\setcounter{table}{0}
\setcounter{figure}{0}

\tableofcontents

\clearpage

\section{Related Work}\label{sec:related_work}
In all three areas, genomics, proteomics, and chemistry, we observe a similar trend that until around 2018 the language models were based on LSTMs \citep{hochreiter1997lstm}, 
then a large amount of models were based on Transformers \citep{vaswani2017attention}, with different modeling paradigms and styles, and from 2023 onwards \acp{ssm} appeared.

\textbf{Language models for genomic sequence data.}
DNABERT \citep{ji2021dnabert} and its successor DNABERT-2 \citep{zhou2024dnabert} are
Transformer-based models that leverage bidirectional encoder representations and masked language modeling
to capture nucleotide context, achieving high performance in tasks like promoter and splice site prediction. 
LOGO \citep{yang2022integrating}, another Transformer-based model, 
applies self-supervised learning to the human genome 
for sequence labeling and variant prioritization, 
while VIBE \citep{gwak2022vibe} employs a hierarchical BERT model 
to enhance the detection of eukaryotic viruses in metagenomic data. 
Models like LookingGlass \citep{hoarfrost2022deep}, based on \ac{rnn}, and GPN \citep{benegas2023dna}, which uses convolutional neural networks (CNNs), are examples of non-Transformer-based approaches, 
with LookingGlass focusing on microbial genomes and GPN on plant genomes. 
More recent developments include the nucleotide transformer (NT) \citep{dalla2023nucleotide}, 
a Transformer model trained on the human genome and data from the 1000 Genomes Project, 
and SpeciesLM \citep{karollus2024species}, which trains Transformer-based models on 1500 fungal genomes. 
The latest advances, represented by Caduceus \citep{schiff2024caduceus} based on Mamba \citep{gu2024mamba}
and HyenaDNA \citep{nguyen2023hyenadna}, 
introduce \acp{ssm} that allow 
generative modeling and representation learning for long DNA sequences.

\textbf{Language models for protein sequence data.}
Until around 2019, the field was dominated by \acp{rnn} and LSTM-based models trained with \ac{clm}.
Notable examples include UniRep \citep{alley2019unified}, 
which employed multiplicative-LSTM to capture rich protein representations, and SSA \citep{bepler2019learning}, which used bidirectional \acp{rnn} for structural similarity prediction.
Since then the field has shifted towards Transformer-based models and \ac{mlm}, driven by their success in natural language processing. Early adopters of this shift included the TAPE benchmark for protein downstream tasks \citep{rao2019evaluating}, which evaluated both an LSTM and a Transformer architecture trained with \ac{clm} and \ac{mlm}, respectively. 
\citet{elnaggar2021prottrans} further expanded the use of Transformers with large-scale \ac{mlm}, setting new benchmarks in protein sequence analysis with Prot-T5. 
ESM \citep{rives2021biological} applied \ac{mlm} to a Transformer on a massive scale, capturing evolutionary patterns across diverse protein sequences. 
Other significant Transformer-based models include MSA-Transformer \citep{rao2021msa}, which applied \ac{mlm} to multiple-sequence alignments (MSA), and ProGen \citep{madani2023large}, which used \ac{clm} and Transformers for controlled protein sequence generation. 
Additionally, models like ProtGPT2 \citep{ferruz2022protgpt2} and ProteinBERT \citep{brandes2022proteinbert} utilized the power of Transformer architectures in generating novel protein sequences and functional predictions. 
Furthermore, \citep{su2024saprot} introduced a "structure-aware vocabulary" which they use as input for Transformer-based models. 
The recently proposed PoET \citep{truong2023poet} is an autoregressive Transformer model trained on non-aligned homologous sequences, providing a novel approach for conditional protein design and protein fitness prediction.
Building on the concept of non-aligned homologous sequences, ProtMamba \citep{sgarbossa2024protmamba} leverages emerging \acp{ssm} to manage long-context conditioning on proteins, effectively utilizing autoregressive and \ac{fim} strategies.
For a more comprehensive review of these advancements, 
including their applications in functional protein design, 
see \citet{notin2024machine} and \citet{hu2022protein}.

\textbf{Language models for chemical sequence data.}
The first language model for chemical sequences was
an LSTM-based, autoregressive method developed by \citet{segler2018generating}, which demonstrated that the \ac{smiles} syntax \citep{weininger1988smiles} and generation of realistic organic molecules can be learned.
\citet{honda2019smiles} introduced a Transformer model for this task, showing that this leads to informative representations of molecules. 
The Molecular Transformer \citep{schwaller2019molecular} consists of a Transformer-based encoder and decoder, trained on chemical reaction data to translate between reactants and products.
More recently, \acp{ssm} have been used for generative modeling of \ac{smiles} strings \citep{ozccelik2024chemical}.
Subsequent models such as MolGPT \citep{bagal2021molgpt} and cMolGPT \citep{wang2023cmolgpt} utilized the GPT architecture to generate \ac{smiles} strings, with MolGPT conditioning on chemical properties and scaffolds, and cMolGPT focusing on biomolecular targets. Transformer-based approaches have also been employed to optimize the properties of small molecules in a reinforcement-learning setting \citep{mazuz2023molecule}. 
Encoder-style language models for chemistry, 
such as SmilesLSTM \citep{mayr2018large}, ChemNet \citep{preuer2018frechet}, and CNN-based models \citep{jastrzkebski2016learning}, initially used activity and property prediction as pre-training or training objectives. 
Later, these encoder-style language models were trained with the masking language modeling objective, as seen in ChemBERTA \citep{chithrananda2020chemberta},  Chemberta-2 \citep{ahmad2022chemberta}, SMILES-BERT \citep{wang2019smiles}, 
MolFormer \citep{ross2022large} and rxnfp-BERT \citep{schwaller2021mapping}.
Some models have also adopted contrastive objectives \citep{seidl2023enhancing}.
Large language models for 
molecules have also been shown to learn complex 
molecular distributions \citep{flam2022language}. 
For a more thorough and comprehensive overview, we
refer to the surveys by \citet{bran2023transformers}
and \citet{zhang2024scientific}

\clearpage

\section{xLSTM Architecture Details}
\label{sec:architecture_details}

\subsection{xLSTM and Bio-xLSTM Blocks}
\label{app-sec:blocks_details}
\citet{beck2024xlstm} suggested xLSTM blocks, which are residual \citep{srivastava2015highway,he2016resnet} block modules, into which the sLSTM and mLSTM layers can be integrated. The two basic blocks can in principle be characterized by either applying post-sLSTM/mLSTM up- and down-projections (similar to \citet{vaswani2017attention})  or by applying pre-sLSTM/mLSTM up-projections and post-sLSTM/mLSTM down-projections (similar to \citet{dao2023flashattention2}). An sLSTM block integrates the sLSTM layer into the up- and down-projection block, while the mLSTM block integrates the mLSTM layer into the pre-up-projection and post-down-projection block. The two basic xLSTM blocks also make use of neural network modules like layer normalization \citep{ba2016layernorm}, short causal convolutions,  
and, group normalization \citep{wu2020groupnorm}. For the exact architecture of the blocks, we refer to \citet[Sec.2.4]{beck2024xlstm}. 
An xLSTM architecture is constructed by residually stacking the suggested xLSTM blocks. For that, the most commonly used pre-LayerNorm residual backbone is used.

\begin{figure}
    \centering
    \includegraphics[width=0.9\textwidth]{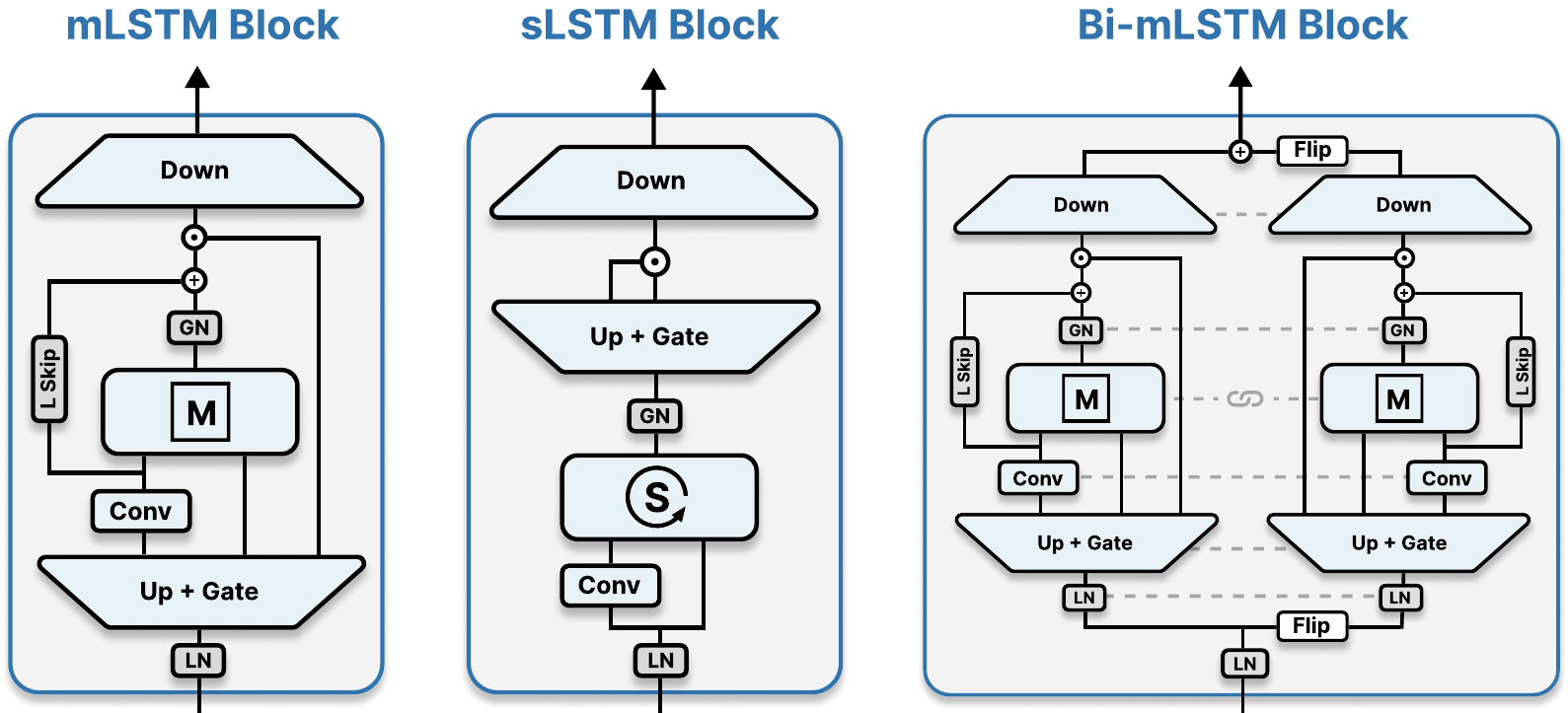}
    \caption{xLSTM and Bio-xLSTM blocks. \textbf{Left: mLSTM block.} \textit{LN} (Layer Normalization) and \textit{GN} (Group Normalization) refer to normalization modules, while \textit{L Skip} represents learnable skip connections and \textit{Conv} denotes causal 1D convolutions. The mLSTM block utilizes a gated pre-up-projection structure, akin to modern State-Space Models, with gates activated by the Swish function. \textbf{Middle: sLSTM block.} The sLSTM block features a GELU-gated post-up-projection structure, similar to Transformer architectures. \textbf{Right: Bidirectional mLSTM block.} For bidirectional processing, the xLSTM applies each block to the input sequence twice before combining the outputs: once left-to-right and once right-to-left.}
    \label{fig:xlstm_block_diagram}
\end{figure}

For Bio-xLSTM we keep the basic xLSTM building blocks and the basic xLSTM architecture template, but adjust them to the respective domains. Figure \ref{fig:xlstm_block_diagram} depicts sLSTM and mLSTM blocks, as well as, a bidirectional mLSTM configuration with weight-tied layers.

\subsection{Modes of Operation: Parallel, Chunkwise or Recurrent}
\label{sec:mlstm_forms}

Similar to linear attention variants \citep{katharopoulos2020transformers, yang2024gated}, the mLSTM has three possible formulations: parallel, recurrent or chunkwise.
The presentation in section~\ref{sec:mLSTM} \citep[and][]{beck2024xlstm} focuses on the recurrent form:
\begin{align*}
    \BC_t &= \sigma(\tilde{f}_t) \, \BC_{t-1} + \exp(\tilde{i}_t) \odot \Bv_t \Bk_t^\mathsf{T} \\
    \Bn_t &= \sigma(\tilde{f}_t) \, \Bn_{t-1} + \exp(\tilde{i}_t) \, \Bk_t \\
    \Bh_t &= \sigma(\tilde{\Bo}_t) \odot \frac{\BC_t \Bq_t}{\max\bigl(|\Bn_t \Bq_t|, 1\bigr)}.
\end{align*}
This form is especially useful for inference when samples arrive one time-step at a time.

The omission of the recurrent connections in mLSTM allows for a parallel formulation \citep[Appendix]{beck2024xlstm}:
\begin{align*}
    \tilde{\BF} &= \begin{bmatrix}
        0 & 0 & 0 & \ldots & 0 \\ 
        \ln\sigma(\tilde{f}_2) & 0 & 0 & \ldots & 0 \\
        \ln\sigma(\tilde{f}_2) + \ln\sigma(\tilde{f}_3) & \ln\sigma(\tilde{f}_3) & 0 & \ldots & 0 \\
        \vdots & \vdots & \vdots & \ddots & \vdots \\
        \sum_{t=2}^T \ln\sigma(\tilde{f}_t) & \sum_{t=3}^T \ln\sigma(\tilde{f}_t) & \sum_{t=4}^T \ln\sigma(\tilde{f}_t) & \ldots & 0
    \end{bmatrix} \\
    \BD &= \exp\bigl(\tilde{\BF} + \BOn \otimes \tilde{\Bi}\bigr) \odot \BM \\
    \BH &= \sigma(\tilde{\BO}) \odot \frac{\BD \odot \BQ \BK^\mathsf{T}}{\max\bigl(|\BD \odot \BQ \BK^\mathsf{T}| \cdot \BOn, \BOn\bigr)} \BV,
\end{align*}
where $\BQ, \BK, \BV, \tilde{\BO} \in \reals^{T \times d}$, $\tilde{\Bi} \in \reals^T$ and $\BM \in \{0, 1\}^{T \times T}$ is a causal (i.e.~lower-triangular) masking matrix.
The $\otimes$ refers to an outer product, while $\odot$ is a Hadamard (i.e.~element-wise) product.
The fraction, $\max$ and other non-linear functions are also applied element-wise.
This parallel form enables an efficient training regime, similar to Transformers.

The chunkwise formulation is a 
hybrid of the recurrent and parallel forms:
\begin{align*}
    \tilde{\BF} &= \begin{bmatrix}
        0 & 0 & 0 & \ldots & 0 \\ 
        \ln\sigma(\tilde{f}_{t-C+2}) & 0 & 0 & \ldots & 0 \\
        \ln\sigma(\tilde{f}_{t-C+2}) + \ln\sigma(\tilde{f}_{t-C+3}) & \ln\sigma(\tilde{f}_{t-C+3}) & 0 & \ldots & 0 \\
        \vdots & \vdots & \vdots & \ddots & \vdots \\
        \sum_{\tau=2}^C \ln\sigma(\tilde{f}_{t-C+\tau}) & \sum_{\tau=3}^C \ln\sigma(\tilde{f}_{t-C+\tau}) & \sum_{\tau=4}^C \ln\sigma(\tilde{f}_{t-C+\tau}) & \ldots & 0
    \end{bmatrix} \\
    \BD &= \exp\bigl(\tilde{\BF} + \BOn \otimes \tilde{\Bi}\bigr) \odot \BM \\
    \Bf &= \bigg(\sigma(\tilde{f}_{t-C+1}), \sigma(\tilde{f}_{t-C+1}) \, \sigma(\tilde{f}_{t-C+2}), \ldots, \prod_{\tau=1}^C \sigma(\tilde{f}_{t-C+\tau}) \bigg) \\
    \BH &= \sigma(\tilde{\BO}) \odot \frac{(\BD \odot \BQ \BK^\mathsf{T}) \, \BV + \diag(\Bf) \, \BQ \BC_{t-C}^\mathsf{T}}{\max\bigl(|(\BD \odot \BQ \BK^\mathsf{T}) \cdot \BOn + \diag(\Bf) \, \BQ \Bn_{t-C}|, \BOn\bigr)} \\
    \BC_t &= \bigg(\prod_{\tau=1}^{C} \sigma(\tilde{f}_{t-C+\tau})\bigg) \, \BC_{t-C} + \BV^\mathsf{T} \diag(\Bd_C) \, \BK \\
    \Bn_t &= \bigg(\prod_{\tau=1}^{C} \sigma(\tilde{f}_{t-C+\tau})\bigg) \, \Bn_{t-C} + \BK^\mathsf{T} \Bd_C,
\end{align*}
with $\BQ, \BK, \BV, \tilde{\BO} \in \reals^{C \times d}$ and $\tilde{\Bi} \in \reals^C$ the pre-activations from $t - C + 1$ to $t$. Furthermore, $\BM \in \{0, 1\}^{C \times C}$, is a local causal (i.e.~lower-triangular) masking matrix, $\Bd_C$ denotes the last row of $\BD$, $\diag$ transforms a vector into a diagonal matrix, and $C$ is the chunk size. The chunk-wise formulation allows for 
implementing hardware-aware
efficient kernels \citep{beck2025unlocking}.
For $C = 1$, we recover the recurrent form, whereas for $C = T$, we obtain the parallel form.

\subsection{Efficient Bidirectional Modeling for Weight-Tied Layers of Bio-xLSTM}
\label{sec:bidirectional_modelling}
Bidirectional modeling is often required to learn
representations of biological and chemical sequences, for 
example with the MLM paradigm. The default approach for bidirectional modeling would be
to use an mLSTM layer on the usual sequence $\BX_{1:T}=(\Bx_1,\Bx_2,\ldots,\Bx_T)$
and then applying a weight-tied layer on the reversed sequence $\BX_{T:1}=(\Bx_T,\Bx_{T-1},\ldots,\Bx_1)$
and subsequently summing those outputs:
\begin{align}
    \BH^+ &=\mathrm{mLSTM}(\BX_{1:T}; \Bw)\label{eq:fwd_mlstm} \\
    \BH^- &=\mathrm{mLSTM}(\BX_{T:1}; \Bw) \label{eq:rev_mlstm}\\
    \BH &= \BH^+  + \BH^-_{T:1},
\end{align}
where $\BH^-_{T:1}$ indicates that the sequence is reversed again, 
and $\Bw$ are the parameters of the LSTM-layer 
$\mathrm{mLSTM}(\BX_{1:T}; \Bw)$ which are assumed to be 
the same for both directions, i.e. weight-tied. This approach is schematically depicted in Figure \ref{fig:xlstm_block_diagram} (Right).
However, this approach is inefficient 
with respect to memory and
operations because it has to calculate and store 
all internal quantities, such as the cell states, 
twice for the backward pass. 
A variant of this approach is to perform 
the forward direction in one block (Eq.~\ref{eq:fwd_mlstm})
and the reverse direction
in a consecutive block (Eq.~\ref{eq:rev_mlstm}) 
of the architecture \citep{alkin2024vision}.

\textbf{We propose an efficient bidirectional modeling approach.} 
Because of the parallelism of mLSTM and its gates depending only on the current time step, the weighted cumulative sum required for the cell state (Eq.~\ref{eq:mlstm_recurrent_begin}), can be done backwards to obtain the representations for the reversed sequence 
\begin{align}
    \BC^+_{t} &= \Rf_t  \BC^+_{t-1}  + \Ri_t  \Bv_t \ \Bk_t^\top &
    \BC^-_{t} &= \Rf_t  \BC^-_{t+1}  + \Ri_t  \Bv_t \ \Bk_t^\top \\
    \Bn^+_{t} &= \Rf_t  \Bn^+_{t-1}  + \Ri_t  \Bk_t &
    \Bn^-_{t} &= \Rf_t  \Bn^-_{t+1}  + \Ri_t  \Bk_t.
\end{align}

The resulting $\Bh_t=\Bh_t^+ + \Bh_t^-$ is a 
bidirectional representation of the input sequence, whereby this 
variant is more efficient with respect to memory usage because of 
shared quantities. Note that the two variants, the default approach, and the efficient approach, are mathematically equivalent.

\clearpage

\section{DNA-xLSTM: Details and Additional Results}
\label{sec:app_dna}
In this section, we provide further details regarding the 
architecture, training setup, and evaluation metrics for the DNA-xLSTM models.

\subsection{Pre-Training}

\textbf{Experimental setup.} We followed the experimental protocol established in \citet{schiff2024caduceus} and \citet{nguyen2023hyenadna}. The human reference genome \citep{church2011modernizing} was used as the training dataset for two main tasks: \textbf{a)} causal language modeling (CLM) and \textbf{b)} masked language modeling (MLM). We employed context lengths of 1,024 and 32,000 tokens for these tasks.

To ensure a fair comparison with previous methods, such as \citet{schiff2024caduceus}, we used character- or base pair-level tokenization, training models with parameter sizes ranging from 500k to 4M. This experimental setup enabled us to evaluate model performance for both \textbf{a)} generative modeling of DNA sequences and \textbf{b)} learning rich DNA sequence representations—core tasks in this domain.

\textbf{Methods and hyperparameters.} In our pre-training experiments, we compared several architectures: a Transformer variant based on the Llama architecture, referred to as Transformer++ \citep{touvron2023llama}, DNA-xLSTM, HyenaDNA \citep{nguyen2023hyenadna}, and DNA-Mamba (also known as Caduceus) \citep{schiff2024caduceus}. Each architecture was trained under both CLM and MLM settings. Additionally, we assessed two types of reverse-complement (RC) equivariant models when applicable: DNA-Mamba-PH and DNA-Mamba-PS, as well as DNA-xLSTM-PH and DNA-xLSTM-PS. For non-equivariant models, reverse-complement augmentation was applied, following the approach described in \citet{schiff2024caduceus}. Further details on RC-equivariant modeling can be found in Section \ref{sec:app_dna_reverse_complement}. The hyperparameters for DNA-xLSTM and Transformer++ were optimized using a validation set, with the final configurations reported in Appendix~Tables \ref{tab:dna_xlstm_hyperparameters} and \ref{tab:dna_transformer_hyperparameters}.

\textbf{Metrics.} We report cross-entropy loss on a held-out test set for both CLM and MLM pre-training experiments.

\textbf{Results.}
Our experiments show that the sLSTM-based DNA-xLSTM-2M model, trained with a context size of 1,024 and reverse-complement augmentation, outperforms DNA-Mamba \citep{schiff2024caduceus}, HyenaDNA \citep{nguyen2023hyenadna}, and Transformer++ across both CLM and MLM tasks. Notably, the performance gap between DNA-xLSTM and the baseline models increases in the MLM setting. See Figure~\ref{fig:dna_pretraining:ntp}.

\begin{figure}
        \centering
        \includegraphics[width=0.49\textwidth]{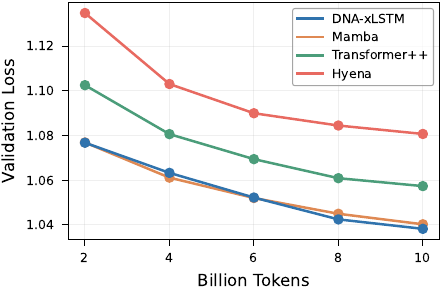}
        \includegraphics[width=0.49\textwidth]{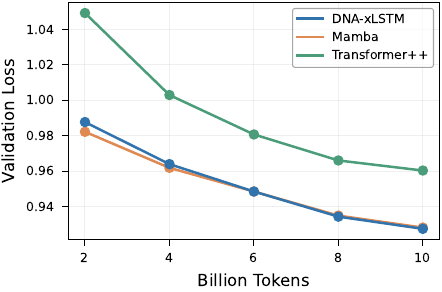}
        \caption{Pre-training of 4M-parameter DNA models on the human reference genome (GRCh38). The models are trained on the human reference genome at single-nucleotide resolution with a context length of 32k bases. \textbf{Left: causal language modeling.} Learning curves display \textbf{\ac{clm} loss} ($\downarrow$) on a held-out test set, plotted against the number of tokens processed. \textbf{Right: masked language modeling.} Learning curves for bidirectional models trained with the \textbf{\ac{mlm}} objective ($\downarrow$). The DNA-xLSTM-4M model outperforms both Transformer++ and Hyena-DNA models of similar size, and matches the performance of Caduceus-4M.}
        \label{fig:dna_pretraining:masked_32k}
    \end{figure}

\begin{figure}
        \centering
        \includegraphics[width=0.49\textwidth]{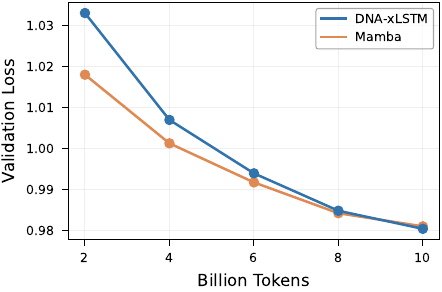}
        \includegraphics[width=0.49\textwidth]{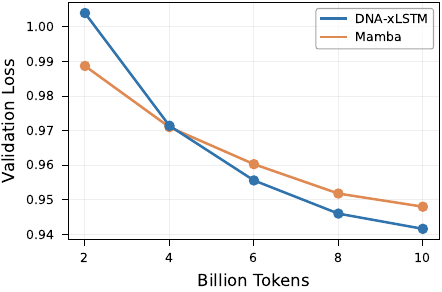}
        \caption{Pre-training of RC-equivariant xLSTM-DNA-PS and Caduceus-PS models with 500k and 2M parameters trained on the human reference genome. Models were trained on 1k context windows using the MLM objective. \textbf{Left:} \textbf{MLM losses} ($\downarrow$) for models with 500k parameters. \textbf{Right:} \textbf{MLM losses} ($\downarrow$) for models in the 2M parameter range. DNA-xLSTM-PS outperforms Caduceus-PS in both settings, with the performance gap widening at larger scales.}
        \label{fig:dna_pretraining:ps}
    \end{figure}

We further enhanced DNA-xLSTM-500k and DNA-xLSTM-2M models by incorporating reverse-complement equivariance via parameter sharing. For smaller models, we achieved MLM losses comparable to DNA-Mamba-PS, with a significant improvement over DNA-Mamba-PS as model size scaled to 2M parameters (Figure \ref{fig:dna_pretraining:ps}). Additionally, we pre-trained a long-range DNA-xLSTM model based on mLSTM, with a context size of 32k, using both CLM and MLM objectives. This model achieved the lowest cross-entropy loss in both tasks, outperforming Transformers and HyenaDNA, while performing comparably to Mamba (Figure \ref{fig:dna_pretraining:masked_32k}).

    \subsection{Downstream Tasks}
    \textbf{Experimental setup.} Two sets of downstream tasks were used
    for evaluating the learned representations: 
    the Genomic benchmark \citep{grevsova2023genomic} and
    the Nucleotide Transformers Tasks \citep{dalla2023nucleotide}, 
    which is a collection of 18 datasets derived from
    five peer-reviewed studies \citep{phaml2005qualitatively,oubounyt2019deepromoter,
    wang2019splicefinder, scalzitti2021spliceator, geng2022deep}. These classification tasks were selected to determine how rich the learned representations of the architectures are. To extract representations from the pre-trained xLSTM-DNA models, we perform average pooling on the activations from the final xLSTM block. For each downstream dataset, these representations served as inputs to a task-specific classification head that were jointly fine-tuned with the pre-trained model parameters.

    \textbf{Methods and hyperparameters.}
    For Nucleaotide Transformer tasks, we compared HyenaDNA, DNA-Mamba, and xLSTM-based 
    models pre-trained with 2M parameters. For Genomic benchmark tasks, we compare the smaller xLSTM-500k against Mamba. In both settings, models were pre-trained with a context size of 1,024. 

    \textbf{Metrics.} For the Nucleotide Transformer downstream tasks different metrics are used depending on the type of task: MCC was used for histone markers and enhancer annotation, F1-score was used for promoter annotation and splice site acceptor/donor, and accuracy was used for the splice site.
    The downstream performance on the Genomic benchmark was evaluated using the Top-1 accuracy. 
    
    \textbf{Results.} On the extensive set of downstream tasks, DNA-xLSTM is the best model with fewer than 2M parameters outperforming other small models on 12 of 18 tasks. In a comparison with much larger models, 
    DNA-xLSTM and is on par with the 500M parameter
    model Nucleotide Transformer (NT-v2) winning 8 of 
    18 tasks (see Table~\ref{tab:dna_nucleotide_downstream_table_full}). On the Genomic benchmark, DNA-xLSTM is overall on par with Mamba-DNA and shows especially strong results with posthoc conjoining, winning 5 of 8 tasks compared to Mamba-DNA-PH. Results are reported in Table \ref{tab:dna_genomics_downstream_direct_comparision}. 

    \begin{table}
        \centering
        \caption{Downstream adaption of DNA models (extended version). The test set performance of DNA models with 2M parameters and models with over 100M parameters, fine-tuned on Nucleotide Transformer classification tasks, is shown. Models marked with \ac{ps} or \ac{ph} are trained to be \ac{rc} equivariant. The used metric is provided in the \textit{Metric} column and best values are highlighted in green Results are averaged over 10 random seeds, with error bars representing the difference between the maximum and minimum values across the runs. The best scores are highlighted in green. xLSTM-DNA-PH with 2M parameters outperforms similarly sized Hyena- and Mamba-based models, while achieving comparable results to the much larger Nucleotide Transformer. Scores for all models except xLSTM were obtained from \citet{schiff2024caduceus}.}
        \label{tab:dna_nucleotide_downstream_table_full}
        \begin{threeparttable}
        \resizebox{\columnwidth}{!}{%

\input{tables/dna_nucleotide_transformer_downstream}

        }
        \begin{tablenotes}
        \end{tablenotes}
        \end{threeparttable}
    \end{table}

    \begin{table}
    \centering
         \caption{Downstream adaption of DNA language models 
         on the Genomics Benchmarks. The Top-1 \textbf{accuracy} ($\uparrow$) for RC-equivariant \ac{ps} and \ac{ph} xLSTM and Mamba-based Caduceus models, both with 500k parameters, are shown. Error bars represent the range of scores across five random seeds. xLSTM achieves comparable overall performance to Mamba and demonstrates superior accuracy when both models employ post-hoc conjoining. Scores for all models except xLSTM were obtained from \citet{schiff2024caduceus}.}
         \label{tab:dna_genomics_downstream_direct_comparision}
    \small 
    \resizebox{\textwidth}{!}{%
    \input{tables/dna_genomics_downstream_direct_comparision}
    }
    \end{table}

    \subsection{Architecture and Hyperparameters}
    The hyperparameters and composition of the DNA-xLSTM and DNA-Transformer++ models for pre-training with context size 1k and 32k are reported in Tables~\ref{tab:dna_xlstm_hyperparameters} and \ref{tab:dna_transformer_hyperparameters}.
    The hyperparameters were selected on a separate validation set using manual hyperparameter selection due to limited computational resources.

\begin{table}
    \centering
    \caption{Pre-training hyperparameters for DNA-Transformer++ models with 2M and 4M parameters. Comma-separated values represent hyperparameter sweeps, with the chosen values indicated in bold.}
    \label{tab:dna_transformer_hyperparameters}
    \begin{tabular}{lcc}
        \toprule
        Hyperparameters          & DNA-Transformer++-2M & DNA-Transformer++-4M \\
        \midrule
        \gcol Embedding Dimension      & 256      & 256 \\
        Number of Blocks         & 4        & 6   \\
        \gcol Number of Heads          & 8        & 8   \\
        Up-Projection Ratio  & 1.25:1      & 2:1 \\
        \gcol Norm Bias and Linear Bias & false    & false \\
        Context Length           & 1,024    & 32,768 \\
        \gcol Position Embeddings      & RoPE     & RoPE \\
        Learning Rate            & 6e-3, 8e-3, \textbf{1e-2} & 6e-3, 8e-3, \textbf{1e-2} \\
        \bottomrule
    \end{tabular}
\end{table}

    \subsection{Reverse-Complement Invariance}
    \label{sec:app_dna_reverse_complement}

        We develop an xLSTM version that is invariant
        to the \ac{rc} of an input sequence which 
        is relevant for DNA applications following \citet{schiff2024caduceus}.
        In double-helix DNA structures, both strands are semantically equivalent, 
        as one strand is the \ac{rc} of the other. 
        Given a strand, $\square$, its \ac{rc}, $\overline{\square}$, is
        oriented in the opposite direction with a base conversion from \texttt{A} to
        \texttt{T} and \texttt{C} to \texttt{G} \citep{schiff2024caduceus}.
        \citet{shrikumar2017reverse} show that a data-driven approach to learning the equivalence between reverse-complement sequences can fail, which is why \citet{schiff2024caduceus} 
        propose to enforce \ac{rc}-equivariance by design,
        making use of two different inductive 
        biases in the model architecture: \ac{ph} \citep{zhou2022towards} and \ac{ps}.
        For \ac{ph} models, 
        sequence-to-sequence models --- in our case realized by the xLSTM ---
        learn to handle both DNA sequences and their \ac{rc} during pre-training by applying \ac{rc} augmentation to the inputs. 
        \ac{rc} augmentation refers to the process of randomly replacing input sequences by their \acp{rc}.
        For downstream tasks \ac{ph} models are applied once to the original sequence and once to the \ac{rc} and eventually outputs are summed:
        \begin{align}
            \BY &= \mathrm{xLSTM}(\BX)+ \mathrm{xLSTM}(\overline \BX) \label{eq:phmodels}. 
        \end{align}
        For \ac{ps} models --- we assume models are realized by xLSTM architectures and therefore a $\mathrm{block}$ refers to a single mLSTM or sLSTM block --- both the DNA sequence and its \ac{rc} are provided simultaneously to each block in the architecture (for both pre-training and downstream task fine-tuning). Precisely, a joint representation, originating from combining a sequence representation and its RC representation, is split into $\BX \in \reals^{D \times t}$ and $\overline \BX \in \reals^{D \times t}$ and fed into the mLSTM or sLSTM block:
        \begin{align}
            \left[\BH, \overline \BH \right] &= \left[ \mathrm{block}(\BX), \mathrm{RC}(\mathrm{block}(\mathrm{RC}(\overline \BX))) \right].\label{eq:psmodel_block}
        \end{align}
        Notably, for each block the reverse-complement input is built by the $\mathrm{RC}$-function which flips both dimensions of $\overline \BX$ and $ \left[ \cdot,\cdot \right] $ indicates concatenation along the first dimension. Eventually, logits for the input sequence and its reverse complement are combined. For more details, we refer to \citet{schiff2024caduceus}.

\subsection{Implementation Details}
\label{app-sec:dna-xlstm-training}

For both CLM and MLM pre-training we perform 10,000 update steps holding the number of tokens per step constant at $2^{20}$. CLM models are trained using autoregressive next-token prediction. For MLM pre-training, we follow the methodology presented by \citet{devlin2019bert}, where 15\% of the input tokens are masked and the model is tasked to predict the corrupted tokens. Concretely, 80\% of the masked tokens are replaced by a special [MASK] token, 10\% are replaced by random tokens sampled from the vocabulary and 10\% remain unchanged. For MLM settings, we use weight-tied bidirectionality as a default (see Section \ref{sec:bidirectional_modelling}). For long-context bidirectional modeling, we use unidirectional xLSTM cells and alternate the modeling direction at each block.

To fine-tune pre-trained models on downstream tasks, we follow the framework from \citet{schiff2024caduceus}. Pre-trained models are augmented with a task-specific classification head, which is trained on average-pooled activations from a model's final block. During fine-tuning, all model parameters are unfrozen. For the Genomic benchmark, we perform five randomly seeded train-validation splits, fine-tune models for 10 epochs, and use early-stopping on validation performance. Final test results are reported as the mean performance $\pm$ max/min over the 5 seeds on a held-out test set. For the Nucleotide Transformer tasks, we use 20 epochs and 10 seeds. For both the Genomic benchmarks and the Nucleotide Transformer tasks, we performed a hyperparameter search for both PH and PS models over batch sizes $\{64, 128, 256, 512\},$ and learning rates $\{\text{4e-4}, \text{6e-4}, \text{8e-4}, \text{1e-3}, \text{2e-3}\}$. The best results for each Nucleotide Transformer task can be found in Table~\ref{tab:dna_nucleotide_hyperparams}.

    \begin{table}
        \centering
        \caption{Pre-training hyperparameters of DNA-xLSTM Models from 500k to 4M parameters. Comma-separated values represent hyperparameter sweeps, with the chosen values indicated in bold.}
        \label{tab:dna_xlstm_hyperparameters}
        \resizebox{\columnwidth}{!}{
        \begin{tabular}{lccc}
            \toprule
            Hyperparameters          & DNA-xLSTM-500k & DNA-xLSTM-2M & DNA-xLSTM-4M \\
            \midrule
  \gcol           Embedding Dimension      & 128      & 256      & 256 \\
            Number of Blocks         & 5        & 6        & 9   \\
   \gcol          Conv 1D Kernel Size      & 4        & 4        & 4   \\
            Number of Heads          & 4        & 4        & 4   \\
   \gcol          Up-Projection Ratio  & 1.25:1      & 1:1      & 2:1 \\
            Bidirectionality         & alternating, \textbf{blockwise} & alternating, \textbf{blockwise} & \textbf{alternating}, native, blockwise \\
   \gcol          Norm Bias and Linear Bias& true, \textbf{false} & true, \textbf{false} & true \\
            QKV Projection Blocksize & -        & -        & 4   \\
  \gcol           m/sLSTM ratio           & \textbf{[0:1]}, [1:0]      & \textbf{[0:1]}, [1:0]      & [0:1], \textbf{[1:0]} \\
            Context Length           & 1,024    & 1,024    & 32,768 \\
  \gcol           Position Embeddings      & None     & None     & RoPE \\
            Optimizer                & AdamW $\beta=(0.9, 0.95)$ & AdamW $\beta=(0.9, 0.95)$ & AdamW $\beta=(0.9, 0.95)$ \\
  \gcol           Learning Rate            & 6e-3, \textbf{8e-3}, 1e-2 & 6e-3, \textbf{8e-3}, 1e-2 & 6e-3, \textbf{8e-3}, 1e-2 \\
            Learning Rate Schedule   & Cosine Decay & Cosine Decay & Cosine Decay \\
  \gcol           Learning Rate Warmup Steps & 1,000    & 1,000    & 1,000 \\
            Weight Decay             & 0.1      & 0.1      & 0.1 \\
 \gcol            Dropout                  & 0        & 0        & 0   \\
            Batch Size               & 1,024    & 1,024    & 32  \\
 \gcol            Update Steps             & 10,000   & 10,000   & 10,000 \\
            \bottomrule
         \end{tabular}
        }
    \end{table}

    \begin{table}
        \centering
        \caption{Hyperparameter selection for DNA-xLSTM-PS and DNA-xLSTM-Ph on Nucleotide Transformer tasks. Fine-tuning hyperparameters were chosen based on best scores averaged over ten train-validation splits.}
        \label{tab:dna_nucleotide_hyperparams} 
        \input{tables/dna_nucleotide_downstream_hyperparams}
    \end{table}

\clearpage

\section{Prot-xLSTM: Details and Additional Results}\label{sec:app_prot}
    This section provides further details regarding the 
    architecture, training setup, and evaluation metrics for the Prot-xLSTM models. Additionally, we present supplementary results that complement the main findings discussed in Section~\ref{sec:results_prot}.  We adopted the experimental protocols outlined in \citet{sgarbossa2024protmamba} to train and evaluate our Prot-xLSTM models. We conducted three key experiments to assess the models' capabilities: \textbf{a) protein language modeling} (Section \ref{app-sec:prot_pretrain}), \textbf{b) homology-conditioned protein design} (Section \ref{app-sec:prot_generation}), and \textbf{c) protein variant fitness prediction} (Section \ref{app-sec:prot_fitness}).

    \subsection{Homology-Aware Training}\label{app-sec:prot_pretrain}
    
    \textbf{Data.} 
    The protein language model training data was derived from the filtered OpenProteinSet \citep{ahdritz2023openproteinset}, comprising 270k UniClust30 MSA clusters that included a total of 508M sequences and 110B residues. We used the ProtMamba pipeline to construct the training data, which is illustrated in Figure 1 of \citet{sgarbossa2024protmamba} and involved two key steps: (i) creating homology-aware but alignment-free training inputs by concatenating unaligned homologous sequences, and (ii) masking patches of tokens in each sequence and concatenating the unmasked patches at the end of each sequence to train the model autoregressively with the \ac{fim} strategy. We also use the train, validation (192 clusters), and test (500 clusters) split provided by ProtMamba. 

    \textbf{Methods and hyperparameters.}  
    We trained two versions of the model: Prot-xLSTM-26M and Prot-xLSTM-102M. The larger model was designed to match the architecture and scale of the original ProtMamba model (ProtMamba-107M) in terms of layer count and embedding size. We optimized the Prot-xLSTM architecture, including block types and positional encodings, on the smaller Prot-xLSTM-26M model, and then applied these optimized architectural choices to the larger Prot-xLSTM-102M model. For comparison, we also trained a smaller ProtMamba model (ProtMamba-28M with an embedding dimension of 512) and implemented a LLaMA-based model (Prot-Transformer++-26M) \citep{touvron2023llama}. Both models incorporate Absolute Positional Encodings (AbsPE) as implemented in ProtMamba for xLSTM blocks, with \ac{rope} applied specifically to the mLSTM blocks. As the sLSTM blocks (or Mamba) lack a $QK$-formulation, RoPE cannot be directly applied to them. The results of the hyperparameter search are provided in
    The results of the hyperparameter search are reported in Table~\ref{tab:prot_xlstm_hp_loss}, and the composition of the Prot-xLSTM and Prot-Transformer++ models are reported in Table~\ref{tab:prot_xlstm_hyperparameters}.

    \begin{table}[h]
    \centering
    \begin{threeparttable}[b]
    \caption{Hyperparameter space considered for the Prot-xLSTM and Prot-Transformer++ at different sizes. The selected values are marked in bold.}
    \label{tab:prot_xlstm_hyperparameters}
    \rowcolors{1}{}{jku_grey!10}
    \input{tables/proteins/protein_xlstm_hparams}
    \begin{tablenotes}
        \item[a] Context length was increased during training. \\
        \item[b] sLSTM blocks at position 1 and 15. \\
    \end{tablenotes}
    \end{threeparttable}
    \end{table}

    \begin{table}[h]
        \centering
        \caption{Prot-xLSTM hyperparameter search: Training loss comparison across different protein language model architectures after 4B training tokens.}
        \label{tab:prot_xlstm_hp_loss}
        \input{tables/proteins/protein_small_model_train_loss}
    \end{table}

    \textbf{Training details.}      
    We trained our models using the ProtMamba pipeline with \ac{clm} with the \ac{fim} strategy. 
    The pipeline efficiently handles long, concatenated sequences by extending the context length up to $T=2^{18}$, supported by a context-length scheduling strategy. For the Prot-xLSTM-102M model, we adhered to the ProtMamba protocol, gradually increasing the context length from $2^{11}$ to $T=2^{18}$, doubling $T$ at each stage when the loss plateaued. In contrast, for the smaller models (Prot-xLSTM-26M and ProtMamba-28M), as recommended in previous work \citep{devlin2019bert,press2021shortformer}, we initially trained with $T = 2^{11}$ for 20B tokens, then switched to $T = 2^{17}$ for an additional 10B tokens. Due to the quadratic scaling of Transformer architectures, Prot-Transformer++-26M was only trained with $T = 2^{11}$, as it could not handle the computational demands of $T = 2^{17}$. Given the substantial computational resources required, we did not fine-tune the training parameters. Instead, we used the default settings established by ProtMamba, which are reported in Table \ref{tab:prot_train_hyperparameters}. 

    \begin{table}[h]
    \centering
    \begin{threeparttable}[b]
    \caption{Hyperparameters for training protein sequence models.}
    \label{tab:prot_train_hyperparameters}
    \rowcolors{1}{}{jku_grey!10}
    \begin{tabular}{l>{\centering\arraybackslash}p{7cm}}
        \toprule 
        Effective batch size\tnote{a,b} & 64-1 \\
        Optimizer           & AdamW $\beta=(0.9, 0.95)$ \\
        Learning rate\tnote{b,c}         & 6e-4  \\
        Learning rate scheduler & constant  \\
        Learning rate warmup steps & 500 \\
        Weight decay & 0.1  \\
        Dropout & 0  \\
        \bottomrule    
    \end{tabular}
    \begin{tablenotes}
        \item[a] Decreased with context size to maintain a fixed total number of tokens per batch. For the larger model, the rule was relaxed for $T>=2^{16}$ to enable multi-GPU training, with the batch size set to the number of GPUs.  \\
        \item[b] Prot-Transformer++ was trained on 6 GPUs with an effective batch size of 96 and a learning rate of 9e-4. \\
        \item[c] Due to unstable training of the larger model at $T=2^{17}$ and $2^{18}$ the learning rate was reduced to 1e-4.
    \end{tablenotes}
    \end{threeparttable}
    \end{table}
   
    \textbf{Metrics.} During training, we evaluated the next-token prediction capabilities of the models using negative log-likelihood and token perplexity. 
    The perplexity was calculated for four subset of tokens of the concatenated-\ac{fim} sequence: the first protein sequence, the last protein sequence the \ac{fim} token and the entire concatenated sequence. 
    Once the models were trained we evaluated their performance on the independent test set.

    \subsection{ICL: Homology-Conditioned Protein Generation}\label{app-sec:prot_generation}

    \textbf{Experimental setup.} To evaluate the capacity of Prot-xLSTM to autoregressively generate novel protein sequences given a context of known homologs, we follow the protocol outlined in Section 3.4 of \citet{sgarbossa2024protmamba}. For a subset of 19 homology clusters from the test set, we generate sequences with contexts consisting of 10, 100, 500, 1000, and $N$ (total number of sequences in the cluster) sequences. For each context length, we generate 100 sequences each with the following parameter combinations of generation temperature ($\tau$), $\text{top-}k$, which restricts the output selection to the $k$ most probable tokens, and $\text{top-}p$, which limits the output to tokens reaching a cumulative probability $p$: $(\tau, \text{top-}k, \text{top-}p) \in \{(0.8, 10, 0.9), (0.9, 10, 0.95), (1, 10, 0.95), (1, 10, 1), (1, 15, 1)\}$ \citep{ferruz2022protgpt2}. This results in a total of 2,500 sequences per cluster.

    \textbf{Methods compared.} We compare both Prot-xLSTM models to ProtMamba models with a similar number of parameters.
    
    \textbf{Metrics.} We evaluate the novelty of generated sequences by calculating the Hamming distance to the closest natural sequence in the cluster using pairwise Smith-Waterman alignment. Additionally, we measure sequence similarity to homologs with the HMMER score from a Hidden Markov Model (HMM) trained on the cluster’s MSA. The generated sequences are also folded using ESMFold \citep{lin2023evolutionary} and assessed by pTM and average pLDDT confidence scores. To compare these scores with natural sequences, we compute Kolmogorov-Smirnov test statistics between the scores of 100 natural sequences and the 100 generated sequences with the lowest perplexity.

\begin{figure}[htb]
        \centering
        \includegraphics[width=\textwidth]{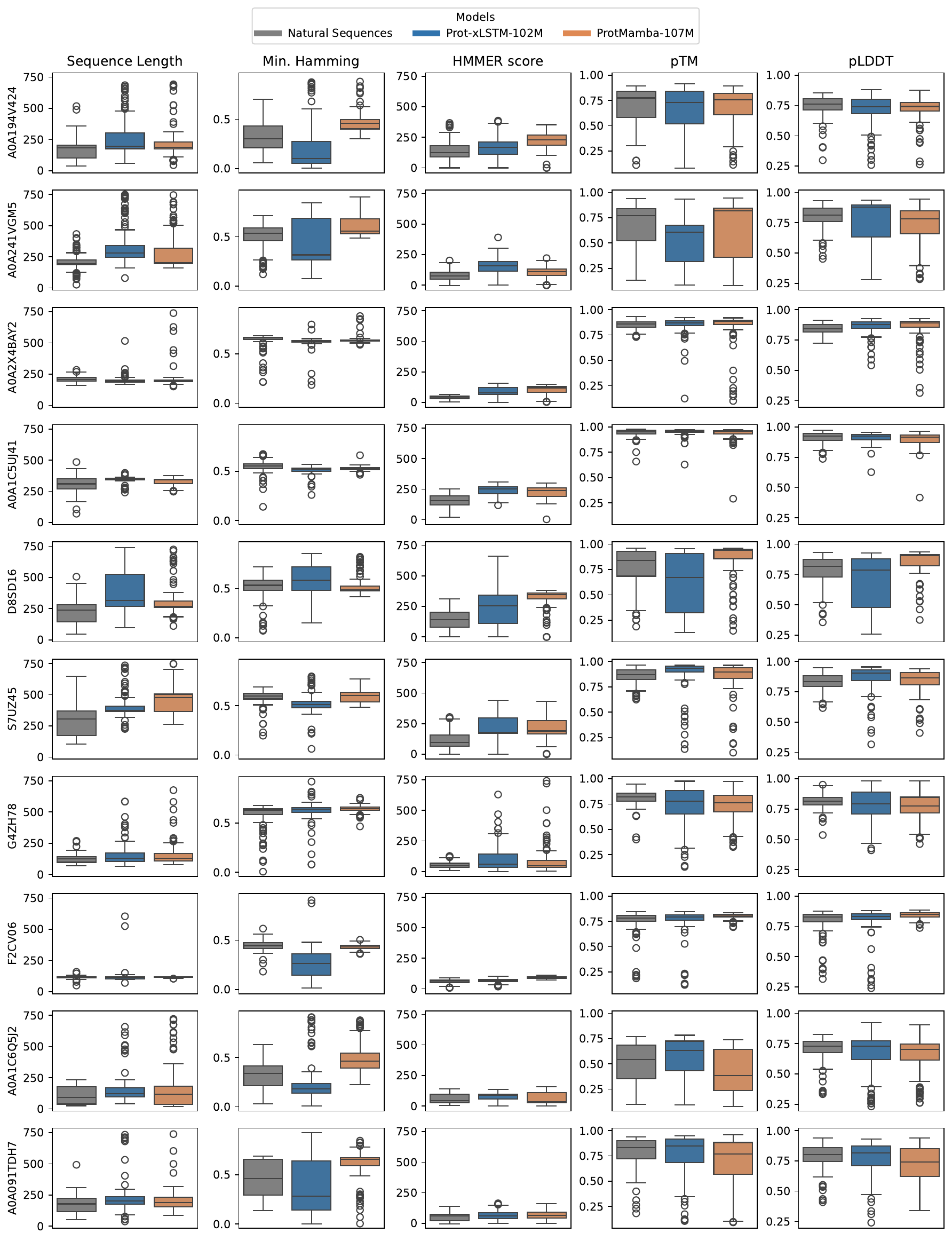}
        \caption{Scores of natural and generated proteins.
        Boxplots of score distributions (sequence length, Hamming distance to the closest natural neighbor, HMMER score, pLDDT, and pTM) for 10 randomly selected clusters evaluated for 100 randomly chosen natural sequences and 100 generated sequences with lowest perplexity values for large Prot-xLSTM and ProtMamba models.}
        \label{fig:prot-boxplost-scores}
    \end{figure}

    \textbf{Results.} Figure \ref{fig:prot-boxplost-scores} displays the distribution of scores for 100 randomly sampled natural sequences from each cluster as well as the 100 sequences with the lowest perplexity generated by Prot-xLSTM and ProtMamba models for 10 randomly selected clusters. The averages across all 19 evaluated test clusters are shown in Table \ref{tab:prot-abs-scores}. Sequences generated by Prot-xLSTM-102M were on average longer, more similar to other proteins in the cluster (measured by Hamming distance), and got a higher HMMER score and higher folding confidence scores compared to ProtMamba-generated sequences. Notably, these observations mostly also hold when compared to natural sequences.

    \begin{table}[htb]
        \centering
        \begin{threeparttable}[b]
        \caption{Score comparison of natural and generated proteins. Average scores (sequence length, Hamming distance to the closest natural neighbor, HMMER score, pLDDT, and pTM) across 19 test clusters for sequences generated with Prot-xLSTM and ProtMamba models. Error bars indicate 95\% confidence intervals across clusters.}
        \small
        \label{tab:prot-abs-scores}
        \centering
        \input{tables/proteins/protein_sequence_evaluation_abs}
        \end{threeparttable}
    \end{table}
    
    Table \ref{tab:prot-ppl-r} demonstrates that Hamming distance, HMMER score, pTM, and pLDDT correlate well with sequence perplexity for both, Prot-xLSTM and ProtMamba, with an average Pearson correlation coefficient of 0.57 and 0.58, respectively, for the large models.

    \begin{table}[htb]
        \centering
        \begin{threeparttable}[b]
        \caption{Score distribution comparison of natural and generated proteins. Average Pearson correlation between model perplexity and sequence scores (sequence length, Hamming distance to the closest natural neighbor, HMMER score, pLDDT, and pTM) for sequences generated with Prot-xLSTM and ProtMamba models. Error bars indicate 95\% confidence intervals across 19 test clusters.}
        \small
        \label{tab:prot-ppl-r}
        \centering
        \input{tables/proteins/protein_sequence_evaluation_ppl_r}
        \end{threeparttable}
    \end{table}

    \subsection{Protein Variant Fitness Prediction}\label{app-sec:prot_fitness}

    \textbf{Experimental setup.}
    We evaluate Prot-xLSTM's ability to predict mutational effects by leveraging its inpainting capabilities from the \ac{fim} training objective. This assessment follows the protocol described in Section 3.2 of \citet{sgarbossa2024protmamba} for the ProteinGym DMS substitution benchmark \citep{notin2023proteingym}, which consists of 217 datasets of single and multiple substitutions in protein sequences, allowing comparison with state-of-the-art methods for protein variant fitness prediction. 
    Briefly, for each wild-type sequence, three sets of 200 homologs were obtained by subsampling MSAs following the ColabFold protocol \citep{mirdita2022colabfold} to be used as context. The context sequences are ordered from the least similar to the most similar one. The wild-type sequence is then concatenated with the context, the mutated residue is masked, and this residue is predicted using the \ac{fim} method. Fitness is evaluated as the difference in likelihood between the concatenated sequence with the wild-type and the mutated amino acid and averaged over the triplicate. For multiple mutations, fitness is approximated as the sum of the likelihoods of single mutations.

    \textbf{Results.} ProteinGym’s main metric is the average Spearman correlation between the fitness predictions and the experimental DMS results.
    Table \ref{tab:prot-gym-perf} summarizes reports ProteinGym’s main metric, the average Spearman correlation between the fitness predictions and the experimental DMS results, for Prot-xLSTM and several other well-known protein models and the current top of the 

    \begin{table}[htb]
        \centering
        \caption{ProteinGym zero-shot DMS substitution benchmark. The average \textbf{Spearman correlation} ($\uparrow$) between predicted fitness scores and experimental measures over 217 DMS assays is shown. While even small Prot-xLSTM models already yield high scores, models which use additional structure tokens, such 
        as SaProt and ProSST, perform best.}

\input{tables/proteins/proteingym}
        \label{tab:prot-gym-perf}
    \end{table}

\clearpage
\section{Chem-xLSTM: Details and Additional Results}
\label{app-sec:chem}

    For chemical sequences, we perform two sets of experiments: \textbf{a) unconditional molecule generation} 
    where we follow the experimental protocol of \citet{ozccelik2024chemical}. 
    Additionally, we propose a new and more challenging task: 
    \textbf{b)  conditional generation with \ac{icl}}, in which we generate new compounds conditional based on provided in-context compounds.
    
\subsection{Unconditional Molecule Generation}\label{app-sec:chem-unconditional}
    \emph{Unconditional molecule generation} is the task of generating valid small molecules without imposing constraints on their characteristics or properties.
    Generative models aim to learn a general distribution by processing many examples of desirable results.
    To this end, models are trained on large training sets of arbitrary small molecules without particular conditions or constraints \citep{segler2018generating, gomez2018automatic,ozccelik2024chemical}.
    Following this approach, we compared the ability of {xLSTM} and several other models to generate valid and diverse molecules.

    \textbf{Experimental setup.}
    For comparability, we aligned our experiments with the setting and dataset of \citet{ozccelik2024chemical}.
    This means that all models are trained to generate molecules as \ac{smiles} strings \citep{weininger1988smiles} using a \ac{clm} paradigm.
    The dataset used in \citep{ozccelik2024chemical} is derived from ChEMBL with a random split in 1.9M~training, 100k~validation, and 23k~test molecules, which have been encoded as \ac{smiles}.
    Before training, all \ac{smiles} strings were tokenized using a regular expression, containing all elements. This results in atoms being represented as one token as well as additional SMILES symbols.

    \textbf{Methods and hyperparameters.}
    We compared {xLSTM} with several other model classes.
    The first baseline is the default LSTM \citep{hochreiter1997lstm} in PyTorch, which includes a forget gate \citep{gers1999learning}.
    This can be considered the direct predecessor of the {xLSTM} architecture.
    We also included a variant GPT-2 \citep{radford2019gpt2} model based on the Transformer architecture \citep{vaswani2017attention} with causal masking.
    Finally, we included two \acp{ssm} in our comparison.
    On one side, we considered an S4 model with the implementation from \citet{gu2022s4}, following \citep{ozccelik2024chemical}. 
    On the other side, we incorporated a Mamba model, using the official repository provided with \citep{gu2024mamba}.
    For our Chem-xLSTM, we used an xLSTM using only mLSTM blocks \citep{beck2024xlstm}.
    The 15M-parameter model consists of 9~layers with a hidden dimension of~512 and 8 heads.
    We trained the model for up to 100~epochs with a batch size of~1,024, a context length of~100, a dropout rate of~0.25, and a learning rate of~0.005.
    All models were trained using the Adam optimizer \citep{kingma2015adam} using $\beta = (0.9, 0.999)$, $\epsilon = {1e^{-8}}$, and a learning-rate schedule with warm-up and cosine decay. 
    We selected the best model based on the minimum validation loss observed at the end of each epoch. 
    The hyperparameters were manually tuned to match the model parameter count for a fair comparison.
    Detailed training information and learning curves can be found in Appendix~Section~\ref{app-sec:unconditional-xlstm-training}.

    \textbf{Metrics.}
    We evaluated each model with the next token perplexity next token, and the \ac{fcd} \citep{preuer2018frechet}.
    The \ac{fcd} has been introduced as an alternative to the FID, which is used to evaluate image generation, for molecule generation. 
    Auxiliary metrics that measure the syntactic correctness, novelty, diversity, or synthetic accessibility are reported in Appendix~Section~\ref{app-sec:chem}.

    \textbf{Results.}
    Our proposed Chem-xLSTM model achieved the best results, with the lowest \ac{fcd} (0.13) and a perplexity (1.68) that is competitive with that of GPT-based models. 
    This indicates that Chem-XLSTM is able to generate realistic chemical structures that match the target distribution well. 
    
    All models in our comparison were able to produce valid, unique, and novel molecules.
    Even though these models have not been optimized for these properties.
    This is evidenced by the auxiliary metrics surpassing practical thresholds (see Table~\ref{tab:chem_auxiliary}).

\subsection{Conditional Molecule Generation with In-Context Learning}\label{app-sec:chem_conditional}
Conditional molecule generation with in-context learning (ICL) leverages contextual information to guide the design of novel molecules tailored for specific domains. By incorporating a sequence of molecules as the input, models can conditionally generate new compounds of the same distribution, without the need for fine-tuning. 

    \textbf{Experimental setup.}
Similar to the unconditional setup, the input consists of \ac{smiles} strings.
In the conditional setup, we additionally model sets of molecules 
from the same molecular domain as a sequence.
Molecules from one molecular domain are serialized and concatenated, 
separated with the "\texttt{.}" token.
During training, the order of the molecules is permuted to improve generalization and robustness.
We construct a novel dataset derived from a variety of molecular domains:
\begin{itemize}
    \item We consider \texttt{natural-products} as domain and utilize the Coconut \citep{nainala2024coconut} as source dataset.
    \item \texttt{Kinase inhibitors}, 
    \texttt{withdrawn}, 
    \texttt{malaria}, 
    \texttt{tool compounds}, 
    \texttt{pathogen}, 
    \texttt{NIH mechanistic}, 
    \texttt{lopac},
    \texttt{natural product-based probes and drugs},
    \texttt{zinc tool},
    \texttt{axon medchem},
    \texttt{adooq bioactive}, 
    \texttt{novartis chemogenetic},
    \texttt{drug matrix},
    \texttt{PROTACs},    
    \texttt{covalentIn db},
    \texttt{DrugBank compounds},
    \texttt{reframe},
    \texttt{cayman bioactive} all from the Probes~\&~Drugs portal \citep{skuta2017probes},
    \item \texttt{product molecules} from the reaction dataset USPTO-50k \citep{lowe2012extraction} split into 10 reaction classes. 
    \item The domains \texttt{bio}, 
    \texttt{diversity}, \texttt{green}, \texttt{yellow}, 
    \texttt{orange}, and \texttt{red}, 
    from ZINClick \citep{levre2018zinclick}.
    \item Active molecules from 
    the domains \texttt{BACE}, \texttt{BBBP}, 
    \texttt{Clintox}, \texttt{HIV}, \texttt{SIDER}, 
    \texttt{Tox21}, \texttt{Tox21-10k}, and \texttt{Toxcast}
    from MoleculeNet \citep{wu2018moleculenet}.
    \item Active molecules from 95 bioassays from FS-MOL \citep{stanley2021fs} considered each as separate domain.
    \item Active molecules from 109 bioassays from PubChem \citep{kim2023pubchem} considered each as separate domain.
    \item A subset of active molecules from the BELKA challenge \citep{quigley2024belka} is modeled as a domain.
\end{itemize}

For the domains that are defined by the active molecules from a particular 
bioassay, we selected assays with at least 300 active molecules 
and only use the active compounds. For the dataset each of the total 249 domains is 
limited to 100,000 compounds, where compounds are selected at random. 
The final dataset is split at 8:1:1 into train-, validation- and test-domains, sorted by their character length in descending order.

    \textbf{Methods and hyperparameters.}
     We benchmark and orient our choices for the model classes as well as hyperparameters based on the unconditional molecule generation results, %
     We consider a context length of 4,096 and adjust batch sizes as well as accumulation steps to accommodate GPU memory constraints. 
     For the S4 model, we were only able to fit a context length of 2,048.

    \textbf{Metrics.}
    To evaluate conditional molecule generation we evaluate NTP loss. This metric quantifies how well the model predicts the next token in a sequence, thus assessing whether a model is able 
    to generate molecules from an unseen, and potentially very special, 
    molecular domain given only a few molecules from that domain.

    \subsection{Architecture and Hyperparameter Selection}
    Considered and selected hyperparameters for Chem-xLSTM
    are given in~\ref{tab:hyperparameter_chem}.

     \begin{table}
            \centering
            \begin{threeparttable}[b]
            \caption{Hyperparameter space considered for the 
            Chem-xLSTM at different sizes. The selected
            values are marked in bold. \label{tab:hyperparameter_chem}}
            \begin{tabular}{lcc}
            \toprule Hyperparameter & Chem-xLSTM-15Mn & Chem-xLSTM-15M-icl \\
                \midrule
           \gcol     Number of layers & \textbf{9} & \textbf{9} \\
                Number of heads & \textbf{8} & \textbf{8} \\
            \gcol    Embedding dimension & \textbf{512} & \textbf{512} \\
                Hidden dimension & \textbf{512}&  \textbf{512} \\
            \gcol    Batch size & 16, \textbf{32}, 64, 128 &  16, \textbf{32} \\
                Proj. factor & \textbf{1.3} & \textbf{1.3} \\
           \gcol     Learning rate & 1e-4, \textbf{2e-4}, 3e-4, 5e-4 &  16,  1e-4, \textbf{2e-4}, 3e-4, 5e-4 \\
                Optimizer &  \textbf{Adam}, AdamW &  \textbf{Adam} \\
            \bottomrule    
            \end{tabular}
            \end{threeparttable}
        \end{table}

\subsection{Implementation Details}
    \label{app-sec:unconditional-xlstm-training}
    Unlike \citet{ozccelik2024chemical}, we do not backpropagate the loss for \texttt{[PAD]} tokens, nor do we interpret them for decoding. We observed that not ignoring \texttt{[EOS]} and \texttt{[PAD]} token leads to more diversity but is not the standard way of decoding in e.g. NLP. Padding tokens are not typically generated during decoding. They are primarily a pre-processing step to handle batches of data efficiently. In our implementation, we end decoding the \ac{smiles} string with the \texttt{[EOS]} token. %
    Further, we do not use \ac{smiles} augmentation, which could further 
    improve the performance of all architectures.

\subsection{Additional Results}
    Practical thresholds are defined based on several key metrics. First, a high percentage of generated SMILES strings must correspond to chemically valid molecules, with a threshold typically set above 90\% to ensure reliability. Additionally, a practical threshold for uniqueness might require that over 80\% of the generated molecules are unique, ensuring diversity in the explored chemical space. For novelty, at least 50-70\% of the generated molecules should be novel compared to known chemical databases, indicating the model's ability to explore new regions of chemical space. Finally, all models exhibit favorable synthetic accessibility (SA) scores, typically ranging between 2.5 and 5, ensuring that the generated molecules are feasible for synthesis. Further metrics and details are provided in the appendix.

\begin{table}
    \caption{Diversity and correctness metrics for the 15M parameter models for small molecules (SMILES). The table reports the percentage of valid, unique, and novel molecules, the synthetic accessibility (SA), and the diversity metric by the percentage of unique Murcko scaffolds divided by the total number of generated molecules.
    \label{tab:chem_auxiliary}
    }
    \centering
    \begin{threeparttable}
    \footnotesize
    \begin{tabular}{lllllll}
    \toprule
     Model & valid \% & unique \% & novel \% & SA ↓ & diverse \% \\
    \midrule
   \gcol  SMILES-LSTM \citep{segler2018generating} & ${90.11^{\pm{10.7}}}$ & ${56.72^{\pm{3.4}}}$ & ${56.66^{\pm{3.6}}}$ & ${2.85^{\pm{0.0}}}$ & ${44.71^{\pm{1.1}}}$ \\
    SMILES-GPT \citep{adilov2021generative} & ${99.05^{\pm{0.5}}}$ & ${62.09^{\pm{12.1}}}$ & ${61.81^{\pm{12.0}}}$ & ${2.90^{\pm{0.0}}}$ & ${48.82^{\pm{9.7}}}$ \\
     \gcol SMILES-S4 \citep{ozccelik2024chemical} & ${97.48^{\pm{0.0}}}$ & ${61.47^{\pm{0.0}}}$ & ${61.34^{\pm{0.0}}}$ & ${2.86^{\pm{0.0}}}$ & ${48.49^{\pm{0.0}}}$ \\
    Chem-Mamba\tnote{a}  & ${91.41^{\pm{8.9}}}$ & ${57.75^{\pm{3.2}}}$ & ${57.63^{\pm{3.8}}}$ & ${2.84^{\pm{0.0}}}$ & ${45.65^{\pm{7.2}}}$ \\
     \gcol Chem-xLSTM (ours) & ${97.08^{\pm{0.7}}}$ & ${61.09^{\pm{8.9}}}$ & ${60.84^{\pm{9.6}}}$ & ${2.83^{\pm{0.0}}}$ & ${45.97^{\pm{5.5}}}$ \\
    \bottomrule
    \end{tabular}
    \begin{tablenotes}
        \item[a] adapted to \ac{smiles} in this work
    \end{tablenotes}
    \end{threeparttable}

\end{table}

\end{document}

%% file: tables/dna_NTv2_smallModels_v2.tex
\begin{tabular}{l c c c c c c} 
\toprule 
Task & Metric & HyenaDNA & Mamba-PS\tnote{a} & Mamba-PH\tnote{a} & xLSTM-PS & xLSTM-PH \\ 
\midrule 
\multicolumn{7}{l}{\textit{Histone Markers}} \\ 
\cellcolor{jku_grey!10}H3 & \cellcolor{jku_grey!10}MCC $\uparrow$ & \cellcolor{jku_grey!10}$0.779^{\pm 0.037}$ & \cellcolor{jku_grey!10}$0.799^{\pm 0.029}$ & \cellcolor{jku_grey!10}$0.815^{\pm 0.048}$ & \cellcolor{jku_grey!10}$0.796^{\pm 0.014}$ & \cellcolor{jku_green!20}$0.824^{\pm 0.010}$ \\ 
H3K14AC & MCC $\uparrow$ & $0.612^{\pm 0.065}$ & $0.541^{\pm 0.212}$ & \cellcolor{jku_green!20}$0.631^{\pm 0.026}$ & $0.570^{\pm 0.008}$ & $0.598^{\pm 0.017}$ \\ 
\cellcolor{jku_grey!10}H3K36ME3 & \cellcolor{jku_grey!10}MCC $\uparrow$ & \cellcolor{jku_grey!10}$0.613^{\pm 0.041}$ & \cellcolor{jku_grey!10}$0.609^{\pm 0.109}$ & \cellcolor{jku_grey!10}$0.601^{\pm 0.129}$ & \cellcolor{jku_grey!10}$0.588^{\pm 0.019}$ & \cellcolor{jku_green!20}$0.625^{\pm 0.010}$ \\ 
H3K4ME1 & MCC $\uparrow$ & $0.512^{\pm 0.024}$ & $0.488^{\pm 0.102}$ & $0.523^{\pm 0.039}$ & $0.490^{\pm 0.012}$ & \cellcolor{jku_green!20}$0.526^{\pm 0.001}$ \\ 
\cellcolor{jku_grey!10}H3K4ME2 & \cellcolor{jku_grey!10}MCC $\uparrow$ & \cellcolor{jku_grey!10}$0.455^{\pm 0.095}$ & \cellcolor{jku_grey!10}$0.388^{\pm 0.101}$ & \cellcolor{jku_grey!10}$0.487^{\pm 0.170}$ & \cellcolor{jku_grey!10}$0.489^{\pm 0.024}$ & \cellcolor{jku_green!20}$0.504^{\pm 0.012}$ \\ 
H3K4ME3 & MCC $\uparrow$ & \cellcolor{jku_green!20}$0.549^{\pm 0.056}$ & $0.440^{\pm 0.202}$ & $0.544^{\pm 0.045}$ & $0.520^{\pm 0.019}$ & $0.537^{\pm 0.012}$ \\ 
\cellcolor{jku_grey!10}H3K79ME3 & \cellcolor{jku_grey!10}MCC $\uparrow$ & \cellcolor{jku_grey!10}$0.672^{\pm 0.048}$ & \cellcolor{jku_grey!10}$0.676^{\pm 0.026}$ & \cellcolor{jku_grey!10}$0.697^{\pm 0.077}$ & \cellcolor{jku_grey!10}$0.662^{\pm 0.011}$ & \cellcolor{jku_green!20}$0.697^{\pm 0.007}$ \\ 
H3K9AC & MCC $\uparrow$ & $0.581^{\pm 0.061}$ & $0.604^{\pm 0.048}$ & $0.622^{\pm 0.030}$ & $0.622^{\pm 0.013}$ & \cellcolor{jku_green!20}$0.627^{\pm 0.008}$ \\ 
\cellcolor{jku_grey!10}H4 & \cellcolor{jku_grey!10}MCC $\uparrow$ & \cellcolor{jku_grey!10}$0.763^{\pm 0.044}$ & \cellcolor{jku_grey!10}$0.789^{\pm 0.020}$ & \cellcolor{jku_grey!10}$0.811^{\pm 0.022}$ & \cellcolor{jku_grey!10}$0.793^{\pm 0.011}$ & \cellcolor{jku_green!20}$0.813^{\pm 0.008}$ \\ 
H4AC & MCC $\uparrow$ & $0.564^{\pm 0.038}$ & $0.525^{\pm 0.240}$ & \cellcolor{jku_green!20}$0.621^{\pm 0.054}$ & $0.558^{\pm 0.018}$ & $0.583^{\pm 0.014}$ \\ 
\midrule 
\multicolumn{7}{l}{\textit{Regulatory Annotation}} \\ 
\cellcolor{jku_grey!10}Enhancer & \cellcolor{jku_grey!10}MCC $\uparrow$ & \cellcolor{jku_grey!10}$0.517^{\pm 0.117}$ & \cellcolor{jku_grey!10}$0.491^{\pm 0.066}$ & \cellcolor{jku_green!20}$0.546^{\pm 0.073}$ & \cellcolor{jku_grey!10}$0.375^{\pm 0.030}$ & \cellcolor{jku_grey!10}$0.545^{\pm 0.024}$ \\ 
Enhancer Types & MCC $\uparrow$ & $0.386^{\pm 0.185}$ & $0.416^{\pm 0.095}$ & $0.439^{\pm 0.054}$ & $0.444^{\pm 0.046}$ & \cellcolor{jku_green!20}$0.466^{\pm 0.011}$ \\ 
\cellcolor{jku_grey!10}Promoter: All & \cellcolor{jku_grey!10}F1 $\uparrow$ & \cellcolor{jku_grey!10}$0.960^{\pm 0.005}$ & \cellcolor{jku_grey!10}$0.967^{\pm 0.004}$ & \cellcolor{jku_green!20}$0.970^{\pm 0.004}$ & \cellcolor{jku_grey!10}$0.962^{\pm 0.002}$ & \cellcolor{jku_grey!10}$0.967^{\pm 0.001}$ \\ 
NonTATA & F1 $\uparrow$ & $0.959^{\pm 0.011}$ & $0.968^{\pm 0.006}$ & $0.968^{\pm 0.010}$ & $0.963^{\pm 0.002}$ & \cellcolor{jku_green!20}$0.970^{\pm 0.001}$ \\ 
\cellcolor{jku_grey!10}TATA & \cellcolor{jku_grey!10}F1 $\uparrow$ & \cellcolor{jku_grey!10}$0.944^{\pm 0.040}$ & \cellcolor{jku_green!20}$0.957^{\pm 0.015}$ & \cellcolor{jku_grey!10}$0.953^{\pm 0.016}$ & \cellcolor{jku_grey!10}$0.948^{\pm 0.006}$ & \cellcolor{jku_grey!10}$0.952^{\pm 0.005}$ \\ 
\midrule 
\multicolumn{7}{l}{\textit{Splice Site Annotation}} \\ 
\cellcolor{jku_grey!10}All & \cellcolor{jku_grey!10}Accuracy $\uparrow$ & \cellcolor{jku_grey!10}$0.956^{\pm 0.011}$ & \cellcolor{jku_grey!10}$0.927^{\pm 0.021}$ & \cellcolor{jku_grey!10}$0.940^{\pm 0.027}$ & \cellcolor{jku_grey!10}$0.965^{\pm 0.006}$ & \cellcolor{jku_green!20}$0.974^{\pm 0.004}$ \\ 
Acceptor & F1 $\uparrow$ & $0.958^{\pm 0.010}$ & $0.936^{\pm 0.077}$ & $0.937^{\pm 0.033}$ & \cellcolor{jku_green!20}$0.970^{\pm 0.005}$ & $0.953^{\pm 0.008}$ \\ 
\cellcolor{jku_grey!10}Donor & \cellcolor{jku_grey!10}F1 $\uparrow$ & \cellcolor{jku_grey!10}$0.949^{\pm 0.024}$ & \cellcolor{jku_grey!10}$0.874^{\pm 0.289}$ & \cellcolor{jku_grey!10}$0.948^{\pm 0.025}$ & \cellcolor{jku_green!20}$0.962^{\pm 0.004}$ & \cellcolor{jku_grey!10}$0.951^{\pm 0.005}$ \\ 
\bottomrule 
\end{tabular}

%% file: tables/proteins/protein_testset_results.tex
\begin{threeparttable}
    \footnotesize
        \begin{tabular}{lcccc}
        \toprule
        &  Prot-xLSTM-26M & ProtMamba-28M & Prot-xLSTM-102M\tnote{a} & ProtMamba-107M\tnote{b} \\
        \midrule
        \rowcolor{jku_grey!10}All tokens & $8.73^{\pm{0.31}}$ & $10.15^{\pm{0.32}}$ & \cellcolor{jku_green!20}$6.83^{\pm0.25}$ ($6.65^{\pm0.25}$) & $7.47^{\pm0.26}$ \\
        First seq tokens & $15.40^{\pm{0.26}}$ & $15.28^{\pm{0.26}}$ & $13.36^{\pm0.35}$ ($13.41^{\pm0.35}$) & \cellcolor{jku_green!20}$13.04^{\pm0.36}$\\
        \rowcolor{jku_grey!10}Last seq tokens & $9.19^{\pm{0.30}}$ & $11.08^{\pm{0.27}}$ & \cellcolor{jku_green!20}$7.32^{\pm0.29}$ ($7.14^{\pm0.29}$)  & $8.37^{\pm0.29}$ \\
        FIM tokens & $6.77^{\pm{0.25}}$ & $7.96^{\pm{0.27}}$ & \cellcolor{jku_green!20}$5.52^{\pm0.20}$ ($5.50^{\pm0.20}$)  & $6.47^{\pm0.23}$  \\
        \bottomrule
        \end{tabular}
    \begin{tablenotes}
        \item[a] Trained for 60B tokens with $T$ up to $2^{17}$. Trained for 115B tokens with $T$ up to $2^{18}$ in parentheses.
        \item[b] Trained for 195B tokens with $T$ up to $2^{17}$.
    \end{tablenotes}
\end{threeparttable}

%% file: tables/proteins/protein_sequence_evaluation_ks.tex
\begin{tabular}{lcccc} 
    \toprule 
    & Prot-xLSTM-26M & ProtMamba-28M &  Prot-xLSTM-102M & ProtMamba-107M  \\
    \midrule 
    \rowcolor{jku_grey!10} Sequence Length & $0.41^{\pm 0.09}$ & $0.52^{\pm 0.09}$ & $0.40^{\pm 0.08}$ & \cellcolor{jku_green!20}$0.36^{\pm 0.08}$ \\
    Min. Hamming & $0.43^{\pm 0.08}$ & $0.60^{\pm 0.11}$ & $0.47^{\pm 0.09}$ & \cellcolor{jku_green!20}$0.42^{\pm 0.07}$ \\ 
    \rowcolor{jku_grey!10} HMMER & $0.57^{\pm 0.10}$ & $0.54^{\pm 0.11}$ & \cellcolor{jku_green!20}$0.44^{\pm 0.09}$ & $0.49^{\pm 0.10}$ \\
    pLDDT & $0.40^{\pm 0.09}$ & $0.68^{\pm 0.12}$ & \cellcolor{jku_green!20}$0.27^{\pm 0.05}$ & $0.30^{\pm 0.07}$ \\
    \rowcolor{jku_grey!10} pTM & $0.38^{\pm 0.08}$ & $0.72^{\pm 0.10}$ & \cellcolor{jku_green!20}$0.26^{\pm 0.05}$ & $0.28^{\pm 0.05}$ \\
    \bottomrule 
\end{tabular}
\begin{tablenotes}
\end{tablenotes}

%% file: tables/dna_nucleotide_transformer_downstream.tex
\begin{tabular}{l c | c c c | c c c c c}
    \toprule
    Task & Metric & \multicolumn{3}{c|}{> 100M Param. Models} & \multicolumn{5}{c}{2M Param. Models} \\
    & & Enformer (252M) & DNABERT-2 (117M) & NT-v2 (500M) & HyenaDNA & Mamba-PS & Mamba-PH & xLSTM-PS & xLSTM-PH \\
    
    \midrule
    
    \multicolumn{10}{l}{\textit{Histone Markers}} \\
    
    \cellcolor{jku_grey!10}H3 & \cellcolor{jku_grey!10}MCC $\uparrow$ &
    \cellcolor{jku_grey!10}$0.719^{\pm 0.048}$ &
    \cellcolor{jku_grey!10}$0.785^{\pm 0.033}$ &
    \cellcolor{jku_grey!10}$0.784^{\pm 0.047}$ &
    \cellcolor{jku_grey!10}$0.779^{\pm 0.037}$ &
    \cellcolor{jku_grey!10}$0.799^{\pm 0.029}$ &
    \cellcolor{jku_grey!10}$0.815^{\pm 0.048}$ &
    \cellcolor{jku_grey!10}$0.796^{\pm 0.014}$ &
    \cellcolor{jku_green!20}$0.824^{\pm 0.010}$ \\

    H3K14AC & MCC $\uparrow$ &
    $0.288^{\pm 0.077}$ & 
    $0.516^{\pm 0.028}$ & 
    $0.551^{\pm 0.021}$ & 
    $0.612^{\pm 0.065}$ & 
    $0.541^{\pm 0.212}$ & 
    \cellcolor{jku_green!20}$0.631^{\pm 0.026}$ &
    $0.570^{\pm 0.008}$ &
    $0.598^{\pm 0.017}$ \\

    \cellcolor{jku_grey!10}H3K36ME3 & \cellcolor{jku_grey!10}MCC $\uparrow$ & 
    \cellcolor{jku_grey!10}$0.344^{\pm 0.055}$ &
    \cellcolor{jku_grey!10}$0.591^{\pm 0.020}$ & 
    \cellcolor{jku_grey!10}$0.625^{\pm 0.030}$ &
    \cellcolor{jku_grey!10}$0.613^{\pm 0.041}$ & 
    \cellcolor{jku_grey!10}$0.609^{\pm 0.109}$ & 
    \cellcolor{jku_grey!10}$0.601^{\pm 0.129}$ &
    \cellcolor{jku_grey!10}$0.588^{\pm 0.019}$ &
    \cellcolor{jku_green!20}$0.625^{\pm 0.010}$ \\

    H3K4ME1 & MCC $\uparrow$ &
    $0.291^{\pm 0.061}$ & 
    $0.511^{\pm 0.028}$ & 
    \cellcolor{jku_green!20}$0.550^{\pm 0.021}$ & 
    $0.512^{\pm 0.024}$ & 
    $0.488^{\pm 0.102}$ &
    $0.523^{\pm 0.039}$ &
    $0.490^{\pm 0.012}$ &
    $0.526^{\pm 0.001}$ \\

    \cellcolor{jku_grey!10}H3K4ME2 & \cellcolor{jku_grey!10}MCC $\uparrow$ &
    \cellcolor{jku_grey!10}$0.211^{\pm 0.069}$ & 
    \cellcolor{jku_grey!10}$0.336^{\pm 0.040}$ & 
    \cellcolor{jku_grey!10}$0.319^{\pm 0.045}$ & 
    \cellcolor{jku_grey!10}$0.455^{\pm 0.095}$ & 
    \cellcolor{jku_grey!10}$0.388^{\pm 0.101}$ & 
    \cellcolor{jku_grey!10}$0.487^{\pm 0.170}$ &
    \cellcolor{jku_grey!10}$0.489^{\pm 0.024}$ &
    \cellcolor{jku_green!20}$0.504^{\pm 0.012}$ \\

    H3K4ME3 & MCC $\uparrow$ &
    $0.158^{\pm 0.072}$ & 
    $0.352^{\pm 0.077}$ & 
    $0.410^{\pm 0.033}$ &
    \cellcolor{jku_green!20}$0.549^{\pm 0.056}$ & 
    $0.440^{\pm 0.202}$ & 
    $0.544^{\pm 0.045}$ &
    $0.520^{\pm 0.019}$ &
    $0.537^{\pm 0.012}$ \\

    \cellcolor{jku_grey!10}H3K79ME3 & \cellcolor{jku_grey!10}MCC $\uparrow$ &
    \cellcolor{jku_grey!10}$0.496^{\pm 0.042}$ & 
    \cellcolor{jku_grey!10}$0.613^{\pm 0.030}$ & 
    \cellcolor{jku_grey!10}$0.626^{\pm 0.046}$ & 
    \cellcolor{jku_grey!10}$0.672^{\pm 0.048}$ & 
    \cellcolor{jku_grey!10}$0.676^{\pm 0.026}$ & 
    \cellcolor{jku_grey!10}$0.697^{\pm 0.077}$  &
    \cellcolor{jku_grey!10}$0.662^{\pm 0.011}$ &
    \cellcolor{jku_green!20}$0.697^{\pm 0.007}$ \\

    H3K9AC & MCC $\uparrow$ &
    $0.420^{\pm 0.063}$ &
    $0.542^{\pm 0.029}$ &
    $0.562^{\pm 0.040}$ &
    $0.581^{\pm 0.061}$ & 
    $0.604^{\pm 0.048}$ & 
    $0.622^{\pm 0.030}$ &
    $0.622^{\pm 0.013}$ &
    \cellcolor{jku_green!20}$0.627^{\pm 0.008}$ \\

    \cellcolor{jku_grey!10}H4 & \cellcolor{jku_grey!10}MCC $\uparrow$ &
    \cellcolor{jku_grey!10}$0.732^{\pm 0.076}$ &
    \cellcolor{jku_grey!10}$0.796^{\pm 0.027}$ & 
    \cellcolor{jku_grey!10}$0.799^{\pm 0.025}$ & 
    \cellcolor{jku_grey!10}$0.763^{\pm 0.044}$ & 
    \cellcolor{jku_grey!10}$0.789^{\pm 0.020}$ & 
    \cellcolor{jku_grey!10}$0.811^{\pm 0.022}$  &
    \cellcolor{jku_grey!10}$0.793^{\pm 0.011}$ &
    \cellcolor{jku_green!20}$0.813^{\pm 0.008}$ \\

    H4AC & MCC $\uparrow$ &
    $0.273^{\pm 0.063}$ &
    $0.463^{\pm 0.041}$ & 
    $0.495^{\pm 0.032}$ & 
    $0.564^{\pm 0.038}$ &
    $0.525^{\pm 0.240}$ & 
    \cellcolor{jku_green!20}$0.621^{\pm 0.054}$  &
    $0.558^{\pm 0.018}$ &
    $0.583^{\pm 0.014}$ \\

    \midrule

    \multicolumn{10}{l}{\textit{Regulatory Annotation}} \\

    \cellcolor{jku_grey!10}Enhancer & \cellcolor{jku_grey!10}MCC $\uparrow$ &
    \cellcolor{jku_grey!10}$0.451^{\pm 0.108}$ & 
    \cellcolor{jku_grey!10}$0.516^{\pm 0.098}$ & 
    \cellcolor{jku_green!20}$0.548^{\pm 0.144}$ & 
    \cellcolor{jku_grey!10}$0.517^{\pm 0.117}$ & 
    \cellcolor{jku_grey!10}$0.491^{\pm 0.066}$ & 
    \cellcolor{jku_grey!10}$0.546^{\pm 0.073}$ &
    \cellcolor{jku_grey!10}$0.375^{\pm 0.030}$ &
    \cellcolor{jku_grey!10}$0.545^{\pm 0.024}$ \\

    Enhancer Types & MCC $\uparrow$ &
    $0.309^{\pm 0.134}$ & 
    $0.423^{\pm 0.051}$ & 
    $0.424^{\pm 0.132}$ & 
    $0.386^{\pm 0.185}$ & 
    $0.416^{\pm 0.095}$ & 
    $0.439^{\pm 0.054}$  &
    $0.444^{\pm 0.046}$ &
    \cellcolor{jku_green!20}$0.466^{\pm 0.011}$ \\

    \cellcolor{jku_grey!10}Promoter: All & \cellcolor{jku_grey!10}F1 $\uparrow$ &
    \cellcolor{jku_grey!10}$0.954^{\pm 0.006}$ & 
    \cellcolor{jku_grey!10}$0.971^{\pm 0.006}$ & 
    \cellcolor{jku_green!20}$0.976^{\pm 0.006}$ & 
    \cellcolor{jku_grey!10}$0.960^{\pm 0.005}$ & 
    \cellcolor{jku_grey!10}$0.967^{\pm 0.004}$ & 
    \cellcolor{jku_grey!10}$0.970^{\pm 0.004}$ &
    \cellcolor{jku_grey!10}$0.962^{\pm 0.002}$ &
    \cellcolor{jku_grey!10}$0.967^{\pm 0.001}$ \\

    NonTATA & F1 $\uparrow$ &
    $0.955^{\pm 0.010}$ & 
    $0.972^{\pm 0.005}$ &
    \cellcolor{jku_green!20}$0.976^{\pm 0.006}$ & 
    $0.959^{\pm 0.011}$ &
    $0.968^{\pm 0.006}$ &
    $0.968^{\pm 0.010}$  &
    $0.963^{\pm 0.002}$  &
    $0.970^{\pm 0.001}$ \\

    \cellcolor{jku_grey!10}TATA & \cellcolor{jku_grey!10}F1 $\uparrow$ &
    \cellcolor{jku_grey!10}$0.960^{\pm 0.023}$ & 
    \cellcolor{jku_grey!10}$0.955^{\pm 0.021}$ & 
    \cellcolor{jku_green!20}$0.966^{\pm 0.013}$ & 
    \cellcolor{jku_grey!10}$0.944^{\pm 0.040}$ & 
    \cellcolor{jku_grey!10}$0.957^{\pm 0.015}$ & 
    \cellcolor{jku_grey!10}$0.953^{\pm 0.016}$  &
    \cellcolor{jku_grey!10}$0.948^{\pm 0.006}$ &
    \cellcolor{jku_grey!10}$0.952^{\pm 0.005}$ \\

    \midrule

    \multicolumn{10}{l}{\textit{Splice Site Annotation}} \\

    \cellcolor{jku_grey!10}All & \cellcolor{jku_grey!10}Accuracy $\uparrow$ &
    \cellcolor{jku_grey!10}$0.848^{\pm 0.019}$ & 
    \cellcolor{jku_grey!10}$0.939^{\pm 0.009}$ & 
    \cellcolor{jku_green!20}$0.983^{\pm 0.008}$ & 
    \cellcolor{jku_grey!10}$0.956^{\pm 0.011}$ & 
    \cellcolor{jku_grey!10}$0.927^{\pm 0.021}$ & 
    \cellcolor{jku_grey!10}$0.940^{\pm 0.027}$  &
    \cellcolor{jku_grey!10}$0.965^{\pm 0.006}$ &
    \cellcolor{jku_grey!10}$0.974^{\pm 0.004}$ \\

    Acceptor & F1 $\uparrow$ &
    $0.914^{\pm 0.028}$ &
    $0.975^{\pm 0.006}$ & 
    \cellcolor{jku_green!20}$0.981^{\pm 0.011}$ & 
    $0.958^{\pm 0.010}$ & 
    $0.936^{\pm 0.077}$ & 
    $0.937^{\pm 0.033}$  &
    $0.970^{\pm 0.005}$  &
    $0.953^{\pm 0.008}$ \\

    \cellcolor{jku_grey!10}Donor & \cellcolor{jku_grey!10}F1 $\uparrow$ &
    \cellcolor{jku_grey!10}$0.906^{\pm 0.027}$ & 
    \cellcolor{jku_grey!10}$0.963^{\pm 0.006}$ & 
    \cellcolor{jku_green!20}$0.985^{\pm 0.022}$ & 
    \cellcolor{jku_grey!10}$0.949^{\pm 0.024}$ & 
    \cellcolor{jku_grey!10}$0.948^{\pm 0.025}$ & 
    \cellcolor{jku_grey!10}$0.874^{\pm 0.289}$  &
    \cellcolor{jku_grey!10}$0.962^{\pm 0.004}$ &
    \cellcolor{jku_grey!10}$0.951^{\pm 0.005}$ \\

    \bottomrule
\end{tabular}

%% file: tables/dna_genomics_downstream_direct_comparision.tex
\begin{adjustbox}{angle=0}
\begin{tabular}{lcc||cc}
    \toprule
    \makecell{} & 
    \makecell{Mamba-PH-500k\\} & 
    \makecell{xLSTM-PH-500k\\} &
    \makecell{Mamba-PS-500k\\} & 
    \makecell{xLSTM-PS-500k\\} \\
    
    \midrule
    
    \cellcolor{jku_grey!10}Mouse Enhancers & 
    \cellcolor{jku_grey!10}$0.754 ^{\pm 0.074}$ &  
    \cellcolor{jku_green!20}$0.780 ^{\pm 0.018}$& 
    \cellcolor{jku_green!20}$0.793 ^{\pm 0.058}$ & 
    \cellcolor{jku_grey!10}$0.778^{\pm 0.007}$\\
    
    {Coding. vs. Intergenomic} & 
    $0.915 ^{\pm 0.003}$ & 
    \cellcolor{jku_green!20}$0.931^{\pm 0.001}$ &
    $0.910 ^{\pm 0.003}$ & 
    \cellcolor{jku_green!20}$0.934^{\pm 0.002}$\\
    
    \cellcolor{jku_grey!10}{Human vs. Worm} & 
    \cellcolor{jku_green!20}$0.973 ^{\pm 0.001}$ & 
    \cellcolor{jku_grey!10}$0.965 ^{\pm 0.001}$&
    \cellcolor{jku_green!20}$0.968 ^{\pm 0.002}$ & 
    \cellcolor{jku_grey!10}$0.956 ^{\pm 0.001}$\\
    
    {Human Enhancers Cohn} & 
    \cellcolor{jku_green!20}$0.747 ^{\pm 0.004}$ &
    $0.742 ^{\pm 0.005}$ &
    \cellcolor{jku_green!20}$0.745 ^{\pm 0.007}$ &
    $0.734 ^{\pm 0.005}$\\
    
    \cellcolor{jku_grey!10}{Human Enhancers Ensemble} &
    \cellcolor{jku_grey!10}$0.893 ^{\pm 0.008}$ &
    \cellcolor{jku_green!20}$0.920^{\pm 0.001}$ &
    \cellcolor{jku_grey!10}$0.900 ^{\pm 0.006}$ & 
    \cellcolor{jku_green!20}$0.902^{\pm 0.004}$\\
    
    {Human Regulatory} & 
    $0.872 ^{\pm 0.011}$ & 
    \cellcolor{jku_green!20}$0.872^{\pm 0.002}$ &
    \cellcolor{jku_green!20}$0.873 ^{\pm 0.007}$ &
    $0.869^{\pm 0.005}$ \\
    
    \cellcolor{jku_grey!10}{Human OCR Ensembl} &
    \cellcolor{jku_green!20}$0.828 ^{\pm 0.006}$ &
    \cellcolor{jku_grey!10}$0.826^{\pm 0.002}$&
    \cellcolor{jku_green!20}$0.818 ^{\pm 0.006}$ & 
    \cellcolor{jku_grey!10}$0.800^{\pm 0.002}$\\
    
    {Human NonTATA Promoters} & 
    $0.946 ^{\pm 0.007}$ & 
    \cellcolor{jku_green!20}$0.951^{\pm 0.004}$&
    $0.945 ^{\pm 0.010}$ & 
    \cellcolor{jku_green!20}$0.949^{\pm 0.001}$\\
    \bottomrule

\end{tabular}
\end{adjustbox}

%% file: tables/dna_nucleotide_downstream_hyperparams.tex
\begin{tabular}{lcccccccc}
    \toprule
    & \multicolumn{2}{c}{DNA-xLSTM-Ph} & \multicolumn{2}{c}{DNA-xLSTM-PS} \\
    & \makecell{Learning Rate} & \makecell{Batch Size} & \makecell{Learning Rate} & \makecell{Batch Size} \\
    
    \midrule
    
    \textit{Histone Markers} \\
    
    \cellcolor{jku_grey!10}H3 & 
    \cellcolor{jku_grey!10}8e-4 & \cellcolor{jku_grey!10}128 & \cellcolor{jku_grey!10}4e-4 & \cellcolor{jku_grey!10}64 \\
    
    H3K14AC & 
    6e-4 & 
    128 & 
    4e-4 & 
    64 \\
    
    \cellcolor{jku_grey!10}H3K36ME3 & 
    \cellcolor{jku_grey!10}6e-4 & 
    \cellcolor{jku_grey!10}64 & 
    \cellcolor{jku_grey!10}4e-4 & 
    \cellcolor{jku_grey!10}64 \\
    
    H3K4ME1 & 
    8e-4 & 
    128 & 
    1e-3 & 
    128 \\
    
    \cellcolor{jku_grey!10}H3K4ME2 & 
    \cellcolor{jku_grey!10}6e-4 & 
    \cellcolor{jku_grey!10}64 & 
    \cellcolor{jku_grey!10}2e-3 & 
    \cellcolor{jku_grey!10}512 \\
    
    H3K4ME3 & 
    8e-4 & 
    128 & 
    1e-3 & 
    512 \\
    
    \cellcolor{jku_grey!10}H3K79ME3 & 
    \cellcolor{jku_grey!10}1e-3 & 
    \cellcolor{jku_grey!10}128 & 
    \cellcolor{jku_grey!10}4e-4 & 
    \cellcolor{jku_grey!10}64 \\
    
    H3K9AC & 
    4e-4 & 
    64 & 
    1e-3 & 
    128 \\
    
    \cellcolor{jku_grey!10}H4 & 
    \cellcolor{jku_grey!10}8e-4 & 
    \cellcolor{jku_grey!10}64 & 
    \cellcolor{jku_grey!10}6e-4 & 
    \cellcolor{jku_grey!10}64 \\
    
    H4AC & 
    4e-4 & 
    64 & 
    1e-3 & 
    128 \\
    
    \midrule
    
    \textit{Regulatory Annotation} \\
    
    \cellcolor{jku_grey!10}Enhancers & 
    \cellcolor{jku_grey!10}2e-3 & 
    \cellcolor{jku_grey!10}512 & 
    \cellcolor{jku_grey!10}2e-3 & 
    \cellcolor{jku_grey!10}512 \\
    
    Enhancers Types & 
    2e-3 & 
    512 & 
    2e-3 & 
    512 \\
    
    \cellcolor{jku_grey!10}Promoter All & 
    \cellcolor{jku_grey!10}4e-4 & 
    \cellcolor{jku_grey!10}64 & 
    \cellcolor{jku_grey!10}1e-3 & 
    \cellcolor{jku_grey!10}128 \\
    
    Promoter No TATA & 
    1e-3 & 
    128 & 
    1e-3 & 
    128 \\
    
    \cellcolor{jku_grey!10}Promoter TATA & 
    \cellcolor{jku_grey!10}3e-3 & 
    \cellcolor{jku_grey!10}128 & 
    \cellcolor{jku_grey!10}1e-3 & 
    \cellcolor{jku_grey!10}128 \\
    
    \midrule
    
    \textit{Splice Site Annotation} \\
    
    \cellcolor{jku_grey!10}Splice Sites All & 
    \cellcolor{jku_grey!10}8e-4 & 
    \cellcolor{jku_grey!10}64 & 
    \cellcolor{jku_grey!10}2e-3 & 
    \cellcolor{jku_grey!10}128 \\
    
    Splice Sites Acceptor & 
    2e-3 & 
    128 & 
    2e-3 & 
    128 \\
    
    \cellcolor{jku_grey!10}Splice Sites Donors & 
    \cellcolor{jku_grey!10}3e-3 & 
    \cellcolor{jku_grey!10}128 & 
    \cellcolor{jku_grey!10}2e-3 & 
    \cellcolor{jku_grey!10}128 \\
    
    \bottomrule
\end{tabular}

%% file: tables/proteins/protein_xlstm_hparams.tex
\begin{tabular}{lccc}
    \toprule 
    Hyperparameter & Prot-xLSTM-26M & Prot-xLSTM-102M & Prot-Transformer++-26M \\
    \midrule
    Embedding dimension    & 512   & 1024 & 512 \\
    Context length & 2\textsuperscript{11},2\textsuperscript{17}\tnote{a} & 2\textsuperscript{11-17}\tnote{a} & 2\textsuperscript{11} \\
    Number of blocks       & 16    & 16 & 6 \\
    m/sLSTM ratio         & [0:1], \textbf{[1:0]}, [1:7]\tnote{b}  & [1:0] & - \\
    Conv 1D kernel size    & 4     & 4  & - \\
    QKV projection blocksize & 4 & 4 & - \\
    Number of heads        & 4     & 4 & 8 \\
    Up projection dimension&  1024 & 2048 & 2176 \\
    Norm bias and linear bias & False & False & False \\
    Position embeddings  & -, AbPE, \textbf{RoPE} & RoPE & RoPE \\
    \bottomrule    
\end{tabular}

%% file: tables/proteins/protein_small_model_train_loss.tex
    \footnotesize
    \centering

    \begin{tabular}{cccc}
        \toprule
        \textbf{Model type} & \textbf{\#p (M)} & \textbf{Positional Encodings} & \textbf{Train loss}\\
        \midrule
        \textbf{Mamba} & \cellcolor{jku_grey!10}27.7 & \cellcolor{jku_grey!10}AbPE & \cellcolor{jku_grey!10}2.623 \\
        \midrule
        \textbf{Transformer++} & 26.4 & RoPE & 2.568 \\
        \midrule
        \multirow{2}{*}{\textbf{sLSTM}}
        & \cellcolor{jku_grey!10}25.8 & \cellcolor{jku_grey!10}- & \cellcolor{jku_grey!10}2.694\\
        & 26.3 & AbPE & 2.688 \\
        \midrule
        \multirow{4}{*}{\textbf{mLSTM}}
        & \cellcolor{jku_grey!10}25.9 & \cellcolor{jku_grey!10}- & \cellcolor{jku_grey!10}2.569 \\
        & 26.4 & AbPE & 2.545 \\
        & \cellcolor{jku_grey!10}25.9 & \cellcolor{jku_grey!10}RoPE & \cellcolor{jku_green!20}\textbf{2.524}\\
        & \textit{102} & \textit{RoPE} & \cellcolor{jku_green!20}\textit{\textbf{2.482}} \\
        \midrule
        \multirow{2}{*}{\textbf{xLSTM}}
        & \cellcolor{jku_grey!10}25.9 & \cellcolor{jku_grey!10}- & \cellcolor{jku_grey!10}2.554 \\
        & 26.4 & AbPE & 2.551 \\    
        \bottomrule
    \end{tabular}

%% file: tables/proteins/protein_sequence_evaluation_abs.tex
\begin{tabular}{l|c|cccc} 
    \toprule 
    & Natural Seqences & Prot-xLSTM & ProtMamba &  Prot-xLSTM & ProtMamba  \\
    & & -26M & -28M & -102M & -107M \\ 
    \midrule 
    Sequence length & $211^{\pm 28}$ & $290^{\pm 36}$ & $326^{\pm 43}$ & $286^{\pm 38}$ & $276^{\pm 40}$ \\
     \rowcolor{jku_grey!10} Min. Hamming $\downarrow$ & $0.51^{\pm 0.04}$ &	$0.55^{\pm 0.05}$ & $0.64^{\pm 0.04}$ & \cellcolor{jku_green!20} $0.44^{\pm 0.07}$ & $0.56^{\pm 0.03}$ \\ 
    HMMER $\uparrow$ & $96^{\pm 25}$ & \cellcolor{jku_green!20} $182^{\pm 56}$ & $122^{\pm 50}$ & $165^{\pm 45}$ & $163^{\pm 45}$ \\
    \rowcolor{jku_grey!10} pLDDT $\uparrow$ & $0.81^{\pm 0.03}$ & $0.79^{\pm 0.04}$ & $0.67^{\pm 0.07}$ & \cellcolor{jku_green!20} $0.80^{\pm 0.03}$ & \cellcolor{jku_green!20} $0.80^{\pm 0.03}$ \\
    pTM $\uparrow$ &  $0.77^{\pm 0.06}$ & $0.74^{\pm 0.06}$ & $0.54^{\pm 0.10}$ & \cellcolor{jku_green!20} $0.75^{\pm 0.06}$ & $0.74^{\pm 0.06}$ \\
    \bottomrule 
\end{tabular}

%% file: tables/proteins/protein_sequence_evaluation_ppl_r.tex
\begin{tabular}{lcccc} 
    \toprule 
    & Prot-xLSTM-26M & ProtMamba-28M &  Prot-xLSTM-102M & ProtMamba-107M  \\ 
    \midrule 
    \rowcolor{jku_grey!10} Min. Hamming & $0.53^{\pm 0.10}$ & $0.41^{\pm 0.10}$ & \cellcolor{jku_green!20} $0.59^{\pm 0.08}$	& $0.57^{\pm 0.11}$  \\ 
    HMMER Score & \cellcolor{jku_green!20} $0.59^{\pm 0.06}$ &	$0.54^{\pm 0.07}$ &	$0.54^{\pm 0.07}$ &	$0.57^{\pm 0.09}$ \\
    \rowcolor{jku_grey!10}pLDDT & \cellcolor{jku_green!20} $0.66^{\pm 0.05}$ & $0.53^{\pm 0.07}$	& $0.60^{\pm 0.08}$	& $0.62^{\pm 0.08}$\\
    pTM & \cellcolor{jku_green!20} $0.59^{\pm 0.06}$ &	$0.44^{\pm 0.08}$ &	$0.55^{\pm 0.07}$ & $0.57^{\pm 0.07}$\\
    \bottomrule 
\end{tabular}

%% file: tables/proteins/proteingym.tex
    \rowcolors{1}{}{jku_grey!10}
    \footnotesize
    \centering
    \begin{threeparttable}
        \begin{tabular}{lllcc}
            \toprule
            Model Type & Model & Reference & \#Params & Spearman $\rho$  \\
            \midrule
             Alignment-based & Site-Independant & \cite{hopf2017mutation} & - & 0.359 \\
             & EVE & \cite{frazer2021disease} & -\tnote{a} & 0.432 \\
             & GEMME & \cite{laine2019gemme}  & - & 0.455 \\
             \midrule
             Protein language model
             & Tranception L (w/o R) & \cite{notin2022tranception} & 700M & 0.374 \\
             (PLM) & VespaG & \cite{marquet2024expert-guided}& 3B & 0.458 \\ 
             & ProGen2 XL & \cite{nijkamp2023progen2}& 6B & 0.391 \\
             & ESM-2 & \cite{lin2023evolutionary} & 15B & 0.401 \\
             \midrule
             Alignment + PLM 
             & MSA-Transformer & \cite{rao2021msa} & 100M & 0.432 \\
             & Tranception L (w/ R) & \cite{notin2022tranception} & 700M & 0.434 \\
             & TranceptEVE L & \cite{notin2022trancepteve} & >700M\tnote{a} & 0.456 \\
             \midrule
             Homology-aware PLM
             & Prot-xLSTM & Ours & 26M & 0.411\tnote{b} \\
             & ProtMamba & \cite{sgarbossa2024protmamba} & 28M & 0.360\tnote{b} \\ 
             & Prot-xLSTM & Ours & 102M & 0.416\tnote{b} \\
             & ProtMamba & \cite{sgarbossa2024protmamba} & 107M & 0.415\tnote{b} \\
             & PoET & \cite{truong2023poet} & 201M & 0.470 \\    
             \midrule
             Inverse folding & ESM-IF1 & \cite{hsu2022learning} & 142M & 0.422 \\
             \midrule
             Structure + PLM 
             & SaProt & \cite{su2024saprot} & 35M & 0.407 \\
             & ProSST & \cite{li2024prosst} & 110M & \cellcolor{jku_green!20}0.507 \\
             & SaProt & \cite{su2024saprot} & 650M & 0.457 \\ 
             \bottomrule
        \end{tabular}
    \end{threeparttable}
    \begin{tablenotes}
        \item[a] EVE parameters depend on the size of a given MSA.\\ 
        \item[b] This work. All other values are retrieved from \href{https://github.com/OATML-Markslab/ProteinGym/blob/main/benchmarks/DMS_zero_shot/substitutions/Spearman/Summary_performance_DMS_substitutions_Spearman.csv}{ProteinGym} on 03/11/2024.
        
    \end{tablenotes}

%% file: main_arxiv.bbl
\begin{thebibliography}{}

\bibitem[{1000 Genomes Project Consortium}, 2010]{1000genomes2010}
{1000 Genomes Project Consortium} (2010).
\newblock A map of human genome variation from population scale sequencing.
\newblock {\em Nature}, 467(7319):1061.

\bibitem[Acosta et~al., 2022]{acosta2022multimodal}
Acosta, J.~N., Falcone, G.~J., Rajpurkar, P., and Topol, E.~J. (2022).
\newblock Multimodal biomedical ai.
\newblock {\em Nature Medicine}, 28(9):1773--1784.

\bibitem[Adilov, 2021]{adilov2021generative}
Adilov, S. (2021).
\newblock Generative pre-training from molecules.
\newblock {\em ChemRxiv preprint chemrxiv-2021-5fwjd}.

\bibitem[Ahdritz et~al., 2023]{ahdritz2023openproteinset}
Ahdritz, G., Bouatta, N., Kadyan, S., Jarosch, L., Berenberg, D., Fisk, I., Watkins, A., Ra, S., Bonneau, R., and AlQuraishi, M. (2023).
\newblock Openproteinset: Training data for structural biology at scale.
\newblock In {\em Advances in Neural Information Processing Systems (NeurIPS)}, volume~36, pages 4597--4609. Curran Associates, Inc.

\bibitem[Ahmad et~al., 2022]{ahmad2022chemberta}
Ahmad, W., Simon, E., Chithrananda, S., Grand, G., and Ramsundar, B. (2022).
\newblock {ChemBERTa-2}: Towards chemical foundation models.
\newblock {\em arXiv preprint arXiv:2209.01712}.

\bibitem[Alkin et~al., 2024]{alkin2024vision}
Alkin, B., Beck, M., P{\"o}ppel, K., Hochreiter, S., and Brandstetter, J. (2024).
\newblock Vision-{LSTM}: {xLSTM} as generic vision backbone.
\newblock {\em arXiv preprint arXiv:2406.04303}.

\bibitem[Alley et~al., 2019]{alley2019unified}
Alley, E.~C., Khimulya, G., Biswas, S., AlQuraishi, M., and Church, G.~M. (2019).
\newblock Unified rational protein engineering with sequence-based deep representation learning.
\newblock {\em Nature methods}, 16(12):1315--1322.

\bibitem[Anfinsen, 1973]{anfinsen1973principles}
Anfinsen, C.~B. (1973).
\newblock Principles that govern the folding of protein chains.
\newblock {\em Science}, 181(4096):223--230.

\bibitem[Arnold, 2018]{arnold2018directed}
Arnold, F.~H. (2018).
\newblock Directed evolution: bringing new chemistry to life.
\newblock {\em Angewandte Chemie}, 57(16):4143.

\bibitem[Ashley, 2016]{ashley2016towards}
Ashley, E.~A. (2016).
\newblock Towards precision medicine.
\newblock {\em Nature Reviews Genetics}, 17(9):507--522.

\bibitem[Ba et~al., 2016]{ba2016layernorm}
Ba, J.~L., Kiros, J.~R., and Hinton, G.~E. (2016).
\newblock Layer normalization.
\newblock {\em arXiv preprint arXiv:1607.06450}.

\bibitem[Bagal et~al., 2021]{bagal2021molgpt}
Bagal, V., Aggarwal, R., Vinod, P., and Priyakumar, U.~D. (2021).
\newblock {MolGPT}: molecular generation using a transformer-decoder model.
\newblock {\em Journal of Chemical Information and Modeling}, 62(9):2064--2076.

\bibitem[Bavarian et~al., 2022]{bavarian2022efficienttraininglanguagemodels}
Bavarian, M., Jun, H., Tezak, N., Schulman, J., McLeavey, C., Tworek, J., and Chen, M. (2022).
\newblock Efficient training of language models to fill in the middle.
\newblock {\em arXiv preprint arXiv:2207.14255}.

\bibitem[Beck et~al., 2024]{beck2024xlstm}
Beck, M., P{\"o}ppel, K., Spanring, M., Auer, A., Prudnikova, O., Kopp, M., Klambauer, G., Brandstetter, J., and Hochreiter, S. (2024).
\newblock {xLSTM}: Extended long short-term memory.
\newblock {\em Neural Infomation Processing Systems (NeurIPS)}.

\bibitem[Beck et~al., 2025]{beck2025unlocking}
Beck, M., Pöppel, K., and Hochreiter, S. (2025).
\newblock Unlocking the power of recurrence for efficient xlstm kernels.
\newblock {\em Under preparation}.

\bibitem[Benegas et~al., 2023]{benegas2023dna}
Benegas, G., Batra, S.~S., and Song, Y.~S. (2023).
\newblock {DNA} language models are powerful predictors of genome-wide variant effects.
\newblock {\em Proceedings of the National Academy of Sciences}, 120(44):e2311219120.

\bibitem[Bepler and Berger, 2019]{bepler2019learning}
Bepler, T. and Berger, B. (2019).
\newblock Learning protein sequence embeddings using information from structure.
\newblock {\em International Conference on Learning Representations (ICLR)}, 7.

\bibitem[Bouwman and de~Laat, 2015]{bouwman2015getting}
Bouwman, B.~A. and de~Laat, W. (2015).
\newblock Getting the genome in shape: the formation of loops, domains and compartments.
\newblock {\em Genome Biology}, 16(1):154.

\bibitem[Bran and Schwaller, 2023]{bran2023transformers}
Bran, A.~M. and Schwaller, P. (2023).
\newblock Transformers and large language models for chemistry and drug discovery.
\newblock {\em arXiv preprint arXiv:2310.06083}.

\bibitem[Brandes et~al., 2023]{brandes2023genome}
Brandes, N., Goldman, G., Wang, C.~H., Ye, C.~J., and Ntranos, V. (2023).
\newblock Genome-wide prediction of disease variant effects with a deep protein language model.
\newblock {\em Nature Genetics}, 55(9):1512--1522.

\bibitem[Brandes et~al., 2022]{brandes2022proteinbert}
Brandes, N., Ofer, D., Peleg, Y., Rappoport, N., and Linial, M. (2022).
\newblock {ProteinBERT}: a universal deep-learning model of protein sequence and function.
\newblock {\em Bioinformatics}, 38(8):2102--2110.

\bibitem[Brown et~al., 2020]{brown20gpt3}
Brown, T., Mann, B., Ryder, N., Subbiah, M., Kaplan, J.~D., Dhariwal, P., Neelakantan, A., Shyam, P., Sastry, G., Askell, A., Agarwal, S., Herbert-Voss, A., Krueger, G., Henighan, T., Child, R., Ramesh, A., Ziegler, D., Wu, J., Winter, C., Hesse, C., Chen, M., Sigler, E., Litwin, M., Gray, S., Chess, B., Clark, J., Berner, C., McCandlish, S., Radford, A., Sutskever, I., and Amodei, D. (2020).
\newblock Language models are few-shot learners.
\newblock In {\em Advances in Neural Information Processing Systems (NeurIPS)}, volume~33, pages 1877--1901. Curran Associates, Inc.

\bibitem[Bubeck et~al., 2023]{bubeck2023sparks}
Bubeck, S., Chandrasekaran, V., Eldan, R., Gehrke, J., Horvitz, E., Kamar, E., Lee, P., Lee, Y.~T., Li, Y., Lundberg, S., et~al. (2023).
\newblock Sparks of artificial general intelligence: Early experiments with {GPT}-4.
\newblock {\em arXiv preprint arXiv:2303.12712}.

\bibitem[Chithrananda et~al., 2020]{chithrananda2020chemberta}
Chithrananda, S., Grand, G., and Ramsundar, B. (2020).
\newblock {ChemBERTa}: large-scale self-supervised pretraining for molecular property prediction.
\newblock {\em arXiv preprint arXiv:2010.09885}.

\bibitem[Choromanski et~al., 2021]{choromanski2021rethinking}
Choromanski, K.~M., Likhosherstov, V., Dohan, D., Song, X., Gane, A., Sarlos, T., Hawkins, P., Davis, J.~Q., Mohiuddin, A., Kaiser, L., Belanger, D.~B., Colwell, L.~J., and Weller, A. (2021).
\newblock Rethinking attention with performers.
\newblock In {\em International Conference on Learning Representations (ICLR)}, volume~9.

\bibitem[Church et~al., 2011]{church2011modernizing}
Church, D.~M., Schneider, V.~A., Graves, T., Auger, K., Cunningham, F., Bouk, N., Chen, H.-C., Agarwala, R., McLaren, W.~M., Ritchie, G.~R., et~al. (2011).
\newblock Modernizing reference genome assemblies.
\newblock {\em PLoS biology}, 9(7):e1001091.

\bibitem[Dalla-Torre et~al., 2023]{dalla2023nucleotide}
Dalla-Torre, H., Gonzalez, L., Mendoza-Revilla, J., Carranza, N.~L., Grzywaczewski, A.~H., Oteri, F., Dallago, C., Trop, E., de~Almeida, B.~P., Sirelkhatim, H., et~al. (2023).
\newblock The nucleotide transformer: Building and evaluating robust foundation models for human genomics.
\newblock {\em BioRxiv}, pages 2023--01.

\bibitem[Dao, 2024]{dao2023flashattention2}
Dao, T. (2024).
\newblock Flashattention-2: {F}aster attention with better parallelism and work partitioning.
\newblock In {\em International Conference on Learning Representations (ICLR)}, volume~12.

\bibitem[Devlin et~al., 2019]{devlin2019bert}
Devlin, J., Chang, M.-W., Lee, K., and Toutanova, K. (2019).
\newblock {BERT}: Pre-training of deep bidirectional transformers for language understanding.
\newblock In {\em Proceedings of the 2019 Conference of the North {A}merican Chapter of the Association for Computational Linguistics: Human Language Technologies, Volume 1 (Long and Short Papers)}, pages 4171--4186, Minneapolis, Minnesota. Association for Computational Linguistics.

\bibitem[Elnaggar et~al., 2021]{elnaggar2021prottrans}
Elnaggar, A., Heinzinger, M., Dallago, C., Rehawi, G., Wang, Y., Jones, L., Gibbs, T., Feher, T., Angerer, C., Steinegger, M., et~al. (2021).
\newblock {ProtTrans}: Toward understanding the language of life through self-supervised learning.
\newblock {\em IEEE Transactions on Pattern Analysis and Machine Intelligence}, 44(10):7112--7127.

\bibitem[Ferruz et~al., 2022]{ferruz2022protgpt2}
Ferruz, N., Schmidt, S., and Höcker, B. (2022).
\newblock {ProtGPT2} is a deep unsupervised language model for protein design.
\newblock {\em Nature Communications}, 13:4348.

\bibitem[Flam-Shepherd et~al., 2022]{flam2022language}
Flam-Shepherd, D., Zhu, K., and Aspuru-Guzik, A. (2022).
\newblock Language models can learn complex molecular distributions.
\newblock {\em Nature Communications}, 13(1):3293.

\bibitem[Frazer et~al., 2021]{frazer2021disease}
Frazer, J., Notin, P., Dias, M., Gomez, A., Min, J.~K., Brock, K., Gal, Y., and Marks, D.~S. (2021).
\newblock Disease variant prediction with deep generative models of evolutionary data.
\newblock {\em Nature}, 599(7883):91--95.
\newblock Publisher: Nature Publishing Group.

\bibitem[Geng et~al., 2022]{geng2022deep}
Geng, Q., Yang, R., and Zhang, L. (2022).
\newblock A deep learning framework for enhancer prediction using word embedding and sequence generation.
\newblock {\em Biophysical Chemistry}, 286:106822.

\bibitem[Gers et~al., 1999]{gers1999learning}
Gers, F.~A., Schmidhuber, J., and Cummins, F. (1999).
\newblock Learning to forget: continual prediction with {LSTM}.
\newblock In {\em 9th {International} {Conference} on {Artificial} {Neural} {Networks} {ICANN} '99}, pages 850--855. IET.

\bibitem[G{\'o}mez-Bombarelli et~al., 2018]{gomez2018automatic}
G{\'o}mez-Bombarelli, R., Wei, J.~N., Duvenaud, D., Hern{\'a}ndez-Lobato, J.~M., S{\'a}nchez-Lengeling, B., Sheberla, D., Aguilera-Iparraguirre, J., Hirzel, T.~D., Adams, R.~P., and Aspuru-Guzik, A. (2018).
\newblock Automatic chemical design using a data-driven continuous representation of molecules.
\newblock {\em ACS Central Science}, 4(2):268--276.

\bibitem[Gre{\v{s}}ov{\'a} et~al., 2023]{grevsova2023genomic}
Gre{\v{s}}ov{\'a}, K., Martinek, V., {\v{C}}ech{\'a}k, D., {\v{S}}ime{\v{c}}ek, P., and Alexiou, P. (2023).
\newblock Genomic benchmarks: a collection of datasets for genomic sequence classification.
\newblock {\em BMC Genomic Data}, 24(1):25.

\bibitem[Gu and Dao, 2023]{gu2024mamba}
Gu, A. and Dao, T. (2023).
\newblock Mamba: {L}inear-time sequence modeling with selective state spaces.
\newblock {\em arXiv preprint arXiv:2312.00752}.

\bibitem[Gu et~al., 2022]{gu2022s4}
Gu, A., Goel, K., and R\'{e}, C. (2022).
\newblock Efficiently modeling long sequences with structured state spaces.
\newblock In {\em International Conference on Learning Representations (ICLR)}, volume~10.

\bibitem[Gwak and Rho, 2022]{gwak2022vibe}
Gwak, H.-J. and Rho, M. (2022).
\newblock {ViBE}: a hierarchical {BERT} model to identify eukaryotic viruses using metagenome sequencing data.
\newblock {\em Briefings in Bioinformatics}, 23(4):bbac204.

\bibitem[He et~al., 2016]{he2016resnet}
He, K., Zhang, X., Ren, S., and Sun, J. (2016).
\newblock Deep residual learning for image recognition.
\newblock In {\em Proceedings CVPR}, pages 770--778.

\bibitem[Hoarfrost et~al., 2022]{hoarfrost2022deep}
Hoarfrost, A., Aptekmann, A., Farfa{\~n}uk, G., and Bromberg, Y. (2022).
\newblock Deep learning of a bacterial and archaeal universal language of life enables transfer learning and illuminates microbial dark matter.
\newblock {\em Nature Communications}, 13(1):2606.

\bibitem[Hochreiter and Schmidhuber, 1997]{hochreiter1997lstm}
Hochreiter, S. and Schmidhuber, J. (1997).
\newblock Long short-term memory.
\newblock {\em Neural Computation}, 9(8):1735--1780.

\bibitem[Honda et~al., 2019]{honda2019smiles}
Honda, S., Shi, S., and Ueda, H.~R. (2019).
\newblock {SMILES} transformer: Pre-trained molecular fingerprint for low data drug discovery.
\newblock {\em arXiv preprint arXiv:1911.04738}.

\bibitem[Hopf et~al., 2017]{hopf2017mutation}
Hopf, T.~A., Ingraham, J.~B., Poelwijk, F.~J., Sch{\"a}rfe, C.~P., Springer, M., Sander, C., and Marks, D.~S. (2017).
\newblock Mutation effects predicted from sequence co-variation.
\newblock {\em Nature Biotechnology}, 35(2):128--135.

\bibitem[Hsu et~al., 2022]{hsu2022learning}
Hsu, C., Verkuil, R., Liu, J., Lin, Z., Hie, B., Sercu, T., Lerer, A., and Rives, A. (2022).
\newblock Learning inverse folding from millions of predicted structures.
\newblock {\em bioRxiv}.

\bibitem[Hu et~al., 2022]{hu2022protein}
Hu, B., Xia, J., Zheng, J., Tan, C., Huang, Y., Xu, Y., and Li, S.~Z. (2022).
\newblock Protein language models and structure prediction: Connection and progression.
\newblock {\em arXiv preprint arXiv:2211.16742}.

\bibitem[Jastrz{\k{e}}bski et~al., 2016]{jastrzkebski2016learning}
Jastrz{\k{e}}bski, S., Le{\'s}niak, D., and Czarnecki, W.~M. (2016).
\newblock Learning to {SMILE(S)}.
\newblock {\em arXiv preprint arXiv:1602.06289}.

\bibitem[Ji et~al., 2021]{ji2021dnabert}
Ji, Y., Zhou, Z., Liu, H., and Davuluri, R.~V. (2021).
\newblock {DNABERT}: pre-trained bidirectional encoder representations from transformers model for {DNA}-language in genome.
\newblock {\em Bioinformatics}, 37(15):2112--2120.

\bibitem[Jumper et~al., 2021]{jumper2021highly}
Jumper, J., Evans, R., Pritzel, A., Green, T., Figurnov, M., Ronneberger, O., Tunyasuvunakool, K., Bates, R., Zidek, A., Potapenko, A., et~al. (2021).
\newblock Highly accurate protein structure prediction with alphafold.
\newblock {\em Nature}, 596(7873):583--589.

\bibitem[Karollus et~al., 2024]{karollus2024species}
Karollus, A., Hingerl, J., Gankin, D., Grosshauser, M., Klemon, K., and Gagneur, J. (2024).
\newblock Species-aware {DNA} language models capture regulatory elements and their evolution.
\newblock {\em Genome Biology}, 25(1):83.

\bibitem[Katharopoulos et~al., 2020]{katharopoulos2020transformers}
Katharopoulos, A., Vyas, A., Pappas, N., and Fleuret, F. (2020).
\newblock Transformers are {RNN}s: Fast autoregressive transformers with linear attention.
\newblock In {\em Proceedings of the 37th International Conference on Machine Learning (ICML)}, volume 119, pages 5156--5165. PMLR.

\bibitem[Kim et~al., 2023]{kim2023pubchem}
Kim, S., Chen, J., Cheng, T., Gindulyte, A., He, J., He, S., Li, Q., Shoemaker, B.~A., Thiessen, P.~A., Yu, B., et~al. (2023).
\newblock Pubchem 2023 update.
\newblock {\em Nucleic Acids Research}, 51(D1):D1373--D1380.

\bibitem[Kingma and Ba, 2015]{kingma2015adam}
Kingma, D.~P. and Ba, J. (2015).
\newblock Adam: A method for stochastic optimization.
\newblock In {\em International Conference on Learning Representations (ICLR)}, volume~3.

\bibitem[Laine et~al., 2019]{laine2019gemme}
Laine, E., Karami, Y., and Carbone, A. (2019).
\newblock {GEMME: A Simple and Fast Global Epistatic Model Predicting Mutational Effects}.
\newblock {\em Molecular Biology and Evolution}, 36(11):2604--2619.

\bibitem[Lander et~al., 2001]{lander2001initial}
Lander, E.~S., Linton, L.~M., Birren, B., Nusbaum, C., Zody, M.~C., Baldwin, J., Devon, K., Dewar, K., Doyle, M., Fitzhugh, W., et~al. (2001).
\newblock Initial sequencing and analysis of the human genome.
\newblock {\em Nature}, 409(6822):860--921.

\bibitem[Levr{\'e} et~al., 2018]{levre2018zinclick}
Levr{\'e}, D., Arcisto, C., Mercalli, V., and Massarotti, A. (2018).
\newblock Zinclick v. 18: expanding chemical space of 1, 2, 3-triazoles.
\newblock {\em Journal of Chemical Information and Modeling}, 59(5):1697--1702.

\bibitem[Li et~al., 2024]{li2024prosst}
Li, M., Tan, P., Ma, X., Zhong, B., Yu, H., Zhou, Z., Ouyang, W., Zhou, B., Hong, L., and Tan, Y. (2024).
\newblock Prosst: Protein language modeling with quantized structure and disentangled attention.
\newblock {\em bioRxiv}.

\bibitem[Lin et~al., 2023]{lin2023evolutionary}
Lin, Z., Akin, H., Rao, R., Hie, B., Zhu, Z., Lu, W., Smetanin, N., Verkuil, R., Kabeli, O., Shmueli, Y., dos Santos~Costa, A., Fazel-Zarandi, M., Sercu, T., Candido, S., and Rives, A. (2023).
\newblock Evolutionary-scale prediction of atomic-level protein structure with a language model.
\newblock {\em Science}, 379(6637):1123--1130.
\newblock Publisher: American Association for the Advancement of Science.

\bibitem[Lowe, 2012]{lowe2012extraction}
Lowe, D.~M. (2012).
\newblock {\em Extraction of chemical structures and reactions from the literature}.
\newblock PhD thesis, University of Cambridge.

\bibitem[Madani et~al., 2023]{madani2023large}
Madani, A., Krause, B., Greene, E.~R., and et~al. (2023).
\newblock Large language models generate functional protein sequences across diverse families.
\newblock {\em Nature Biotechnology}, 41:1099--1106.

\bibitem[Marquet et~al., 2024]{marquet2024expert-guided}
Marquet, C., Schlensok, J., Abakarova, M., Rost, B., and Laine, E. (2024).
\newblock Expert-guided protein language models enable accurate and blazingly fast fitness prediction.
\newblock {\em bioRxiv}.

\bibitem[Mayr et~al., 2018]{mayr2018large}
Mayr, A., Klambauer, G., Unterthiner, T., Steijaert, M., Wegner, J.~K., Ceulemans, H., Clevert, D.-A., and Hochreiter, S. (2018).
\newblock Large-scale comparison of machine learning methods for drug target prediction on {ChEMBL}.
\newblock {\em Chemical science}, 9(24):5441--5451.

\bibitem[Mazuz et~al., 2023]{mazuz2023molecule}
Mazuz, E., Shtar, G., Shapira, B., and Rokach, L. (2023).
\newblock Molecule generation using transformers and policy gradient reinforcement learning.
\newblock {\em Scientific Reports}, 13(1):8799.

\bibitem[Merrill et~al., 2024]{merrill2024illusion}
Merrill, W., Petty, J., and Sabharwal, A. (2024).
\newblock The illusion of state in state-space models.
\newblock In {\em Proceedings of the 41st International Conference on Machine Learning (ICML)}, volume 235, pages 35492--35506. PMLR.

\bibitem[Min et~al., 2022]{min2022rethinking}
Min, S., Lyu, X., Holtzman, A., Artetxe, M., Lewis, M., Hajishirzi, H., and Zettlemoyer, L. (2022).
\newblock Rethinking the role of demonstrations: What makes in-context learning work?
\newblock {\em arXiv preprint arXiv:2202.12837}.

\bibitem[Mirdita et~al., 2022]{mirdita2022colabfold}
Mirdita, M., Schütze, K., Moriwaki, Y., Heo, L., Ovchinnikov, S., and Steinegger, M. (2022).
\newblock {ColabFold}: making protein folding accessible to all.
\newblock {\em Nature Methods}, 19(6):679--682.
\newblock Publisher: Nature Publishing Group.

\bibitem[Nainala et~al., 2024]{nainala2024coconut}
Nainala, V.~C., Rajan, K., Kanakam, S. R.~S., Sharma, N., Wei{\ss}enborn, V., Schaub, J., and Steinbeck, C. (2024).
\newblock Coconut 2.0: A comprehensive overhaul and curation of the collection of open natural products database.
\newblock {\em ChemRxiv preprint chemrxiv-2024-fxq2s}.

\bibitem[Nguyen et~al., 2023]{nguyen2023hyenadna}
Nguyen, E., Poli, M., Faizi, M., Thomas, A., Wornow, M., Birch-Sykes, C., Massaroli, S., Patel, A., Rabideau, C., Bengio, Y., et~al. (2023).
\newblock {HyenaDNA}: Long-range genomic sequence modeling at single nucleotide resolution.
\newblock In {\em Advances in Neural Information Processing Systems (NeurIPS)}, volume~36, pages 43177--43201. Curran Associates, Inc.

\bibitem[Nijkamp et~al., 2023]{nijkamp2023progen2}
Nijkamp, E., Ruffolo, J.~A., Weinstein, E.~N., Naik, N., and Madani, A. (2023).
\newblock {ProGen2}: {Exploring} the boundaries of protein language models.
\newblock {\em Cell Systems}, 14(11):968--978.e3.
\newblock Publisher: Elsevier.

\bibitem[Notin et~al., 2022a]{notin2022tranception}
Notin, P., Dias, M., Frazer, J., Marchena-Hurtado, J., Gomez, A., Marks, D.~S., and Gal, Y. (2022a).
\newblock Tranception: protein fitness prediction with autoregressive transformers and inference-time retrieval.

\bibitem[Notin et~al., 2023]{notin2023proteingym}
Notin, P., Kollasch, A., Ritter, D., van Niekerk, L., Paul, S., Spinner, H., Rollins, N., Shaw, A., Orenbuch, R., Weitzman, R., Frazer, J., Dias, M., Franceschi, D., Gal, Y., and Marks, D. (2023).
\newblock {ProteinGym}: Large-scale benchmarks for protein fitness prediction and design.
\newblock In {\em Advances in Neural Information Processing Systems (NeurIPS)}, volume~36, pages 64331--64379. Curran Associates, Inc.

\bibitem[Notin et~al., 2022b]{notin2022trancepteve}
Notin, P., Niekerk, L.~V., Kollasch, A.~W., Ritter, D., Gal, Y., and Marks, D.~S. (2022b).
\newblock {TranceptEVE}: {Combining} {Family}-specific and {Family}-agnostic {Models} of {Protein} {Sequences} for {Improved} {Fitness} {Prediction}.

\bibitem[Notin et~al., 2024]{notin2024machine}
Notin, P., Rollins, N., Gal, Y., Sander, C., and Marks, D. (2024).
\newblock Machine learning for functional protein design.
\newblock {\em Nature Biotechnology}, 42(2):216--228.
\newblock Publisher: Nature Publishing Group.

\bibitem[Oubounyt et~al., 2019]{oubounyt2019deepromoter}
Oubounyt, M., Louadi, Z., Tayara, H., and Chong, K.~T. (2019).
\newblock Deepromoter: robust promoter predictor using deep learning.
\newblock {\em Frontiers in Genetics}, 10:286.

\bibitem[{\"O}z{\c{c}}elik et~al., 2024]{ozccelik2024chemical}
{\"O}z{\c{c}}elik, R., de~Ruiter, S., Criscuolo, E., and Grisoni, F. (2024).
\newblock Chemical language modeling with structured state space sequence models.
\newblock {\em Nature Communications}, 15(1):6176.

\bibitem[Papadatos et~al., 2010]{papadatos2010lead}
Papadatos, G., Alkarouri, M., Gillet, V.~J., Willett, P., Kadirkamanathan, V., Luscombe, C.~N., Bravi, G., Richmond, N.~J., Pickett, S.~D., Hussain, J., et~al. (2010).
\newblock Lead optimization using matched molecular pairs: inclusion of contextual information for enhanced prediction of herg inhibition, solubility, and lipophilicity.
\newblock {\em Journal of Chemical Information and Modeling}, 50(10):1872--1886.

\bibitem[Phaml et~al., 2005]{phaml2005qualitatively}
Phaml, T.~H., Tran, D.~H., Ho, T.~B., Satou, K., and Valiente, G. (2005).
\newblock Qualitatively predicting acetylation and methylation areas in {DNA} sequences.
\newblock {\em Genome Informatics}, 16(2):3--11.

\bibitem[Poli et~al., 2023]{poli2023hyena}
Poli, M., Massaroli, S., Nguyen, E., Fu, D.~Y., Dao, T., Baccus, S., Bengio, Y., Ermon, S., and R\'{e}, C. (2023).
\newblock Hyena hierarchy: {T}owards larger convolutional language models.
\newblock In {\em Proceedings of the 40th International Conference on Machine Learning (ICML)}, volume 202, pages 28043--28078. PMLR.

\bibitem[Press et~al., 2021]{press2021shortformer}
Press, O., Smith, N.~A., and Lewis, M. (2021).
\newblock Shortformer: Better language modeling using shorter inputs.
\newblock In Zong, C., Xia, F., Li, W., and Navigli, R., editors, {\em Proceedings of the 59th Annual Meeting of the Association for Computational Linguistics and the 11th International Joint Conference on Natural Language Processing}, volume~1, pages 5493--5505. Association for Computational Linguistics.

\bibitem[Preuer et~al., 2018]{preuer2018frechet}
Preuer, K., Renz, P., Unterthiner, T., Hochreiter, S., and Klambauer, G. (2018).
\newblock Fr{\'e}chet chemnet distance: a metric for generative models for molecules in drug discovery.
\newblock {\em Journal of Chemical Information and Modeling}, 58(9):1736--1741.

\bibitem[Quigley et~al., 2024]{quigley2024belka}
Quigley, I.~K., Blevins, A., Halverson, B.~J., and Wilkinson, N. (2024).
\newblock Belka: The big encoded library for chemical assessment.
\newblock In {\em NeurIPS 2024 Competition Track}.

\bibitem[Radford et~al., 2018]{radford2018gpt}
Radford, A., Narasimhan, K., Salimans, T., and Sutskever, I. (2018).
\newblock Improving language understanding by generative pre-training.
\newblock {\em OpenAI Blog}.

\bibitem[Radford et~al., 2019]{radford2019gpt2}
Radford, A., Wu, J., Child, R., Luan, D., Amodei, D., and Sutskever, I. (2019).
\newblock Language models are unsupervised multitask learners.
\newblock {\em OpenAI Blog}.

\bibitem[Rao et~al., 2019]{rao2019evaluating}
Rao, R., Bhattacharya, N., Thomas, N., Duan, Y., Chen, P., Canny, J., Abbeel, P., and Song, Y. (2019).
\newblock Evaluating protein transfer learning with {TAPE}.
\newblock {\em Advances in Neural Information Processing Systems (NeurIPS)}, 32:9689--9701.

\bibitem[Rao et~al., 2021]{rao2021msa}
Rao, R.~M., Liu, J., Verkuil, R., Meier, J., Canny, J., Abbeel, P., Sercu, T., and Rives, A. (2021).
\newblock {MSA} transformer.
\newblock In {\em Proceedings of the 38th International Conference on Machine Learning (ICML)}, volume 139, pages 8844--8856. PMLR.

\bibitem[Riesselman et~al., 2018]{riesselman2018deep}
Riesselman, A.~J., Ingraham, J.~B., and Marks, D.~S. (2018).
\newblock Deep generative models of genetic variation capture the effects of mutations.
\newblock {\em Nature Methods}, 15(10):816--822.

\bibitem[Rives et~al., 2021]{rives2021biological}
Rives, A., Meier, J., Sercu, T., Goyal, S., Lin, Z., Liu, J., Guo, D., Ott, M., Zitnick, C.~L., Ma, J., and Fergus, R. (2021).
\newblock Biological structure and function emerge from scaling unsupervised learning to 250 million protein sequences.
\newblock {\em Proceedings of the National Academy of Sciences}, 118(15).

\bibitem[Ross et~al., 2022]{ross2022large}
Ross, J., Belgodere, B., Chenthamarakshan, V., Padhi, I., Mroueh, Y., and Das, P. (2022).
\newblock Large-scale chemical language representations capture molecular structure and properties.
\newblock {\em Nature Machine Intelligence}, 4(12):1256--1264.

\bibitem[Scalzitti et~al., 2021]{scalzitti2021spliceator}
Scalzitti, N., Kress, A., Orhand, R., Weber, T., Moulinier, L., Jeannin-Girardon, A., Collet, P., Poch, O., and Thompson, J.~D. (2021).
\newblock Spliceator: multi-species splice site prediction using convolutional neural networks.
\newblock {\em BMC Bioinformatics}, 22:1--26.

\bibitem[Schiff et~al., 2024]{schiff2024caduceus}
Schiff, Y., Kao, C.~H., Gokaslan, A., Dao, T., Gu, A., and Kuleshov, V. (2024).
\newblock Caduceus: Bi-directional equivariant long-range {DNA} sequence modeling.
\newblock In {\em Proceedings of the 41st International Conference on Machine Learning (ICML)}, volume 235, pages 43632--43648. PMLR.

\bibitem[Schimunek et~al., 2023]{schimunek2023context}
Schimunek, J., Seidl, P., Friedrich, L., Kuhn, D., Rippmann, F., Hochreiter, S., and Klambauer, G. (2023).
\newblock Context-enriched molecule representations improve few-shot drug discovery.
\newblock In {\em International Conference on Learning Representations (ICLR)}, volume~11.

\bibitem[Schwaller et~al., 2019]{schwaller2019molecular}
Schwaller, P., Laino, T., Gaudin, T., Bolgar, P., Hunter, C.~A., Bekas, C., and Lee, A.~A. (2019).
\newblock Molecular transformer: a model for uncertainty-calibrated chemical reaction prediction.
\newblock {\em ACS Central Science}, 5(9):1572--1583.

\bibitem[Schwaller et~al., 2021]{schwaller2021mapping}
Schwaller, P., Probst, D., Vaucher, A.~C., Nair, V.~H., Kreutter, D., Laino, T., and Reymond, J.-L. (2021).
\newblock Mapping the space of chemical reactions using attention-based neural networks.
\newblock {\em Nature Machine Intelligence}, 3(2):144--152.

\bibitem[Segler et~al., 2018]{segler2018generating}
Segler, M.~H., Kogej, T., Tyrchan, C., and Waller, M.~P. (2018).
\newblock Generating focused molecule libraries for drug discovery with recurrent neural networks.
\newblock {\em ACS Central Science}, 4(1):120--131.

\bibitem[Seidl et~al., 2023]{seidl2023enhancing}
Seidl, P., Vall, A., Hochreiter, S., and Klambauer, G. (2023).
\newblock Enhancing activity prediction models in drug discovery with the ability to understand human language.
\newblock In {\em Proceedings of the 40th International Conference on Machine Learning (ICML)}, volume 202, pages 30458--30490. PMLR.

\bibitem[Sgarbossa et~al., 2024]{sgarbossa2024protmamba}
Sgarbossa, D., Malbranke, C., and Bitbol, A.-F. (2024).
\newblock {ProtMamba}: a homology-aware but alignment-free protein state space model.
\newblock {\em bioRxiv}, pages 2024--05.

\bibitem[Shrikumar et~al., 2017]{shrikumar2017reverse}
Shrikumar, A., Greenside, P., and Kundaje, A. (2017).
\newblock Reverse-complement parameter sharing improves deep learning models for genomics.
\newblock {\em BioRxiv}, page 103663.

\bibitem[Skuta et~al., 2017]{skuta2017probes}
Skuta, C., Popr, M., Muller, T., Jindrich, J., Kahle, M., Sedlak, D., Svozil, D., and Bartunek, P. (2017).
\newblock Probes \& drugs portal: an interactive, open data resource for chemical biology.
\newblock {\em Nature methods}, 14(8):759--760.

\bibitem[Srivastava et~al., 2015]{srivastava2015highway}
Srivastava, R.~K., Greff, K., and Schmidhuber, J. (2015).
\newblock Training very deep networks.
\newblock In {\em Advances in Neural Information Processing Systems (NeurIPS)}, volume~28, pages 2377--2385. Curran Associates, Inc.

\bibitem[Stanley et~al., 2021]{stanley2021fs}
Stanley, M., Bronskill, J.~F., Maziarz, K., Misztela, H., Lanini, J., Segler, M., Schneider, N., and Brockschmidt, M. (2021).
\newblock {FS-Mol}: A few-shot learning dataset of molecules.
\newblock In {\em Proceedings of the Neural Information Processing Systems Track on Datasets and Benchmarks}, volume~1.

\bibitem[Su et~al., 2024a]{su2021roformer}
Su, J., Ahmed, M., Lu, Y., Pan, S., Bo, W., and Liu, Y. (2024a).
\newblock {Roformer}: Enhanced transformer with rotary position embedding.
\newblock {\em Neurocomputing}, 568:127063.

\bibitem[Su et~al., 2024b]{su2024saprot}
Su, J., Han, C., Zhou, Y., Shan, J., Zhou, X., and Yuan, F. (2024b).
\newblock {SaProt}: Protein language modeling with structure-aware vocabulary.
\newblock In {\em International Conference on Learning Representations (ICLR)}, volume~12.

\bibitem[{The UniProt Consortium}, 2023]{uniprot2023uniprot}
{The UniProt Consortium} (2023).
\newblock Uniprot: the universal protein knowledgebase in 2023.
\newblock {\em Nucleic Acids Research}, 51(D1):D523--D531.

\bibitem[Touvron et~al., 2023]{touvron2023llama}
Touvron, H., Lavril, T., Izacard, G., Martinet, X., Lachaux, M.-A., Lacroix, T., Rozi\`{e}re, B., Goyal, N., Hambro, E., Azhar, F., Rodriguez, A., Joulin, A., Grave, E., and Lample, G. (2023).
\newblock {LLaMA}: Open and efficient foundation language models.
\newblock {\em arXiv preprint arXiv:2302.13971}.

\bibitem[Truong~Jr and Bepler, 2023]{truong2023poet}
Truong~Jr, T. and Bepler, T. (2023).
\newblock {PoET}: A generative model of protein families as sequences-of-sequences.
\newblock In {\em Advances in Neural Information Processing Systems (NeurIPS)}, volume~36, pages 77379--77415. Curran Associates, Inc.

\bibitem[Vaswani et~al., 2017]{vaswani2017attention}
Vaswani, A., Shazeer, N., Parmar, N., Uszkoreit, J., Jones, L., Gomez, A.~N., Kaiser, L., and Polosukhin, I. (2017).
\newblock Attention is all you need.
\newblock In {\em Advances in Neural Information Processing Systems (NeurIPS)}, volume~30, pages 5998--6008. Curran Associates, Inc.

\bibitem[Vincent et~al., 2010]{vincent2010denoisingae}
Vincent, P., Larochelle, H., Lajoie, I., Bengio, Y., and Manzagol, P.-A. (2010).
\newblock Stacked denoising autoencoders: {L}earning useful representations in a deep network with a local denoising criterion.
\newblock {\em Journal of Machine Learning Research}, 11(110):3371--3408.

\bibitem[Wang et~al., 2019a]{wang2019splicefinder}
Wang, R., Wang, Z., Wang, J., and Li, S. (2019a).
\newblock Splicefinder: ab initio prediction of splice sites using convolutional neural network.
\newblock {\em BMC Bioinformatics}, 20:1--13.

\bibitem[Wang et~al., 2019b]{wang2019smiles}
Wang, S., Guo, Y., Wang, Y., Sun, H., and Huang, J. (2019b).
\newblock {SMILES-BERT}: large scale unsupervised pre-training for molecular property prediction.
\newblock In {\em Proceedings of the 10th ACM International Conference on Bioinformatics, Computational Biology and Health Informatics}, pages 429--436.

\bibitem[Wang et~al., 2023]{wang2023cmolgpt}
Wang, Y., Zhao, H., Sciabola, S., and Wang, W. (2023).
\newblock {cMolGPT}: A conditional generative pre-trained transformer for target-specific de novo molecular generation.
\newblock {\em Molecules}, 28(11):4430.

\bibitem[Weininger, 1988]{weininger1988smiles}
Weininger, D. (1988).
\newblock {SMILES}, a chemical language and information system. 1. introduction to methodology and encoding rules.
\newblock {\em Journal of Chemical Information and Modeling}, 28(1):31--36.

\bibitem[Wu and He, 2020]{wu2020groupnorm}
Wu, Y. and He, K. (2020).
\newblock Group normalization.
\newblock {\em International Journal of Computer Vision}, 128(3):742--755.

\bibitem[Wu et~al., 2018]{wu2018moleculenet}
Wu, Z., Ramsundar, B., Feinberg, E.~N., Gomes, J., Geniesse, C., Pappu, A.~S., Leswing, K., and Pande, V. (2018).
\newblock Moleculenet: a benchmark for molecular machine learning.
\newblock {\em Chemical science}, 9(2):513--530.

\bibitem[Yang et~al., 2019]{yang2019protein}
Yang, K.~K., Wu, Z., and Arnold, F.~H. (2019).
\newblock Machine-learning-guided directed evolution for protein engineering.
\newblock {\em Nature Methods}, 16(8):687--694.

\bibitem[Yang et~al., 2022]{yang2022integrating}
Yang, M., Huang, L., Huang, H., Tang, H., Zhang, N., Yang, H., Wu, J., and Mu, F. (2022).
\newblock Integrating convolution and self-attention improves language model of human genome for interpreting non-coding regions at base-resolution.
\newblock {\em Nucleic Acids Research}, 50(14):e81--e81.

\bibitem[Yang et~al., 2024]{yang2024gated}
Yang, S., Wang, B., Shen, Y., Panda, R., and Kim, Y. (2024).
\newblock Gated linear attention transformers with hardware-efficient training.
\newblock In {\em Proceedings of the 41st International Conference on Machine Learning (ICML)}, volume 235, pages 56501--56523. PMLR.

\bibitem[Zdrazil et~al., 2023]{zdrazil2023chembl}
Zdrazil, B., Felix, E., Hunter, F., Manners, E.~J., Blackshaw, J., Corbett, S., de Veij, M., Ioannidis, H., Lopez, D.~M., Mosquera, J., Magarinos, M., Bosc, N., Arcila, R., Kizilören, T., Gaulton, A., Bento, A., Adasme, M., Monecke, P., Landrum, G., and Leach, A. (2023).
\newblock The {ChEMBL} database in 2023: a drug discovery platform spanning multiple bioactivity data types and time periods.
\newblock {\em Nucleic Acids Research}, 52(D1):D1180--D1192.

\bibitem[Zhang et~al., 2024]{zhang2024scientific}
Zhang, Q., Ding, K., Lyv, T., Wang, X., Yin, Q., Zhang, Y., Yu, J., Wang, Y., Li, X., Xiang, Z., et~al. (2024).
\newblock Scientific large language models: A survey on biological \& chemical domains.
\newblock {\em arXiv preprint arXiv:2401.14656}.

\bibitem[Zhavoronkov et~al., 2019]{zhavoronkov2019deep}
Zhavoronkov, A., Ivanenkov, Y.~A., Aliper, A., Veselov, M.~S., Aladinskiy, V.~A., Aladinskaya, A.~V., Terentiev, V.~A., Polykovskiy, D.~A., Kuznetsov, M.~D., Asadulaev, A., et~al. (2019).
\newblock Deep learning enables rapid identification of potent {DDR1} kinase inhibitors.
\newblock {\em Nature Biotechnology}, 37(9):1038--1040.

\bibitem[Zhou et~al., 2022]{zhou2022towards}
Zhou, H., Shrikumar, A., and Kundaje, A. (2022).
\newblock Towards a better understanding of reverse-complement equivariance for deep learning models in genomics.
\newblock In {\em Machine Learning in Computational Biology}, pages 1--33. PMLR.

\bibitem[Zhou et~al., 2024]{zhou2024dnabert}
Zhou, Z., Ji, Y., Li, W., Dutta, P., Davuluri, R.~V., and Liu, H. (2024).
\newblock {DNABERT}-2: Efficient foundation model and benchmark for multi-species genomes.
\newblock In {\em International Conference on Learning Representations (ICLR)}, volume~12.

\end{thebibliography}
